\input harvmac.tex

\def\sssec#1{$\underline{\rm #1}$}


\def\unlockat{\catcode`\@=11}

\def\lockat{\catcode`\@=12}

\unlockat


\def\newsec#1{\global\advance\secno by1\message{(\the\secno. #1)}
\global\subsecno=0\global\subsubsecno=0\eqnres@t\noindent
{\bf\the\secno. #1}
\writetoca{{\secsym} {#1}}\par\nobreak\medskip\nobreak}

\global\newcount\subsecno \global\subsecno=0
\def\subsec#1{\global\advance\subsecno by1\message{(\secsym\the\subsecno. #1)}
\ifnum\lastpenalty>9000\else\bigbreak\fi\global\subsubsecno=0
\noindent{\it\secsym\the\subsecno. #1}
\writetoca{\string\quad {\secsym\the\subsecno.} {#1}}
\par\nobreak\medskip\nobreak}

\global\newcount\subsubsecno \global\subsubsecno=0
\def\subsubsec#1{\global\advance\subsubsecno by1
\message{(\secsym\the\subsecno.\the\subsubsecno. #1)}
\ifnum\lastpenalty>9000\else\bigbreak\fi
\noindent\quad{\secsym\the\subsecno.\the\subsubsecno.}{#1}
\writetoca{\string\qquad{\secsym\the\subsecno.\the\subsubsecno.}{#1}}
\par\nobreak\medskip\nobreak}

\def\subsubseclab#1{\DefWarn#1\xdef #1{\noexpand\hyperref{}{subsubsection}%
{\secsym\the\subsecno.\the\subsubsecno}%
{\secsym\the\subsecno.\the\subsubsecno}}%
\writedef{#1\leftbracket#1}\wrlabeL{#1=#1}}
\lockat

\def\IB{\relax\hbox{$\inbar\kern-.3em{\rm B}$}}
\def\IC{\relax\hbox{$\inbar\kern-.3em{\rm C}$}}
\def\ID{\relax\hbox{$\inbar\kern-.3em{\rm D}$}}
\def\IE{\relax\hbox{$\inbar\kern-.3em{\rm E}$}}
\def\IF{\relax\hbox{$\inbar\kern-.3em{\rm F}$}}
\def\IG{\relax\hbox{$\inbar\kern-.3em{\rm G}$}}
\def\IGa{\relax\hbox{${\rm I}\kern-.18em\Gamma$}}
\def\IH{\relax{\rm I\kern-.18em H}}
\def\II{\relax{\rm I\kern-.18em I}}
\def\IK{\relax{\rm I\kern-.18em K}}
\def\IL{\relax{\rm I\kern-.18em L}}
\def\IP{\relax{\rm I\kern-.18em P}}
\def\IQ{\relax{\rm I\kern-.42em Q}}
\def\IR{\relax{\rm I\kern-.18em R}}
\def\IS{\relax{\rm I\kern-.18em S}}

\def\IZ{\relax\ifmmode\mathchoice{\hbox{\cmss
Z\kern-.4em Z}}{\hbox{\cmss Z\kern-.4em Z}}
{\lower.9pt\hbox{\cmsss Z\kern-.4em Z}}
{\lower1.2pt\hbox{\cmsss Z\kern-.4em Z}}\else{\cmss Z\kern-.4em Z}\fi}


\def\CA {{\cal A}}
\def\CB {{\cal B}}
\def\CC {{\cal C}}
\def\CD {{\cal D}}
\def\CE {{\cal E}}
\def\CF {{\cal F}}

\def\CH {{\cal H}}
\def\CI {{\cal I}}

\def\CL {{\cal L}}
\def\CM {{\cal M}}
\def\CN {{\cal N}}
\def\CO {{\cal O}}
\def\CP {{\cal P}}

\def\CR {{\cal R}}
\def\CS {{\cal S}}
\def\CT {{\cal T}}
\def\CU {{\cal U}}
\def\CV {{\cal V}}
\def\CW {{\cal W}}

\def\CZ {{\cal Z}}

\def\p{\partial}



\def\Det{{\rm Det}}

\def\Tr{\rm Tr}

\def\p{{\partial}}
\def\pbar{{\bar\partial}}
\def\eexp{{\bar{\bf e}}}

\font\manual=manfnt \def\dbend{\lower3.5pt\hbox{\manual\char127}}

\def\c{\cdot}

\def\half {{1\over 2}}

\def\ch{{\rm ch}}

\def\Det{{\rm Det}}

\def\inbar{\,\vrule height1.5ex width.4pt depth0pt}


\def\lieg{{\underline{\bf g}}}

\def\lieh{{\underline{\bf h}}}

\def\liet{{\underline{\bf t}}}



\font\cmss=cmss10 \font\cmsss=cmss10 at 7pt

\def\boxit#1{\vbox{\hrule\hbox{\vrule\kern8pt
\vbox{\hbox{\kern8pt}\hbox{\vbox{#1}}\hbox{\kern8pt}}
\kern8pt\vrule}\hrule}}

\def\mathboxit#1{\vbox{\hrule\hbox{\vrule\kern8pt\vbox{\kern8pt
\hbox{$\displaystyle #1$}\kern8pt}\kern8pt\vrule}\hrule}}


\def\inbar{\,\vrule height1.5ex width.4pt depth0pt}

\font\cmss=cmss10 \font\cmsss=cmss10 at 7pt


\def\a1{{\cal A}^{1,1}}

\def\hi{\chi^{2,0}}


%

\lref\pauldan{L.~Alvarez-Gaume, Commun.Math.Phys.{\bf 90} (1983)161\semi
D.~Friedan, P.~Windey, Nucl.Phys. B235 (1984) 395 }
\lref\ashvarz{A.S.~Schwarz, Lett. Math. Phys. {\bf 2} (1978) 247}
\lref\aniemi{A.~Hietamaki, A.~Niemi, A.~Morozov, K.~Palo
``Geometry Of $\CN=1/2$ Supersymmetry And The
Atiyah-Singer
Theorem'', Phys.Lett. {\bf B263} (1991) 417 }
\lref\arnold{V.~Arnol'd, ``Mathematical Methods of Classical Mechanics'',
Springer-Verlag, 1980}
\lref\konman{M.~Kontsevich, Yu.~Manin, ``Gromov-Witten classes and
enumerative geometry'', CMP {\bf 164} (1994) 525}
\lref\konenu{M.~Kontsevich, ``Enumeration of rational curves via
torus action'', hep-th/9405035}
\lref\konrec{M.~Kontsevich, ``Deformation Quantization of Poisson Manifolds'',
dg-ga/  and talk at Strings'97}
\lref\givental{A.~Givental, ``Equivariant Gromov-Witten Invariants'',
 alg-geom/9603021\semi
``A mirror theorem for toric complete intersections'', alg-geom/9701016}
\lref\frmor{R.~Friedman, J.~Morgan, ``Smooth Four-Manifolds
and Complex Surfaces'', Ergebnisse der. Math. und Grenz. {\bf 3},
Springer-Verlag, New York, 1994}
\lref\fulmac{W.~Fulton, R.~MacPherson, ``A Compactification of
Configuration
Spaces'', Annals of Math. (1993)}
\lref\beilginz{A.~Beilinson, V.~Ginzburg, ``Resolution of Diagonals, Homotopy
Algebra and Moduli Spaces'', preprint (1993?)}
\lref\ginzburg{V. Ginzburg, M. Kapranov, and E. Vasserot,
``Langlands Dualtiy for Surfaces,'' IAS preprint}

\lref\gottsh{L.~G\"ottsche, Math. Ann. 286 (1990) 193}
\lref\gottshi{L.~G\"ottsche, ``Modular forms and Donaldson invariants
for $4$-manifolds with $b_{2}^{+}=1$'', alg-geom/9506018}
\lref\gothuy{L.~G\"ottsche and D. Huybrechts,
``Hodge numbers of moduli spaces of stable
bundles on $K3$ surfaces,'' alg-geom/9408001}
\lref\gz{L.~G\"ottsche, D.~Zagier, ``Jacobi forms and the structure
of Donaldson invariants for $4$-manifolds with $b_{2}^{+}=1$'',
 alg-geom/9612020}
\lref\borch{R.~Borcherds, ``Automorphic forms with singularities on
Grassmanians'', alg-geom/9609022}
\lref\fintstern{R.~ Fintushel and R.J.~ Stern,
``The Blowup Formula For Donaldson
Invariants'', alg-geom/9405002; Annals of Math. {\bf 143}(1996) 529}

\lref\GrHa{P.~ Griffiths and J.~ Harris,
{\it Principles of Algebraic geometry},p. 445, J.Wiley and Sons, 1978. }
\lref\grojn{I. Grojnowski, ``Instantons and
affine algebras I: the Hilbert scheme and
vertex operators,'' alg-geom/9506020.}
\lref\adhmfk{I. Grojnowski,
A. Losev, G. Moore, N. Nekrasov, S. Shatashvili,
``ADHM and the Frenkel-Kac construction,'' in preparation}

\lref\hirz{F. Hirzebruch and T. Hofer, Math. Ann. 286 (1990)255}

\lref\gn{Valeri~ A. Gritsenko,  Viacheslav~ V. Nikulin,
``K3 surfaces, Lorentzian Kac--Moody algebras and Mirror Symmetry,''
alg-geom/9510008}

\lref\milnor{J. Milnor, ``A unique decomposition
theorem for 3-manifolds,'' Amer. Jour. Math, (1961) 1}

\lref\milnsta{J. Milnor, J.Stasheff, ``Characteristic Classes'',
Princeton University Press, Princeton, New Jersey, USA, 1974}

\lref\ffeta{S.~ Ferrara et al., ``Prepotential and Monodromies in $N=2$
Heterotic String,'' hep-th/9504034.}
\lref\fhsv{S.~ Ferrara,  J.A.~ Harvey,  A.~ Strominger,  C.~ Vafa,
``Second-Quantized Mirror Symmetry,''
hep-th/9505162.}

\lref\afgnti{
I. Antoniadis, S. Ferrara, E. Gava, K.S. Narain, and
T.R. Taylor,
``Duality Symmetries in $N=2$ Heterotic Superstring,''
hep-th/9510079}
\lref\AFT{I. Antoniadis, S.Ferrara, T.R.Taylor, ``
$N=2$ Heterotic Superstring and its Dual Five Dimensional
Theory'', hep-th/9511108}

\lref\Fre{P.~Fre,
``Lectures on Special Kahler Geometry and Electric--Magnetic Duality
Rotations'' ,
     Nucl.Phys.Proc.Suppl. 45BC (1996) 59-114,
     hep-th/9512043}
\lref\aspinwall{P.S.~ Aspinwall,  J.~ Louis,
``On the Ubiquity of K3 Fibrations in String Duality,''
hep-th/9510234}
\lref\polch{J.~Polchinski, ``Evaluation of the one-loop string
path integral'', Comm. Math. Phys. {\bf 104} (1986) 37\semi
B.~McClain and B.D.B.~Roth, ``Modular invariance for interacting
bosonic strings at finite temperature'', Comm. Math. Phys. {\bf 111} (1987)
539}
\lref\vrldy{R.~ Dijkgraaf, E.~ Verlinde and H.~ Verlinde,
``Counting Dyons in $N=4$ String Theory'', CERN-TH/96-170,
hepth/9607}
\lref\vrlsq{E. Verlinde and H. Verlinde,
``Conformal Field Theory and Geometric Quantization,''
in {\it Strings '89},Proceedings
of the Trieste Spring School on Superstrings,
3-14 April 1989, M. Green, et. al. Eds. World
Scientific, 1990}
\lref\verlabl{E. Verlinde, ``Global Aspects of
Electric-Magnetic Duality,'' hep-th/9506011}
\lref\ver{E. Verlinde, Nucl. Phys. {\bf B} 300 (1988) 360}


\lref\gerasimov{A.~ Gerasimov, ``Localization in GWZW and Verlinde
formula'', UUITP 16/1993, hepth/9305090}


\lref\BlThlgt{M.~ Blau and G.~ Thompson, ``Lectures on 2d Gauge
Theories: Topological Aspects and Path
Integral Techniques", Presented at the
Summer School in Hogh Energy Physics and
Cosmology, Trieste, Italy, 14 Jun - 30 Jul
1993, hep-th/9310144.}

\lref\btverlinde{M.~ Blau, G.~ Thomson,
``Derivation of the Verlinde Formula from Chern-Simons Theory and the
$G/G$
   model'',Nucl. Phys. {\bf B}408 (1993) 345-390 }
\lref\btqmech{M.~ Blau , G.~ Thompson, ``Topological gauge theories
from supersymmetric quantum mechanics on
spaces of connections'', Int. J. Mod. Phys. A8 (1993) 573-586}
\lref\klbck{T.P.~Killingback, ``Two dimensional topological
gravity and intersection theory on the moduli space of holomorphic
bundles'', CERN-TH.5911/1990, November 1990}

\lref\Candelas{P.~ Candelas, X.~ De la Ossa, A.~ Font, S.~ Katz,
and D.~ Morrison,
``Mirror Symmetry for Two Parameter Models - I,'' Nucl.~Phys.~
{\bf B}416 (1994) 481, hep-th/9308083.}

\lref\Cardoso{Gabriel Lopes Cardoso,  Gottfried Curio,  Dieter Lust,
Thomas Mohaupt,  Soo-Jong Rey,
``BPS Spectra and Non--Perturbative Couplings in N=2,4 Supersymmetric String
Theories,''
hep-th/9512129}
\lref\matone{G.~Bonelli, M.~Matone, Phys. Rev. Lett. {\bf 77} (1996)
4712-4715; hep-th/9605090}
\lref\matins{M.~Matone, ``Instantons and recursion relations in
in $\CN=2$ susy gauge theory'', Phys. Lett. {\bf B}357(1995) 342,
hep-th/9506102}
\lref\contchl{
K.~Saito, Publ. RIMS, Kyoto Univ. 19 (1983) 1231\semi
B. Blok, A. Varchenko, Int.J.Mod.Phys.{\bf A7} (1992) 1467-1490 \semi
R.  Dijkgraaf, talk on RIMS conference "Infinite dimensional
analysis", July 1993 \semi
A.~Losev, Theor.Math.Phys. 95 (1993) 595\semi
A.~Losev,  in
"Integrable models
 and Strings ", 172,
Proceedings of 1993 Helsinki Conference,
Springer-Verlag, 1995 \semi
A.~Losev, I.~Polyubin, Int.J.Mod.Phys. {\bf A10} (1995) 4161}


\lref\diss{N.~ Nekrasov, PhD. Thesis, Princeton 1996}


\lref\hm{J.A.~ Harvey,  G.~ Moore,
``Algebras, BPS States, and Strings,'' hep-th/9510182, Nucl. Phys. {\bf B463}
(1996) 315}
\lref\hmii{J.A.~ Harvey,  G.~ Moore,
``On the Algebras of  BPS States,'' hep-th/9608???}
\lref\hmiii{J.A.~ Harvey,  G.~ Moore, ``Fivebrane instantons
and $R^2$ couplings in $\CN =4$ string theory'', hep-th/9610237\semi
``Exact gravitational
threshold corrections in FHSV model'', hep-th/9611176}
\lref\bost{L.~Alvarez-Gaume, J.B.~Bost , G.~Moore, P.~Nelson, C.~Vafa,
``Bosonization on higher genus Riemann surfaces,''
CMP {\bf 112} (1987) 503}
\lref\agmv{L.~Alvarez-Gaum\'e, C.~Gomez, G.~Moore,
and C.~Vafa, ``Strings in the Operator Formalism,''
Nucl. Phys. {\bf B} 303 (1988)455}
\lref\hms{hep-th/9501022,
 Reducing $S$- duality to $T$- duality, J. A. Harvey, G. Moore and A.
Strominger}

\lref\fivedim{N.~Nekrasov, ``Five dimensional gauge theories and relativistic
integrable systems'', hep-th/9609219}
\lref\instsm{A.~Lawrence, N.~Nekrasov, ``Instanton Sums and Five Dimensional
Gauge Theory'', hep-th/9706025}
\lref\lomns{A. Losev, G. Moore, N. Nekrasov,
S. Shatashvili, unpublished.}.
\lref\cssev{L.~Baulieu, A.~Losev, N.~Nekrasov, ``Chern-Simons
and Twisted Supersymmetry in Higher Dimensions'', hep-th/9707174}
\lref\CMR{S.~Cordes, G.~Moore, and S.~Ramgoolam,
`` Lectures on 2D Yang Mills theory, Equivariant
Cohomology, and Topological String Theory,''
hep-th/9411210}
\lref\elitzur{S. Elitzur, G. Moore,
A. Schwimmer, and N. Seiberg,
``Remarks on the Canonical Quantization of the Chern-Simons-
Witten Theory,'' Nucl. Phys. {\bf B326}(1989)108 \semi
G. Moore and N. Seiberg,
``Lectures on Rational Conformal Field Theory'',
in {\it Strings'89}, Proceedings
of the Trieste Spring School on Superstrings,
3-14 April 1989, M. Green, et. al. Eds. World
Scientific, 1990}

\lref\adhmfk{I. Grojnowski,
A. Losev, G. Moore, N. Nekrasov, S. Shatashvili,
``ADHM and the Frenkel-Kac construction,'' in preparation}

\lref\hypvol{G.~Moore, N.~Nekrasov, S.~Shatashvili,
``Volumes of  Hyperkahler Quotients,'' in preparation}
\lref\fdrcft{A. Losev, G. Moore, N. Nekrasov, S. Shatashvili, in
preparation.}
\lref\cenexts{A. Losev, G. Moore, N. Nekrasov, S. Shatashvili,
``Central Extensions of Gauge Groups Revisited,''
hep-th/9511185.}
\lref\taming{G. Moore and N. Seiberg,
``Taming the conformal zoo,'' Phys. Lett.
{\bf 220 B} (1989) 422}


\lref\locmir{S.~Katz, A.~Klemm, C.~Vafa,
``Geometric Engineering of Quantum Field Theories'',
hep-th/9609239, Nucl.Phys. B497 (1997) 173-195}

\lref\KLM{A.~Klemm, W.~Lerche, and P.~Mayr, ``K3-Fibrations and
Heterotic-Type II String Duality,'' hep-th/9506112.}

\lref\KV{S.~Kachru and C.~Vafa,
`` Exact Results for N=2 Compactifications of Heterotic Strings,''
hep-th/9505105; Nucl. Phys. B450 (1995) 69-89}

\lref\KKLMV{S.~Kachru, A.~Klemm, W.~Lerche, P.~Mayr and C.~Vafa,
``Nonperturbative Results on the Point Particle
 Limit of N=2 Heterotic String Compactifications,''
hep-th/9508155.}

\lref\kawai{Toshiya Kawai,
``$N=2$ heterotic string threshold correction, $K3$
surface and generalized Kac-Moody superalgebra,''
hep-th/9512046}


\lref\HET{D. Gross, J. Harvey, E. Martinec, and R. Rohm, ``The
Heterotic String,'' {\it Phys. Rev. Lett.} {\bf {54}} (1985) 502.}

\lref\grjakiw{D.J.~ Gross, R.~ Jackiw, ....
``Towards the Theory of Strong Interactions''}

\lref\grwilczek{D.J.~ Gross, F.~ Wilczek, ``Asymptotically Free
Gauge Theories. 1'', Phys. Rev. D8 (1973) 3633-3652, \semi
``Asymptotically Free
Gauge Theories. 2'', Phys.Rev.D9:980-993,1974. }

\lref\grcurrent{D.J.~ Gross, C.H.~ Lewellyn Smith, ``High-Energy
Neutrino-Nucleon Scattering, Current Algebra and Partons'',
Nucl.Phys.{\bf B}14 (1969) 337-347}

\lref\grcoleman{S.~ Coleman,
								R.~ Jackiw, D.J.~ Gross, ``Fermion
								Avatars of Sugawara Model''}


\lref\recqcd{W.~ A.~ Bardeen, ``Self-Dual Yang-Mills, Integrability
and Multi-Parton Amplitudes'',
Fermilab - Conf - -95-379-T, Aug 1995,
Presented at Yukawa
International Seminar '95:
`From the Standard Model to Grand
Unified Theories', Kyoto, Japan, 21-25 Aug 1995. \semi
D. Cangemi, ``Self-dual Yang-Mills
Theory and One-Loop Like-Helicity QCD
Multi-Gluon Amplitudes,'' hep-th/9605208.}


\lref\gwdzki{K. Gawedzki, ``Topological Actions in Two-Dimensional
Quantum Field Theories,'' in {\it Nonperturbative
Quantum Field Theory}, G. 't Hooft, A. Jaffe, et. al. , eds. ,
Plenum 1988}


\lref\shatashi{S. Shatashvili,
Theor. and Math. Physics, 71, 1987, p. 366}
\lref\gmps{A. Gerasimov, A. Morozov, M. Olshanetskii,
 A. Marshakov, S. Shatashvili, ``Wess-Zumino-Witten model as a theory of
free fields,'' Int. J. Mod. Phys. A5 (1990) 2495-2589}


\lref\vafagas{C. Vafa,
``Gas of D-Branes and Hagedorn Density of BPS States'',
hep-th/9511088}

\lref\vafa{C. Vafa, ``Conformal theories and punctured
surfaces,'' Phys.Lett.199B:195,1987 }

\lref\VaWi{C.~ Vafa and E.~ Witten, ``A Strong Coupling Test of
$S$-Duality",
hep-th/9408074.}

\lref\giveon{hep-th/9502057,
 ``$S$-Duality in $\CN=4$ Yang-Mills Theories with General Gauge Groups'',
 Luciano Girardello, Amit Giveon, Massimo Porrati, and Alberto Zaffaroni}

\lref\OldLG{C. Vafa, ``String Vacua and Orbifoldized LG Models,''
{\it Mod. Phys. Lett.} {\bf A4} (1989) 1169 \semi
K. Intriligator and C. Vafa, ``Landau-Ginzburg Orbifolds,"
{\it Nucl. Phys.}{\bf B339} (1990) 95 \semi
S. Cecotti, L. Girardello, and A. Pasquinucci,
``Non-perturbative Aspects and Exact Results for the $N=2$ Landau-Ginzburg
Models,'' Nucl. Phys. B338 (1989) 701, ``Singularity
Theory and N=2 Supersymmetry,'' Int. J. Mod. Phys. {\bf A6} (1991)
2427.}
\lref\bcov{M.~ Bershadsky, S.~ Cecotti, H.~ Ooguri and C.~ Vafa,
``Kodaira-Spencer Theory of Gravity and Exact Results for Quantum
String Amplitudes'', Comm. Math. Phys. 165(1994):311-428}
\lref\bjsv{M.~ Bershadsky, A.~ Johansen, V.~ Sadov and C.~ Vafa,
``Topological Reduction of 4D SYM to 2D $\sigma$--Models'',
hep-th/9501096}
\lref\ogvf{H. Ooguri and C. Vafa, ``Self-Duality
and $N=2$ String Magic,'' Mod.Phys.Lett. {\bf A5} (1990) 1389-1398\semi
``Geometry
of$N=2$ Strings,'' Nucl.Phys. {\bf B361}  (1991) 469-518.}
\lref\VafaQ{C. Vafa, ``Quantum Symmetries of String Vacua,''
Mod. Phys. Lett. {\bf A4} (1989) 1615. }
\lref\berk{N. Berkovits,
``Super-Poincare Invariant Superstring Field Theory''
hep-th/9503099}
\lref\SING{C. Vafa, ``Strings and Singularities,''  hep-th/9310069}
\lref\ttstar{S.~Cecotti, C.~Vafa, ``Topological Anti-Topological Fusion'',
Nucl. Phys. {\bf B367} (1991) 359}
\lref\gromov{M.~Gromov, Invent. Math. 82 (1985) 307}

\lref\ken{K.~Intriligator, ``Fusion residues'', Mod. Phys. Lett. {\bf A6}
(1991) 3541-3556, hep-th/9108005}

\lref\WitDonagi{R.~ Donagi, E.~ Witten,
``Supersymmetric Yang-Mills Theory and
Integrable Systems'', hep-th/9510101, Nucl.Phys.{\bf B}460 (1996) 299-334}
\lref\Witfeb{E.~ Witten, ``Supersymmetric Yang-Mills Theory On A
Four-Manifold,'' J. Math. Phys. {\bf 35} (1994) 5101.}
\lref\Witr{E.~ Witten, ``Introduction to Cohomological Field
Theories",
Lectures at Workshop on Topological Methods in Physics, Trieste, Italy,
Jun 11-25, 1990, Int. J. Mod. Phys. {\bf A6} (1991) 2775.}
\lref\Witgrav{E.~ Witten, ``Topological Gravity'', Phys.Lett.206B:601, 1988}
\lref\witaffl{I. ~ Affleck, J.A.~ Harvey and E.~ Witten,
	``Instantons and (Super)Symmetry Breaking
	in $2+1$ Dimensions'', Nucl. Phys. {\bf B}206 (1982) 413}
\lref\wittabl{E.~ Witten,  ``On $S$-Duality in Abelian Gauge Theory,''
hep-th/9505186; Selecta Mathematica {\bf 1} (1995) 383}
\lref\wittgr{E.~ Witten, ``The Verlinde Algebra And The Cohomology Of
The Grassmannian'',  hep-th/9312104}
\lref\wittenwzw{E. Witten, ``Nonabelian bosonization in
two dimensions,'' Commun. Math. Phys. {\bf 92} (1984)455 }
\lref\witgrsm{E. Witten, ``Quantum field theory,
grassmannians and algebraic curves,'' Commun.Math.Phys.113:529,1988}
\lref\wittjones{E. Witten, ``Quantum field theory and the Jones
polynomial,'' Commun.  Math. Phys., 121 (1989) 351. }
\lref\witttft{E.~ Witten, ``Topological Quantum Field Theory",
Commun. Math. Phys. {\bf 117} (1988) 353.}
\lref\wittmon{E.~ Witten, ``Monopoles and Four-Manifolds'', hep-th/9411102}
\lref\Witdgt{ E.~ Witten, ``On Quantum gauge theories in two
dimensions,''
Commun. Math. Phys. {\bf  141}  (1991) 153\semi
 ``Two dimensional gauge
theories revisited'', J. Geom. Phys. 9 (1992) 303-368}
\lref\Witgenus{E.~ Witten, ``Elliptic Genera and Quantum Field Theory'',
Comm. Math. Phys. 109(1987) 525. }
\lref\OldZT{E. Witten, ``New Issues in Manifolds of SU(3) Holonomy,''
{\it Nucl. Phys.} {\bf B268} (1986) 79 \semi
J. Distler and B. Greene, ``Aspects of (2,0) String Compactifications,''
{\it Nucl. Phys.} {\bf B304} (1988) 1 \semi
B. Greene, ``Superconformal Compactifications in Weighted Projective
Space,'' {\it Comm. Math. Phys.} {\bf 130} (1990) 335.}

\lref\bagger{E.~ Witten and J. Bagger, Phys. Lett.
{\bf 115B}(1982) 202}

\lref\witcurrent{E.~ Witten,``Global Aspects of Current Algebra'',
Nucl.Phys.B223 (1983) 422\semi
``Current Algebra, Baryons and Quark Confinement'', Nucl.Phys. B223 (1993)
433}
\lref\Wittreiman{S.B. Treiman,
E. Witten, R. Jackiw, B. Zumino, ``Current Algebra and
Anomalies'', Singapore, Singapore: World Scientific ( 1985) }
\lref\Witgravanom{L. Alvarez-Gaume, E.~ Witten, ``Gravitational Anomalies'',
Nucl.Phys.B234:269,1984. }

\lref\CHSW{P.~Candelas, G.~Horowitz, A.~Strominger and E.~Witten,
``Vacuum Configurations for Superstrings,'' {\it Nucl. Phys.} {\bf
B258} (1985) 46.}

\lref\AandB{E.~Witten, in ``Proceedings of the Conference on Mirror Symmetry",
MSRI (1991).}

\lref\phases{E.~Witten, ``Phases of N=2 Theories in Two Dimensions",
Nucl. Phys. {\bf B403} (1993) 159, hep-th/9301042}
\lref\WitKachru{S.~Kachru and E.~Witten, ``Computing The Complete Massless
Spectrum Of A Landau-Ginzburg Orbifold,''
Nucl. Phys. {\bf B407} (1993) 637, hep-th/9307038}

\lref\WitMin{E.~Witten,
``On the Landau-Ginzburg Description of N=2 Minimal Models,''
IASSNS-HEP-93/10, hep-th/9304026.}

\lref\twisted{E.~Witten.,  Comm. Math. Phys. {\bf 118} (1988) 411\semi
E. Witten, Nucl. Phys. {\bf B340} (1990) 281}
\lref\eguchi{T.~Eguchi and S.-K.~Yang,
Mod. Phys. Lett. {\bf A5} (1990) 1693.}

\lref\witseiii{N.~Seiberg, E.~Witten,
``Gauge Dynamics and Compactification to Three Dimensions,''
IASSNS-HEP-96-78, RU-96-55, hep-th/9607163}


\lref\KLT{ Vadim Kaplunovsky  ,  Jan Louis  ,  Stefan Theisen,
``Aspects of Duality in $\CN=2$ String Vacua,''
hep-th/9506110;
{\it Phys. Lett.} {\bf B357} (1995) 71.}
\lref\sty{J.~Sonnenschein, S.~Theisen, S.~Yankielowicz, Phys. Lett.
{\bf 367B} (1996) 145\semi
T.~Eguchi, S.-K.~Yang, Mod. Phys. Lett. {\bf A11} (1996) 131 }

\lref\dWKLL{B.~de Wit, V.~Kaplunovsky, J.~Louis, and D.~Lust,
``Perturbative Couplings of Vector
Multiplets in $\CN=2$ Heterotic String Vacua,''
hep-th/9504006.}

\lref\hkty{S.~Hosono, A.~Klemm, S.~Theisen, and S.T.~Yau, ``Mirror Symmetry,
Mirror Map and Applications to Calabi-Yau Hypersurfaces,''
Comm. Math. Phys. {\bf 167} (1995) 301, hep-th/9308122\semi
S.~Hosono, A.~Klemm, S.~Theisen, and S.T.~Yau,
``Mirror Symmetry, Mirror Map, and Applications to Complete Intersection
																							 Calabi-Yau Spaces,''
Nucl. Phys. {\bf B433} (1995) 501, hep-th/9406055.}


\lref\blzh{A. Belavin, V. Zakharov, ``Yang-Mills Equations as inverse
					 scattering
					 problem''Phys. Lett. B73, (1978) 53}

\lref\bpz{A.A. Belavin, A.M. Polyakov, A.B. Zamolodchikov,
					 ``Infinite conformal symmetry in two-dimensional quantum
					 field theory,'' Nucl.Phys.B241:333,1984}

\lref\atbott{M.~Atiyah, R.~Bott, ``
The Moment Map And
Equivariant Cohomology'', Topology {\bf 23} (1984) 1-28}
\lref\atbotti{M.~Atiyah, R.~Bott, ``The Yang-Mills Equations Over
Riemann Surfaces'', Phil. Trans. R.Soc. London A {\bf 308}, 523-615 (1982)}
\lref\atiyah{M. Atiyah, ``Green's Functions for
Self-Dual Four-Manifolds,'' Adv. Math. Suppl.
{\bf 7A} (1981)129}

\lref\AHS{M.~ Atiyah, N.~ Hitchin and I.~ Singer,
``Self-Duality in Four-Dimensional
Riemannian Geometry", Proc. Royal Soc. (London) {\bf A362} (1978)
425-461.}

\lref\fmlies{M.~ Atiyah and I.~ Singer,
``The index of elliptic operators IV,'' Ann. Math. {\bf 93} (1968) 119}

\lref\hitchin{N.~ Hitchin, ``Polygons and gravitons,''
Math. Proc. Camb. Phil. Soc, (1979){\bf 85} 465}

\lref\hklr{N.~Hitchin, Karlhede, Lindstrom, and M.~Rocek,
``Hyperkahler metrics and supersymmetry,''
Commun. Math. Phys. {\bf 108}(1987) 535}

\lref\hi{N.~ Hitchin, Duke Math. Journal, Vol. 54, No. 1 (1987)}


\lref\banks{T. Banks, ``Vertex Operators in 2D Dimensions,''
																		hep-th/9503145}


\lref\donii{S.~Donaldson, ``Infinite Determinants, Stable
						Bundles, and Curvature,''
						Duke Math. J. , {\bf 54} (1987) 231. }

\lref\BGS{J.-M.~Bismut, H.~Gillet, and C.~Soul\'e,
						``Analytic Torsion and  Holomorphic Determinant
						Bundles, I.II.III''
						CMP {\bf 115}(1988)49-78;79-126;301-351}

\lref\RS{D.~Ray and I.M.~Singer, ``R-torsion and the Laplasian on
						Riemannian Manifolds'', Adv. in Math. 7 (1971), 145-210}

\lref\quillen{D.~Quillen, ``Determinants of Cauchy-Riemann operators
						over a Riemann Surface'', Funk. Anal. i Prilozen. 19, 37-41 (1985) [=
						Func. Anal. Appl. 19, 31-34 (1986)]}

\lref\bismut{J.-M.~Bismut, ``The Atiyah-Singer theorem for families
						of Dirac Operators: two heat kernel proofs'', Invent. Math. 83, 91-151
						(1986)}

\lref\BottChern{R.~ Bott and S.~ Chern, ``Hermitian Vector Bundles
						and the Equidistribution of the zeroes of their holomorphic
						cross sections'', Acta Math. 114, 71-112 (1968)}


\lref\braam{P.J. Braam, A. Maciocia, and A. Todorov,
						``Instanton moduli as a novel map from tori to
						K3-surfaces,'' Inven. Math. {\bf 108} (1992) 419}

\lref\maciocia{A. Maciocia, ``Metrics on the moduli
						spaces of instantons over Euclidean 4-Space,''
						Commun. Math. Phys. {\bf 135}(1991) , 467}

\lref\martinecsix{E.~ Martinec, N.~ Warner, ``Integrability
						in $N=2$ Theories: A Proof'', hep-th/ }
\lref\cllnhrvy{C.~Callan and J.~Harvey, Nucl. Phys. {\bf B250}(1985)427}


\lref\galperin{A. Galperin, E. Ivanov, V. Ogievetsky,
E. Sokatchev, Ann. Phys. {\bf 185}(1988) 1}

\lref\evans{M. Evans, F. G\"ursey, V. Ogievetsky,
``From 2D conformal to 4D self-dual theories:
Quaternionic analyticity,''
Phys. Rev. {\bf D47}(1993)3496}

\lref\devchand{Ch. Devchand and V. Ogievetsky,
``Four dimensional integrable theories,'' hep-th/9410147}
\lref\devchandi{
Ch. Devchand and A.N. Leznov,
``B \"acklund transformation for supersymmetric self-dual theories
for semisimple  gauge groups and a hierarchy of
$A_1$ solutions,'' hep-th/9301098,
Commun. Math. Phys. {\bf 160} (1994) 551}


\lref\fs{L.~Faddeev and S.~Shatashvili, Theor. Math. Fiz., 60 (1984)
206}

\lref\faddeevlmp{L.D.~Faddeev, ``Some Comments on Many Dimensional
Solitons'',
Lett. Math. Phys., 1 (1976) 289-293.}

\lref\nov{ S.~Novikov, ``The Hamiltonian
formalism and many-valued analogue of  Morse theory'',
Russian Math.~Surveys 37:5 (1982),1-56}

\lref\fsi{L.~ Faddeev, Phys. Lett. B145 (1984) 81.}

\lref\mick{J. Mickellsson, CMP, 97 (1985) 361.}
\lref\mickelsson{J. Mickelsson, ``Kac-Moody groups, topology of
the Dirac  determinant bundle and fermionization,''
Commun. Math. Phys. {\bf 110}(1987)173.}


\lref\fz{I.~ Frenkel, I.~ Singer, unpublished.}
\lref\fk{I.~ Frenkel and B.~ Khesin, ``Four dimensional
realization of two dimensional current groups,'' Yale
preprint, July 1995, to appear in Commun. Math. Phys.}

\lref\ff{B. Feigin , E. Frenkel, N. Reshetikhin, ``Gaudin Model,
Critical Level and Bethe Ansatz'', CMP {\bf  166} (1995), 27-62}

\lref\efk{P.~ Etingof, I.~ Frenkel, A.~ Kirillov, Jr.,
``Spherical functions on affine Lie groups'', Yale preprint, 1994}

\lref\etingof{P.I. Etingof and I.B. Frenkel,
``Central Extensions of Current Groups in
Two Dimensions,'' Commun. Math.
Phys. {\bf 165}(1994) 429}


\lref\dnld{S. Donaldson, ``Anti self-dual Yang-Mills
connections over complex  algebraic surfaces and stable
vector bundles,'' Proc. Lond. Math. Soc,
{\bf 50} (1985)1}

\lref\biquard{O. Biquard, ``Sur les fibr\'es paraboliques
sur une surface complexe,'' to appear in J. Lond. Math.
Soc.}

\lref\DoKro{S.K.~ Donaldson and P.B.~ Kronheimer,
``The Geometry of Four-Manifolds'',
Clarendon Press, Oxford, 1990.}

\lref\donii{S. Donaldson, Duke Math. J. , {\bf 54} (1987) 231. }
\lref\doniii{S.~Donaldson, ``Polynomial Invariants For Smooth Four-Manifolds'',
Topology, {\bf 29} , No. 3, (1990) 257-315}

\lref\lbstmrn{J.M.F.~Labastida, P.M.~Llatas,
``Potentials for topological sigma models'', Phys. Lett. {\bf B}271 (1991), 101\semi
J.M.F.~Labastida, M.~Mari\~no, Nucl. Phys. {\bf B}448(1995) 373\semi
J.M.F.~Labastida, M.~Mari\~no, hep-th/9603169}
\lref\baryon{J.M.F.~Labastida, M.~Mari\~no, ``Twisted Baryon Number
in $\CN=2$ Supersymmetric QCD'', hep-th/9702054}


\lref\ShiBeta{V.~ Novikov, M.A. Shifman, A.I. Vainshtein, V.I. Zakharov,
``Exact  Gell-Mann-Low Function of Supersymmetric Yang-Mills
Theories From Instanton
Calculus'', Nucl.Phys.B229:381,1983. \semi

``Beta Function in Supersymmetric Gauge Theories:
Instantons Versus Traditional Approach'', Phys.Lett.166B:329,1986}

\lref\gwdzki{K.~ Gawedzki, ``Topological Actions in Two-Dimensional
Quantum Field Theories,'' in {\it Nonperturbative
Quantum Field Theory}, G. 't Hooft, A. Jaffe, et. al. , eds. ,
Plenum 1988}

\lref\ginzburg{V. Ginzburg, M. Kapranov, and E. Vasserot,
``Langlands Reciprocity for Algebraic Surfaces,'' q-alg/9502013}
\lref\giveon{
 ``S-Duality in N=4 Yang-Mills Theories with General Gauge Groups'',
hep-th/9502057,
 Luciano Girardello, Amit Giveon, Massimo Porrati, and Alberto
Zaffaroni
}


\lref\johansen{A. Johansen, ``Infinite Conformal
Algebras in Supersymmetric Theories on
Four Manifolds,'' hep-th/9407109, Nucl. Phys. B436 (1995) 291-341 \semi
``Realization of $W_{1+\infty}$ and Virasoro Algebras in Supersymmetric
   Theories on
Four Manifolds'',
hep-th/9406156, Mod. Phys. Lett. A9 (1994) 2611-2622\semi
``Twisting of
$N=1$ SUSY Gauge
Theories and Heterotic Topological Theories'',hep-th/9403017}


\lref\kronheimer{P. Kronheimer, ``The construction of ALE spaces as
hyper-kahler quotients,'' J. Diff. Geom. {\bf 28}1989)665}
\lref\kricm{P. Kronheimer, ``Embedded surfaces in
4-manifolds,'' Proc. Int. Cong. of
Math. (Kyoto 1990) ed. I. Satake, Tokyo, 1991}

\lref\krmw{P. Kronheimer and T. Mrowka,
``Gauge theories for embedded surfaces I,''
Topology {\bf 32} (1993) 773\semi
``Gauge theories for embedded surfaces II,''
preprint\semi
``Recurrence relations and asymptotics for four-manifold invariants'',
BAMS {\bf 30} (1994) 215\semi
``Embedded surfaces and the structure of Donaldson's polynomial invariants'',
preprint (1994)}

\lref\rade{J. Rade, ``Singular Yang-Mills fields. Local
theory I. '' J. reine ang. Math. , {\bf 452}(1994)111; {\it ibid}
{\bf 456}(1994)197; ``Singular Yang-Mills
fields-global theory,'' Intl. J. of Math. {\bf 5}(1994)491.}

\lref\biquard{O. Biquard, ``Sur les fibr\'es paraboliques
sur une surface complexe,'' to appear in J. Lond. Math.
Soc.}

\lref\uhlnb{K.~Uhlenbeck, ``Removable singularities in Yang-Mills
fields'', Comm. Math. Phys. {\bf 83} (1982) 11}

\lref\KN{P.~ Kronheimer and H.~ Nakajima,  ``Yang-Mills instantons
on ALE gravitational instantons,''  Math. Ann.
{\bf 288}(1990)263}

\lref\nakajima{H. Nakajima, ``Homology of moduli
spaces of instantons on ALE Spaces. I'' J. Diff. Geom.
{\bf 40}(1990) 105\semi
 ``Instantons on ALE spaces,
quiver varieties, and Kac-Moody algebras,'' preprint,\semi
``Gauge theory on resolutions of simple singularities
and affine Lie algebras,'' preprint.}

\lref\nakheis{H.~Nakajima, ``Heisenberg algebra and Hilbert schemes of
points on
projective surfaces ,'' alg-geom/9507012}


\lref\nair{V.P.Nair, ``K\"ahler-Chern-Simons Theory'', hep-th/9110042}
\lref\ns{V.P.~Nair and J.~Schiff,
``K\"ahler Chern Simons theory and symmetries of
antiselfdual equations'' Nucl.Phys.B371:329-352,1992\semi
``A K\"ahler Chern-Simons theory and quantization of the
moduli of antiselfdual instantons,''
Phys.~Lett.~B246:~423-429, 1990\semi
``Topological gauge theory and twistors,''
Phys.~Lett.~B233:~343, 1989}


\lref\park{J.-S. Park, ``Holomorphic Yang-Mills theory on compact
Kahler
manifolds,'' hep-th/9305095; Nucl. Phys. {\bf B423} (1994) 559;
J.-S.~ Park, ``$\CN=2$ Topological Yang-Mills Theory on Compact
K\"ahler
Surfaces", Commun. Math, Phys. {\bf 163} (1994) 113\semi
S. Hyun and J.-S.~ Park, ``$\CN=2$ Topological Yang-Mills Theories
and Donaldson
Polynomials", hep-th/9404009}
\lref\parki{S.~Hyun and J.-S.~Park,
``Holomorphic Yang-Mills Theory and Variation
of the Donaldson Invariants,'' hep-th/9503036}
\lref\parkii{S.~Hyun, J.~Park, J.-S.~Park, ``Topological QCD'',
hep-th/9503201 }


\lref\pohl{Pohlmeyer, Comm. Math. Phys. {\bf 72}(1980)37}

\lref\prseg{A.~Pressley and G.~Segal, "Loop Groups", Oxford Clarendon
Press, 1986}
\lref\segal{G. Segal, The definition of CFT}


\lref\fr{V.~ Fock, A.~ Rosly, ``Flat connections and polyubles'', Teor.
Math. Fiz. 1992}


\lref\miracles{M. Dine and N. Seiberg, ``Are (0,2) Models
String Miracles?,"
{\it Nucl. Phys.} {\bf B306} (1988) 137.}

\lref\GrSei{M. Green and N. Seiberg, ``Contact interactions in
superstring theory," Nucl. Phys. {\bf B299} (1988) 559.}

\lref\DSWW{M. Dine, N. Seiberg, X.G. Wen and E. Witten,
``Non-Perturbative Effects on the String World Sheet I,'' {\it Nucl.
Phys.}~{\bf B278} (1986) 769, ``Non-Perturbative Effects on the String
World Sheet II,'' {\it Nucl. Phys.}~{\bf B289} (1987) 319. }

\lref\SeObserv{N. Seiberg, Nucl. Phys. {\bf B303} (1988) 286.}

\lref\SeBeta{N.~ Seiberg, ``Supersymmetry and
Non-perturbative Beta-Functions'', Phys.Lett.206B:75,1988}

\lref\SeWi{N. Seiberg, E. Witten, ``Electric-Magnetic Duality,
Monopole Condensation, And Confinement in $N=2$ Supersymmetric
Yang-Mills Theory ''
Nucl. Phys. B426 (1994) 19-52 (and erratum - ibid. B430 (1994) 485-486 )\semi
``Monopoles, Duality and Chiral Symmetry Breaking in
N=2 Supersymmetric QCD'', hep-th/9408099,
Nucl. Phys. B431 (1994) 484-550.}

\lref\seiberg{``Monopole Condensation and Confinement'',
hep-th/9408013,  Nathan Seiberg; hep-th/9408155,
Phases of N=1 supersymmetric gauge theories in four dimensions,
K.~ Intriligator
and N.~ Seiberg; hep-ph/9410203,
Proposal for a Simple Model of Dynamical SUSY Breaking,
by K.~ Intriligator, N.~ Seiberg, and S.~ H.~ Shenker;
hep-th/9411149,
 Electric-Magnetic Duality in Supersymmetric Non-Abelian Gauge
Theories,
 N. Seiberg; hep-th/9503179 Duality, Monopoles, Dyons, Confinement
and Oblique
Confinement in Supersymmetric $SO(N_c)$ Gauge Theories,
K. Intriligator and N. Seiberg}
\lref\ganor{O.~Ganor,
``Toroidal Compactification of Heterotic 6D Non-Critical
     Strings Down to Four
Dimensions'', hep-th/9608109,
          Nucl.Phys.  B488 (1997) 223-235}
\lref\GaMorSe{O.~ Ganor, D.~ Morrison, N.~ Seiberg, ``Branes,
						Calabi-Yau Spaces
						and Compactification of
						Six Dimensional $E_{8}$ Theory'', hep-th/9610251}
\lref\argfar{P.C.~ Argyres, A.E.~ Farragi,
``The Vacuum Structure and Spectrum
of $N=2$ Supersymmetric $SU(n)$ Gauge Theory'', hep-th/9411057\semi
A.~ Klemm, W.~ Lerche, S.~ Theisen and S.~ Yankielowicz,
``Simple Singularities and                                                         $N=2$ Supersymmetric Yang-Mills Theory'', hep-th/9411048}
\lref\argsei{P.C.~ Argyres, M.R.~ Plesser and N.~ Seiberg,
``The Moduli Space of Vacua of   $N=2$ SUSY QCD and Duality in $N=1$
SUSY QCD'', hep-th/9603042}
\lref\plesmor{D.~Morrison, M.~Plesser, hep-th/9412236}
\lref\sen{A. Sen, ``Dyon-Monopole bound states, selfdual harmonic
forms on the multimonopole moduli space and $SL(2,Z)$
invariance,'' hep-th/9402032}
\lref\thooft{G. 't Hooft , ``A property of electric and
magnetic flux in nonabelian gauge theories,''
Nucl.Phys.B153:141,1979}
\lref\wrdhd{R. Ward, Nucl. Phys. {\bf B236}(1984)381}
\lref\ward{Ward and Wells, {\it Twistor Geometry and
Field Theory}, CUP }
\lref\avatar{A.~ Losev, G.~ Moore, N.~ Nekrasov, S.~ Shatashvili,
``Four-Dimensional Avatars of 2D RCFT,''
hep-th/9509151, Talks at Strings'95 and Trieste 1995}
\lref\cocycle{A.~ Losev, G.~ Moore, N.~ Nekrasov, S.~ Shatashvili,
``Central Extensions of Gauge Groups Revisited,''
hep-th/9511185.}
\lref\gncal{A.~ Gorsky, N.~ Nekrasov, ``Hamiltonian
systems of Calogero type and Two Dimensional Yang-Mills
Theory'', Nucl. Phys. {\bf B} (1994) }
\lref\gnell{A.~ Gorsky, N.~ Nekrasov, ``Elliptic Calogero-Moser System from
Two Dimensional Current Algebra'', hep-th/9401021}
\lref\gnru{A.~ Gorsky, N.~ Nekrasov, ``Relativistic Calogero-Moser model as
gauged WZW theory'', Nucl.Phys. {\bf B} 436 (1995) 582-608}
\lref\nhol{N.~ Nekrasov,    ``Holomorphic bundles and many-body
systems'', hep-th/9503157, Comm. Math. Phys. (1997)}
\lref\gnfu{N. Nekrasov, A. Gorsky ,
``Integrable systems on moduli
spaces'', in preparation}
\lref\dual{V.~ Fock, A.~ Gorsky, N.~ Nekrasov, V.~ Rubtsov,
``Duality in Many-Body Systems and Gauge Theories'', in preparation}

\lref\gmmm{A.~ Gorsky,~ I.~ Krichever,~
A.~ Marshakov,~ A.~ Morozov,~ A.~ Mironov,~
``Integrablity and
Seiberg-Witten Exact Solution'',
hep-th/9505035,  Phys.Lett.B355: 466-474, 1995}
\lref\bbkt{O.~ Babelon, E.~ Billey, I.~ Krichever, M.~ Talon,
``Spin generalization of
the Calogero-Moser system and the Matrix KP equation'',
hep-th/9411160}
\lref\kr{ I.~ Krichever, Funk. Anal. and Appl., {\bf 12}
(1978),  1, 76-78; {\bf 14} (1980), 282-290}
\lref\krz{I.~ Krichever, A.~ Zabrodin, ``Spin Generalization of
Ruijsenaars-Schneider model, Non-abelian Toda Chain and Representations
of Sklyanin Algebra'', hep-th/9505039}
\lref\krichever{I.~Krichever, ``Universal $\tau$-function of
Whitham hierarchy'', LPTENS-92/18, hep-th/9205110,
in
``30 years of the Landau
Institute'', 477 , I.M.~Khalatnikov, V.P.~Mineev, V.P. (eds.)}

\lref\fa{G.~ Faltings, ``A proof of Verlinde formula'',
J.Alg.Geom.{\bf 3}, (1994) }
\lref\beil{A.~ Beilinson, V.~ Drinfeld,
``Quantization of Hitchin's Integrable System and Hecke Eigensheaves'',
{\rm A. Beilinson's lectures at IAS, fall 1994}}
\lref\be{D.~ Bernard, Nucl. Phys. {\bf B}303 (1988) 77,
{\bf B}309 (1988), 14}
\lref\c{F.~ Calogero, J.~ Math.~ Phys. {\bf 12} (1971) 419}
\lref\ch{I.~ Cherednik, ``Difference-elliptic operators and root systems'',
hep-th/9410188}
\lref\eq{P.Etingof,
``Quantum integrable systems and
representations of Lie algebras'', hep-th/9311132}
\lref\ek{P.Etingof, A.Kirillov, Jr., ``On the affine
analogue of Jack's and Macdonald's polynomials'', Yale preprint, 1994}
\lref\fv{G. Felder, A. Varchenko, ``Integral formula for the
solutions of the elliptic KZB  equations'', Int. Math. Res. Notices,
1995, vol. 5, pp. 222-233}
\lref\ga{R. Garnier, ``Sur une classe de system\`es differentiels
Abeli\'ens d\'eduits de la theorie des \'equations lin\'eares'',
Rend. del Circ. Matematice Di Palermo, {\bf 43}, vol. 4 (1919)}
\lref\gau{M. Gaudin, Jour. Physique, {\bf 37} (1976), 1087-1098}
\lref\iv{D.~ Ivanov, ``Knizhnik-Zamolodchikov-Bernard equations
on Riemann surfaces'', hep-th/9410091}
\lref\kks{D.~ Kazhdan, B.~ Kostant and S.~ Sternberg, Comm. on Pure and
Appl. Math., vol. {\bf XXXI}, (1978), 481-507}
\lref\lo{A.~ Losev, ``Coset construction and Bernard Equations'',
CERN-TH.6215/91}
\lref\op{M.~ Olshanetsky, A.~ Perelomov, Phys. Rep. {\bf 71} (1981), 313}
\lref\m{J.~ Moser, Adv.Math. {\bf 16} (1975), 197-220; }
\lref\rs{S.N.M. Ruijsenaars,
H. Schneider, Ann. of Physics {\bf 170} (1986),
 370-405}
\lref\r{S.~ Ruijsenaars, CMP {\bf 110}  (1987), 191-213 }
\lref\s{B.~ Sutherland, Phys. Rev. {\bf A5} (1972), 1372-1376;}
\lref\sch{L.~ Schlesinger, `` \"Uber eien Klasse
von Differentialsystemen
beliebiger Ordnung mit festen kritischen Punkten'',
Journal f\"ur die reine und angewandte Mathematik,
Bd. CXLI (1912), pp. 96-145}
\lref\er{B.~ Enriquez, V.~ Rubtsov, ``Hitchin systems, higher
Gaudin operators and {\sl r}-matrices'', alg-geom/9503010}

\lref\promise{A.~Losev, N.~Nekrasov, S.~Shatashvili, ``Seiberg-Witten
Solution and ADHM integration'', in progress}
\lref\khoze{N.~Dorey, V.~Khoze, M.~Mattis, ``Supersymmetry and Multi-Instanton
Measure'', hep-th/9708036}
\lref\polyakov{A.~Polyakov, ``Gauge fields and strings'', Klewer Academic
Press}
\lref\grwt{G.~Moore, E.~Witten, ``Integration over $u$-plane
in Donaldson theory'', hep-th/9709193}
\Title{ \vbox{\baselineskip12pt\hbox{hep-th/9711108}
\hbox{HUTP -97/A064 }\hbox{YCTP -P22-97} \hbox{ITEP - TH.49/97} }}
{\vbox{
\centerline{ISSUES IN TOPOLOGICAL GAUGE THEORY}
}}
\vskip 0.3cm
\centerline{A.~ Losev$^{1,3}$, N.~ Nekrasov$^{1,2}$\footnote{*}{Junior Fellow,
Harvard Society of Fellows}, S.~ Shatashvili$^{3}$\footnote{**}{
on leave of absence from St. Petersburg Steklov Mathematical Institute}}
\vskip 0.3cm
\centerline{$^{1}$ Institute of Theoretical and Experimental Physics,
117259, Moscow, Russia}
\centerline{$^{2}$ Lyman Laboratory of Physics, Harvard University,
Cambridge MA 02138}
\centerline{$^{3}$ Department of Physics, Yale University, New Haven  CT
06520, Box 208120}
\bigskip
\centerline{losev@genesis5.physics.yale.edu, nikita@string.harvard.edu,
samson@euler.physics.yale.edu}
\vskip 0.5cm
We discuss topological  theories,
arising from the general $\CN=2$ twisted gauge theories.
We initiate a program of their study in the Gromov-Witten
paradigm.
We re-examine
the low-energy effective abelian theory in the presence of
sources and study the mixing between the various $p$-observables.
We present the twisted superfield
formalism which makes duality transformations transparent.
We propose a scheme which uniquely fixes all the contact terms.
We derive a formula for the correlation functions
of $p$-observables on the manifolds of generalized
simple type for $0 \leq p \leq 4$ and on some manifolds with
$b_{2}^{+} =1$.
We study the
theories with matter and explore the properties
of universal instanton.  We also discuss
the compactifications of higher dimensional theories.
Some relations to sigma models of type $\bf A$ and
$\bf B$ are pointed out
and exploited.
\vskip 0.1cm

\Date{11/97}
\newsec{\bf Introduction and summary}

\subsec{Motivations}

The study of topological theories has
several important reasons even in the era of string duality.
The first reason has to do with the sensitivity of certain
topological correlation functions to the geometry of the
space of vacua of the physical theory. This
may provide a check on many conjectures on non-perturbative
behavior of supersymmetric
field theories in various dimensions. The second
reason is the desire to understand the proper structure
of the space-time in four dimensional field theories
in the following sense.
The prescription of evaluation of correlations functions
in topological theory relies on a particular way the
moduli space of {\it distinct} points on the space-time manifold is
compactified. In most two dimensional theories
this compactification is that of Deligne-Mumford. The proper
analogue of this construction in four dimensions is unknown
\foot{There exists an analogue of Deligne-Mumford
compactification of the space of configurations of
points on complex surfaces}. In order
to get a hint of how this space might look like
we propose to study the contact terms in the topological
gauge theory. We expect that the algebraic structures arising in the
course of the study would generalize the associativity (WDVV)
equations and Whitham hierarchies in two dimensions.

One of the sources of the duality revolution is the
solution of N.~Seiberg and E.~Witten \SeWi\ of $\CN=2$ supersymmetric
Yang-Mills theory which has been tested in many indirect ways
but never directly. The solution, among other things,
predicts the formula for the effective coupling constant of
the low-energy theory as a function of the order parameter $u$
(for concreteness we talk about $SU(2)$ theory):
\eqn\swcl{{\tau}(u) = {{2i}\over{\pi}}
{\rm log}\left({u\over{\Lambda^{2}}}\right) + \sum_{n=1}^{\infty}
{\tau}_{n} \left( {{\Lambda^{2}}\over{u}} \right)^{2n}}
The coefficients $\tau_{n}$ are claimed to be instanton corrections.
The direct test of \swcl\ would involve integration
over the moduli space of instantons on $\IR^{4}$ of certain
form. This integration can be shown to localize onto the space
of {\it point-like instantons}
\promise\ and has a potential divergence in it
related to the fact that the space of point-like
instantons is non-compact (it is a resolution of singularities
of $S^{k}{\IR}^{4}$). Although this difficulty seems to be
avoidable by appropriate regularization \promise\
so far no substantial success has been achieved on this route
(nevertheless, see the
related discussion in \khoze).
Instead, suppose that one may calculate
exactly some correlation functions in the theory without using
the form of the effective action (we shall make this
statement precise later) and compare it to the computation
using the techniques of effective actions. Then, if the correlators
``probe'' the moduli space of vacua
sufficiently well then the
function $\tau(u)$
can be reconstructed. The tricky point of this idea is that the
exactly computable quantities  are the correlation functions
in the topological  (twisted) theory. The latter has more
freedom in the choice of action then the original physical
theory. In particular, one may take the limit
of zero coupling in the non-abelian theory while
keeping the representatives of the observables being such
that the resulting finite-dimensional counting problem has
a well-defined formulation\foot{It is important
to understand that the theory is non-trivial
even at zero coupling precisely due to
the existence of instantons}. In the context of gauge theory
such a counting  problem is equivalent
 to the way S.~Donaldson formulated his
invariants in the language of problems of ``generic position''
(for the gauge groups $SU(2)$ and $SO(3)$).
The insight coming from the equivalence of Donaldson
theory and twisted version of $\CN=2$ super-Yang-Mills
theory is that the same problem may be addressed in
the infrared limit.
It seems that the well-defined ultraviolet problem
in the field-theoretic formulation involves
the definition of the compactified moduli space of {\it distinct}
points on the space-time manifold $\Sigma$.  The infrared
theory contains the information about this compactification
in the contact terms between the observables. In the context
of gauge theory these contact terms can be studied
using severe constraints of modular invariance and ghost number
anomalies.

\subsec{Extract of Gromov-Witten theory}

The story above has a counterpart
in the context of two dimensional field theories.
Namely, let us consider the {\it conformal} theory which
is type $\bf A$ topological sigma model in the limit of
 infinite volume of the target space.
One can show that such a limit exists by passing to
the first order action of a curved $\beta-\gamma$ system
perturbed by marginal (but not exactly marginal) operator
proportional to {\it inverse} metric. The unperturbed
system is naturally an {\it interacting}\foot{i.e. nonlinear}
conformal theory.
Such a conformal model could be consistently coupled to topological
gravity to produce Gromov-Witten (GW) invariants \twisted\konman.
Recall, that GW invariants are
the maps from the $n$-th tensor power of the space of
zero-observables to the cohomologies of
Deligne-Mumford compactification
 ${\bar M}_{g,n}$
of the moduli space of complex structures
of Riemann surfaces with $n$ {\it distinct} marked points.
The correlators of $p$-observables can
be interpreted as the integrals of
the GW invariants over
some cycles in ${\bar M}_{g,n}$,
corresponding to the fixed complex structure of the curve $\Sigma$
 without marked points\foot{More presicely, let $m \in \bar M_{g,0}$
be the point and $\pi: \bar M_{g,n} \to \bar M_{g,0}$
be the forgetful map. Then the cycles lie in $\pi^{-1}(m)$.}.

The important and rather subtle point is that the
compactification of the certain spaces is necessary
for producing field theoretic interpretation
of the enumerative problems while in  most cases
the problems themselves
are formulated without this complication.

The construction of
Gromov-Witten classes is
possible if the following triple
of compact spaces and maps between them
is presented \konman\konenu.
The first space is the compactification
of the space of $n$ points on the worldsheet: $\bar M_{n, \Sigma}$.
The second space is the compactified
moduli space $\CV_{\Sigma, n; T}$ of the pairs
$({\Sigma}, x_{1}, \ldots, x_{n}; f)$
where $f$ is the holomorphic map (instanton) of the worldsheet
$\Sigma$ to the target space $T$.
The non-trivial part of the compactification is that it
possesses two maps:
\eqn\grwtn{\matrix{& & &\CV_{\Sigma, n;T} & & &\cr
        &  & {\rm p} \swarrow &    & \searrow {\rm  ev}& & \cr
&  \bar M_{n,\Sigma} & &     &  & T^{\times n} &\cr}}
where $T$ is the target space and the restriction
of the map ${\rm ev}= \times_{i} {\rm ev}_{i}$
onto the internal part of $\CV$ is simply
the evaluation of $f$ at the points $x_{i}$: ${\rm ev}_{i}  =
f(x_{i})$,
while that of $p$ is the projection
$p({\Sigma}, x_{1}, \ldots, x_{n}; f) =
({\Sigma}, x_{1}, \ldots, x_{n})$.
Then the construction of the classes proceeds as follows:
take the cohomology elements $g_{1}, \ldots, g_{n} \in
H^{*}(T)$. Then
the corresponding Gromov-Witten class is
\eqn\grwtni{{\CI}(g_{1}\otimes \ldots \otimes g_{n}) =
p_{*}\left( \otimes_{i}
{\rm ev}_{i}^{*} \quad g_{i} \right) \in H^{*}({\bar M}_{n,\Sigma})}

It is desirable that the four dimensional gauge theories also
may be treated within similar paradigm. The first problem is to
find the analogue of $\bar M_{n,\Sigma}$. In the case
where the instantons can be described as holomorphic
bundles the natural candidate is the resolution of diagonals constructed
by W.~Fulton and R.~MacPherson \fulmac. Unfortunately
this appeals to the choice of complex structure on $\Sigma$
and hence violates the Lorentz invaraince.

\subsec{Summary of the results}
This paper makes the first steps in the direction of
GW program. Specifically we discuss the geometry of
the twisted abelian  gauge theories and
present the superfield
formalism which makes duality transformations transparent
(chapter $3$).
We study the deformations of the theory. It turns out
that the proper formulation of the deformation
problem is equivalent to the studies of the deformations
of $\Gamma$-invaraint Lagrangian submanifolds
in a complex vector symplectic space, where $\Gamma$ is a certain
discrete subgroup of linear symplectic group.

We propose the extention of these
results to the case of supermanifolds. In other words,
we include all observables into consideration. In particular,
we derive the formula
for the contact terms between the $p_{1}$- and $p_{2}$-observables
for $p_{1} + p_{2} \geq 4$ (otherwise the contact term vanishes).
We propose the conjecture of universality of contact terms (see
chapter $2$) and
test it in some examples. The validity
of this conjecture is a strong hint about the topology
of the compactified configuration space ${\bar M}_{n, \Sigma}$.
In the chapter $4$  we apply these
results to the computations of generalized
Donaldson invariants
- the correlation functions
of $p$-observables of the functions
$\left( {\Tr}{\phi}^{2} \right)^{r}$
(Donaldson studied the case $r=1$). The subtleties of this definition
are discussed in the chapter $5$.
We
compute all topological correlation functions in the theory on
some manifolds with $b_{2}^{+}=1$
thus generalizing the results of \gz\grwt. They are not exactly
topological invariants. The reason is  that sometimes the enumerative
problem of Donaldson's is not well-behaved under the variations
of parameters, such as the metric. This occurs precisely
when $b_{2}^{+}=1$ due to the abelian instantons\foot{They are
honest solutions of instanton equations with abelian gauge group}.
The jumps of the correlations functions/Donaldson invariants
are under control \gz\borch\grwt.
We use these formulae to test certain conjectures which formed
the ground for the analysis of the chapter $2,3$.

Then we go on and discuss more general theories. We explain
the meaning of the theories with matter, and the compactified
higher dimensional theories, both microscopically
(i.e. from the point of view of intersection
theory on instanton moduli spaces) in chapter $5$ and
macroscopically in chapter $6$. In particular,
we show that the theory with massive matter
can be intepreted as the integration of the equivariant
Euler class of the Dirac  index bundle
over instanton moduli space. The higher
dimensional theories are interpreted as
the integration of $\hat A$-genus and
elliptic genus respectively. We discuss the $A,D,E$
singularities on the Coulomb branches and their contribution to the
correlation functions using equivariant cohomology with respect to the
enhanced global symmetry groups.
In the chapter $7$ we discuss the two
dimensional analogues of several constructions which we considered in the
gauge theory case.  We present the proper analogues of the theories with
matter in the type $\bf A$ sigma model framework. We explain the relations of
gauge theories (both two and four-dimensional) to the type $\bf B$ sigma
models.  The formulae we obtain have very transparent meaning in some cases.
This is the limit of the metric on the manifold $\Sigma$ where it looks like
a two dimensional surface.  For example, for $\IP^{1} \times \IP^{1}$ one may
take one of the spheres much smaller then the other.  In this limit the
theory becomes equivalent to a kind of type $\bf B$ two dimensional sigma
model with Landau-Ginzburg superpotential.  The target space of the sigma
model is a disjoint union of the spaces $M = \amalg_{\lambda \in \Lambda}
M_{\lambda}$, where $\Lambda$ is the weight lattice of $G$, and $M_{\lambda}$
is the moduli space of curves, corresponding to the Seiberg-Witten
theory with the group $G$ together with the choice of cycle
in their first homology, proportional to $\lambda$.
The superpotential can be written universally for all theories
with arbitrary gauge groups:
\eqn\sprptnl{W = \sum_{i=1}^{r} \lambda_{i}a^{i}  + f(u^{k}) }
where $u^{k}$ are the invariants (casimirs) of the group $G$,
$f$ is some holomorphic function corresponding to the $2$-observable,
$\lambda_{i}$ are the components of the weight vector,
$a^{i}$ are the central charges in the $\CN=2$ algebra in four
dimensions, viewed as functions of $u^{k}$.
We compare two dimensional Yang-Mills theory and Landau-Ginzburg
model. We explain the absence of the contact terms in the
two dimensional Yang-Mills theory as a result of its hidden
higher  dimensional nature.

Due to the length of the paper we present a short overview of
explicit formulae and the relevant notations in the chapter $2$.
The material of the  chapters $3$ and $4$ has been reported at
the Cargese conference on ``Strings, Branes and Duality''
in June 1997.

The recent paper \grwt\ has
some overlap with these chapters.
The beautiful paper \grwt\ treated extensively the Donaldson invariants
for the $SO(3)$ group for various $w_{2}(E)$. The reader
should consult it for the detailed formulae for Donaldson
invariants of manifolds with $b_{2}^{+}=1$. Also, some formulae
for the theories with matter are presented there.\foot{G.~Moore has informed
us that he and M.~Mari\~no had some progress in extending the
results of \grwt\ to the case of higher rank groups}

\newsec{Answers for Correlators and Principles of Their Computation
in the Pure Gauge Theory}
The objective of this section is to formulate
certain principles of calculations of the correlation functions
of various observables in topological gauge theory.

We start with
presentation of the explicit formulae for the correlators in the
case of pure Donaldson theory.
We compute the correlation functions
of $p$-observables $\CO_{r}^{(p)}$ ($p = 0,1,2,3,4$) of operators
$\left( - {{{\Tr}{\phi}^{2}}\over{4\pi^{2}}} \right)^{r}$
($r=1, \ldots$) in $SU(2)$ twisted gauge
theory on a oriented closed four-manifold $\Sigma$.

\subsec{The setup}

The metric
$g$ allows to split the bundle of the real two-forms
$\Omega^{2}({\Sigma})$
into the sum $\Omega^{2} = \Omega^{2,+} \oplus \Omega^{2,-}$ of
the self-dual and anti-self-dual forms. This decomposition
descends to cohomology: $H^{2}(\Sigma;\IR) = H^{2,+} \oplus H^{2,-}$.
Let $b_{p} = {\rm dim}H^{p}({\Sigma}; \IR), \, b_{2}^{\pm} =
{\rm dim}H^{2, \pm}$.

Let $e_{\alpha}$ be the base in
the cohomology group of $\Sigma$:
$e_{\alpha} \in H^{*}({\Sigma}; {\IC})$.
Let $d_{\alpha}$ denote the degree of $e_{\alpha}$.

The breakthrough in the physical approach to the Donaldson theory
came from the realization of the fact that the essential
contribution to the correlation functions in the twisted low-energy
effective theory comes from the singularities in the space
of vacua (the ``$u$-plane'') where extra massless particles
appear.
In the paper \wittmon\ the following equations were proposed:
\eqn\mneq{\eqalign{&F^{+}_{ij} = -{i\over{2}} {\bar M}\Gamma_{ij} M\cr
    & \sum_{i} \Gamma^{i}D_{i} M = 0\cr}}
where $M$ is the section of the $Spin^{c}$ bundle $S^{+} \otimes L$
\foot{Here $L^{2}$ is the ordinary complex line bundle.
For two $Spin^{c}$ structures $S_{+}\otimes L$ and
$S_{+}\otimes L^{\prime}$ the ratio $L^{\prime} \otimes L^{-1}$
is also the ordinary line bundle.
Notice that $c_{1}(L^{2}) \equiv w_{2}({\Sigma})
{\rm mod}2$.},
$\bar M$ is the section of $S^{+} \otimes L^{-1}$, $F$ is
the curvature of the connection in the line bundle $L^{2}$,
$\Gamma^{i}$ are the Clifford matrices, and
$\Gamma_{ij} = {\half} [\Gamma_{i}, \Gamma_{j}]$.
Let $\ell \in H^{2}({\Sigma}, {\IZ})$ be the class
such that $\ell \equiv w_{2}({\Sigma})
{\rm mod}2$. Let $\CM({\ell})$ be the moduli
space of the solutions of the equations \mneq\ with
$c_{1}(L^{2}) = \ell$. Its dimension can be computed by virtue of the
index theorem
\wittmon:
\eqn\dmn{d_{\ell} =
{\rm dim}\CM({\ell}) = {{(\ell,\ell) - 2\chi - 3\sigma}\over{4}}}
where $\chi$ and $\sigma$ denote the Euler characteristics and
the signature of the manifold $\Sigma$ respectively,  $(\ell,\ell)$ is
the intersection pairing in $H^{*}({\Sigma})$.
Fix a point $P \in \Sigma$.
One may also form the framed moduli space $\CM({\ell},P)$. It is the
space of
solutions of the equations \mneq\ modulo gauge transformations
equal to identity at $P$.
The space $\CM({\ell},P)$ is the $U(1)$ bundle over $\CM({\ell})$.
Let $c_{1}$ be its first Chern class. It does not depend on $P$.
Define
the Seiberg-Witten invariant corresponding to the $Spin^{c}$ structure
$\ell$ the integral \grwt:
\eqn\dfnsw{SW(\ell) = \int_{\CM({\ell})} c_{1}^{{d_{\ell}}\over{2}}}
It vanishes if the dimension $d_{\ell}$ is odd. The manifold
$\Sigma$ is of {\it generalized simple type }
if there is a finite number $l$ such that $SW(\ell)$
vanishes for $\vert (\ell,\ell) \vert > l$.

\subsec{The explicit formulae}

First we present the answers for the correlators.
The relevant notations
are explained shortly. The conjectures which form the
ground for our computations are formulated afterwards.

The expressions for correlators involve various derivatives
of the following {\it master} function $\CF (a; \{ t_{r} \})$.
Let $(a(u), a_{D}(u))$ be the functions, defined as follows.
Both $a$ and $a_{D}$ are the solutions of the second-order
differential
equation:
\eqn\auu{\Biggl[(1 -u^{2}) {{d^{2}}\over{du^{2}}} + {1\over{4}}
\Biggr]
\pmatrix{ a(u) \cr a_{D}(u)\cr} = 0}
with the asymptotics at $u\to \infty$:
\eqn\auuu{a(u) \sim \sqrt{{u}\over{2}} + \ldots, \quad
a_{D} \sim - {{2a}\over{\pi i}}{\rm log}(u) +\ldots    }

Then $\CF(a; 0)$ is defined as the solution to the equation:
$$ d\CF = a_{D}(u) da(u) $$
Next we define the function
$H(a, a_{D})$ which obeys the following three properties:
\item{1.}{\it Normalization:}
\eqn\prpi{H(a(u), a_{D}(u)) = -{{u}\over{2}}}
\item{2.}{\it Homogeneity.} For $\mu \neq 0$:
\eqn\prpii{H(\mu a, \mu a_{D}) =
\mu^{2} H(a, a_{D})}
\item{3.}{\it Modular invariance.} For any
$\pmatrix{\alpha & \beta\cr \gamma & \delta \cr} \in
\Gamma \subset SL_{2}({\IZ})$:
\eqn\prpiii{H(\gamma a_{D} + \delta a, \alpha a_{D} + \beta a) =
H(a, a_{D}), \quad}

The function $\CF(a; t_{r})$ is the (formal)
solution to the following system of partial differential equations:
\eqn\hmljcb{
{{\p \CF}\over{\p t_{r}}} = - H^{r} (a, {{\p
\CF}\over{\p a}}) }

The master function $\CF$ can be analytically continued by allowing
the parameters $t_{r}$ to take values in any (super)commutative
algebra $\CV$. In our case the algebra $\CV$ is that of cohomologies
of $\Sigma$: $\CV = H^{*}({\Sigma}, {\IC})$. We expand $t_{r}$ as
follows:
$t_{r}  = T^{\alpha}_{r} e_{\alpha} \in \CV \equiv H^{*}({\Sigma})$.
Now, $\CF$ solves the system:
\eqn\shmjb{
{{\p \CF}\over{\p T^{\alpha}_{r}}} = - e_{\alpha} H^{r} (a, {{\p
\CF}\over{\p a}}) }

Consider the formal seria:
\eqn\aui{\pmatrix{a(u; \{ t_{r} \}) \cr
a_{D}(u; \{ t_{r} \}) \cr} = e^{{\lambda(u,t)}\over{2}}
\pmatrix{a( u e^{-{{\lambda(u,t)}\over 4} })\cr
a_{D}( u e^{-{{\lambda(u,t)}\over 4}})\cr}}
where
\eqn\lmb{\lambda (u,t) = \sum_{r=1}^{\infty} rt_{r}u^{r-1}}

Define $a_{1,2}$ as follows:
$$
a_{1} = a_{D}, \, a_{2} = - a_{D} + 2a
$$
Together with $\CF(a, t)$ we also need its Legendre transforms:
\eqn\lgndr{\CF_{1}(a_{1}, t) =  - a a_{D} + \CF(a), \quad
\CF_{2}(a_{2}, t) = - 3 aa_{D} + 3 a^{2} - {\half} a_{D}^{2} +
\CF (a)}

The last notation is the following. Let $\psi$ denote the odd
(fermionic)
variable with values in $H^{1}({\Sigma}; {\IR})$. Let $[d\psi]$ be the
fermionic measure  on $H^{1}({\Sigma}; {\IR})$.

{\bf Claim 1.} for the manifolds $\Sigma$ with $b_{2}^{+} > 1$:
\eqn\frmli{\eqalign{
\CZ ( T^{\alpha, r} ) = \langle \exp
\left(  T^{\alpha,r}
\int_{\Sigma} e_{\alpha} \CO_{r}^{(4-d_{\alpha})} \right)
\rangle
= &\cr
\sum_{i=1}^{2} \oint_{a_{i}=0}
du \bigl[ d \left({{du}\over{da_{i}}} \psi\right) \bigr]
a_{i}^{{{\chi + \sigma}\over{4}}-1}
(u^{2}-1)^{{\sigma}\over{8}}
\left({{du}\over{da_{i}}}\right)^{{b_{2}}\over{2}}
 & \sum_{\ell \in H^{2}({\Sigma}, {\IZ})} SW(\ell)
e^{{1\over{\pi i}}\int_{\Sigma} {\CF}_{i} (a_{i} + \psi + \ell, T) }
\cr}}
Here we substitute $a_{i} = a_{i} (u,t)$.

As an example of the correlator on
the manifold $\Sigma$ with $b_{2}^{+}=1$
we consider the case of $\Sigma_{0}^{g} = {\IP}^1 \times C_{g}$,
where $C_{g}$ is the genus $g$ Riemann surface and that of
$\Sigma_{l}^{g}$,
which is $\Sigma_{0}^{g}$ blown up at $l$ points in generic position
(for $l \leq 8$). The surfaces $\Sigma_{l}^{0}$ are called Del Pezzo
surfaces.
We study them in the limit where the sizes of the two-sphere and
all exceptional divisors are  small compared to that of $C_{g}$.

{\bf Claim 2.} the answer for the correlator \frmli\ is
given by the contour integral
\eqn\frmlii{\sum_{N \in \IZ} \oint B^{l} (u)
{{(du)^{2} [d\tilde \psi] e^{{1\over{\pi
i}}\int_{\Sigma}\CF(a + \psi; T)}}\over{dW_{N}}}, \quad W_{N} = Na(u)
- \int_{C_{g}} \CF(a + \psi; T) } where
$\tilde \psi =
{{du}\over{da}} \psi$ and
$B^{l}(u)$ is the blowup factor
which
is  equal to
$$
B^{l}(u; t) = \prod_{i=1}^{l} {{{\theta}_{00}
(\tau(u), {1\over{\pi i}}\int_{e_{i}} \CF (a;t))}\over{{\theta}_{00}
(\tau (u),0)}}
$$
with $e_{i}$ denoting the exceptional divisors (the definition
of $\theta_{00}, \theta_{01}$ etc. is given below).

{\bf Claim 3.} The formulae \frmli\ are easily modified to
cover the case of the $SU(2)$ theory on spin manifold $\Sigma$ (we relax
this requirement in the sequel) with
$N_{f} \leq 3$ massless multiplets
(its mathematical meaning is explained in
chapter $5$). The modification is as follows.
The ``functions'' $a(u)$ and
$a_{D}(u)$ obey the equations:
\eqn\pfmd{\biggl[ -\Delta
{{d^{2}}\over{du^{2}}} + {{f}\over{4}} \biggr]
\pmatrix{a\cr a_{D}\cr} = 0}
where
\eqn\dscrm{\matrix{N_{f} &\Vert &  \Delta  & \Vert & f &\Vert \cr
1 &\Vert & 1 - u^{3} &\Vert & u & \Vert \cr
2 &\Vert & (u^{2} - 1)^{2} &\Vert & u^{2}-1 &\Vert \cr
3 &\Vert & \left( u+{{2}\over{3}}\right)^{3}({{1}\over{3}}-u) &\Vert &
\left( u+ {{2}\over{3}}\right)^{2} &\Vert \cr }}
The solutions to \pfmd\ are
fixed by their asymptotics at $u \to \infty$:
\eqn\asmt{a (u) \sim {\half}
\sqrt{2u} + \ldots, a_{D}(u) \sim {{N_{f}-4}\over{2\pi i}}
a(u) {\rm log} u +
\ldots} The factor $(u^{2}-1)^{{\sigma}\over{8}}$ is replaced by
$\Delta^{{\sigma}\over{8}}$.
Notice that in the case $N_{f} =3$ one has to make a shift
$u \mapsto u - {2\over{3}}$ while comparing to the formulae
of\foot{We set the scales of \SeWi\ to:
$\Lambda_{3} = 8$, $\Lambda_{2}^{2} = 8$, $\Lambda_{1}^{6} =
-{{256}\over{27}}$} \SeWi\grwt (the formalism of \SeWi\ allowed
for an arbitrary shift in $u$).
The rest of the formalism is unchanged.

\subsec{Geometry of the abelian $\CN =2$ theory}

Here discuss the geometric
representation of the data entering the construction
of the low-energy effective abelian theory. Although the
various parts of it are well-known \SeWi\Fre, we
present our reformulation since in this
setting the deformation problem can be clearly
stated.

\sssec{Embedded \quad vacua}

Consider the complex symplectic vector space $\IC^{2r}$ with the
symplectic form $\omega = \sum_{i=1}^{r} da^{i} \wedge da_{i,D}$.
Let $\theta = \sum_{i=1}^{r} a_{i,D}da^{i}\equiv (a_{D}, da)$.
Let $\Gamma$ be a subgroup of $Sp(2r, {\IZ})$. Let $\CL$ be a
$\Gamma$-invariant Lagrangian submanifold in $\IC^{2r}$. By
definition, the restriction of $\omega$ on $\CL$ vanishes, hence
\eqn\prptn{\theta \vert_{\CL} = d\CF, \quad \CF: \CL \to \IC}
The function $\CF$
is called the generating function \foot{It is globally well-defined
on $\CL$ if $\CL$ is simply-connected}
of the
Lagrangian submanifold. It transforms under the action
of element $g \in \Gamma$ as follows:
\eqn\trpr{g  = \pmatrix{A & B\cr C & D\cr}, \quad
g^{*}\CF(x) = \CF(x) + (Ba, Ca_{D}) + {\half} (Ba, Da)
+ {\half}(A a_{D}, Ca_{D})
+c(g)}
where $c(g)$ is a certain cocycle: $[c(g)] \in H^{1}({\Gamma}, {\IC})$.
If the cocycle $c(g)$ is trivial then one can solve \trpr\ as
follows:
\eqn\slmtn{\CF = {\half} (a, a_{D}) + {{u}\over{\pi i}}}
where $u$ is some $\Gamma$-invariant function on $\CL$.
This property
of the prepotential $\CF$ has
been observed by several authors (cf.
\matins\sty\matone).  It reflects the scaling properties of $\CF$. To see
this one inserts into \slmtn\ that $a_{D} = {\p \CF}/{\p a}$  and use the
equation \hmljcb\ for the evolution generated by $u$ provided that the
extension of $u$ to ${\IC}^{2r}$ is known.  We claim that the
$\Gamma$-invariant Lagrangian submanifold determines an effective abelian
$\CN=2$ gauge theory, whose duality group is precisely $\Gamma$.

{\it \quad A note on definitions.} There is one confusing point
which must be kept in mind. The name {\it generating function}
for the prepotential $\CF$ which comes from symplectic geometry may be confusing with
the notion of generating function for correlators. In fact, in two
dimensional topological field theories the same letter $\CF$
goes under the name of prepotential and the generating
function of correlators. Moreover, sometimes $\CF$ {\it is}
the prepotential of the effective four dimensional
supersymmetric field theory. Throughout this paper we
use the letter $\CF$ only for the prepotential and
when we call it generating function it is only the generating function
of Lagrangian variety which is meant by the name. The generating
functions of the correlators will be called by $Z, \CZ,$ etc.

\sssec{Deformations \quad of \quad the \quad theory}

The symplectomorphisms of ${\IC}^{2r}$ map $\CL$ to another
Lagrangian submanifolds. The symplectomorphisms in the component
of identity are generated by the (time-dependent) Hamiltonians
$H(a, a_{D})$. The generating function $\CF$ changes according to
the Hamilton-Jacobi equation:
\eqn\hmjcb{{{\p \CF}\over{\p t}} = - H (a , {{\p \CF}\over{\p a}}, t)}
where we view $\CF$ as a function of $a$. This description is
local.
The flows which preserve the property of $\Gamma$-invariance
are generated by $\Gamma$-invariant Hamiltonians. Let us denote
the space of all $\Gamma$-invariant holomorphic
functions on ${\IC}^{2r}$ by $\CC$.
The Hamiltonian flows in ${\IC}^{2r}$ which do not change $\CL$
are generated by the Hamiltonians obeying
\eqn\bgphs{\tau_{ij} {{\p H}\over{\p a_{D,i}}} = - {{\p H}\over{\p a^{j}}}}
where
\eqn\prd{\tau_{ij} = {{\p \CF}\over{\p a^{i}\p a^{j}}} =
{{\p a_{D,j}}\over{\p a^{i}}}}
The space of such Hamiltonians is denoted as ${\CC}_{\CL}$.
The quotient ${\CW}_{\CL} = {\CC}/{\CC}_{\CL}$ can be identified with the
space of $\Gamma$-invariant functions on $\CL$.
In general there is no canonical way of extending the
$\Gamma$-invariant function on $\CL$ to the
$\Gamma$-invariant function on $\IC^{2r}$.
Thus, we have two potential difficulties: $1)$ the Hamiltonians
may be time-dependent and $2)$ even if they are time-independent
there are many ways to extend a given function $u \in \CC$ to
the whole $\IC^{2r}$.

Let us dispose of the problems $1)$ and $2)$. In general,
the family $H_{k}(a, a_{D}, t)$ may be used in defining
a consistent system of the type \hmjcb\ if and only
if:
\eqn\shles{ {{\p H_{k}}\over{\p t_{l}}} -
{{\p H_{l}}\over{\p t_{k}}}  +
\{ H_{l}, H_{k} \} = 0 }
We propose to impose the extra condition that {\it the Hamiltonians
are actually time-independent}: ${{\p}\over{\p t_{l}}} H_{k} = 0$.
{\it This is the principle of background independence.}
Clearly, relaxing the background independence principle
allows to add to $\CF(a, t)$ an arbitrary $\Gamma$-invariant function
on $\CL$ at any order in $t$. This freedom would spoil
the uniqueness of the contact terms which we discuss shortly.

The problem $2)$ we shall solve in the important

\sssec{Rank \quad one \quad case.}
Let $r=1$. The group $\IC^{*}$ acts in $\IC^{2}$ in a standard
way. This action commutes with that of $\Gamma$. Suppose
that the basis of $\Gamma$-invariant functions on $\CL$ is chosen.
As $\CL$ is one-dimensional it is sufficient to choose
one function. We call it $u$. Then any other
admissible function is a rational function of $u$. Fix an
integer $p$.

{\it The next principle
is called ``Homogeneity'':}
The function $u$ extends to a function $\Gamma$-invariant
$H(a, a_{D})$ on $\IC^{2}$ with the following
properties:
\eqn\ext{\eqalign{H(\mu a, \mu a_{D}) &= \mu^{\rm d}H(a, a_{D})\cr
H(a, a_{D})\vert_{\CL} &= u\cr}}
We will
see that in asymptotically free theories ${\rm d}=2$.

Once we have fixed a continuation of one function the
extension of the rest is unique once the two principles
are at their power. Indeed, since the Hamiltonians must
Poisson-commute the only possibility is that the higher
Hamiltonians are the functions of the Hamiltonian
corresponding to $u$. Homogeneity fixes this function uniquely.
In particular, for a polynomial $P$
the function $P(u)$ extends to $P(H(a, a_{D}))$.
The deformation problem is well-posed now \arnold.
We have to solve the equation of motion
\eqn\hmem{\eqalign{{\dot a} = {{\p H}\over{\p a_{D}}}, \quad
{\dot a_{D}} = - {{\p H}\over{\p a}} }}
with the initial condition $(a(0), a_{D}(0)) \in \CL$. Then
\eqn\sltn{\CF(a, t) = \CF({\tilde a}, 0) +
\int_{0}^{t} \left( a_{D}(t^{\prime}) {\dot a}(t^{\prime})  -
H (a(t^{\prime}), a_{D}(t^{\prime})) \right) dt^{\prime}}
where the trajectory $(a(t^{\prime}), a_{D}(t^{\prime}))$ is such
that
\eqn\sht{a(0) = {\tilde a}, \quad
a(t) = a }
We introduce the set of ``times'' $t_{1}, t_{2}, \ldots$:
\eqn\nneq{{{\p \CF}\over{\p t^{k}}} = - H^{k}(a, {{\p \CF}\over{\p a}})}
In applications it is sometimes useful to have an expansion of the form:
\eqn\expn{\CF(a, t) = \CF_{0}(a) + \sum_{k>0} t^{k}\CF_{k} +
\sum_{k,l>0} {\half} t^{k}t^{l} \CF_{k,l} + \ldots}
The equations \nneq\ allow to calculate all the
coefficients $\CF_{k_{1}, \ldots, k_{p}}$ term by term.
In particular, the value of $\CF_{k}$ depends only
on the restriction of the function $H^{k}$ to $\CL$.
The next term $\CF_{k,l}$ which we call a {\it pair
contact term} depends on the first jets of $H^{k}$ and $H^{l}$,
etc. It is easy to calculate:
\eqn\cntctrm{\CF_{k,l} =
kl u^{k+l-2} {{\p H}\over{\p a_{D}}} {{du}\over{da}}}
where the last factor is the derivative of the function $u$ along $\CL$.
The quasihomogeneity of $H$ allows to calculate \cntctrm\
quite explicitly:
\eqn\cnctrmi{\CF_{k,l} =
kl u^{k+l-2} \left(
{{{\rm d} u - a {{du}\over{da}}}\over{a_{D}{{da}\over{du}} -
a {{da_{D}}\over{du}}}} \right)}

\sssec{Higher \quad rank.} In the case of higher rank we
do not know how to extend the functions $u_{1}, \ldots, u_{r}$.
However, we can solve the equations for the evolution with respect
to the Hamiltonian $u_{1}$.
It is done again by the homogeneity
principles. Let $d_{k}$ be the degree of the
coordinate $u_{k}$ with respect to the $U(1)$ ghost number symmetry.
For the low-energy
effective theory corresponding to the ultra-violet
gauge theory with
simple group $G$ the degrees of the basic
invariant polynomials $u_{k}$ are  ${\rm deg}u_{k} = d_{k} = m_{k}+1$ where
$m_{k}$ are the corresponding exponents
of the Weyl group $W$ of $G$. Thus, the
ghost number of $u_{k}$ is equal to $2m_{k} + 2$.
In particular, for $G= SU(r+1)$ one has:
$d_{k} = k+1$. Then:
\eqn\hmemsi{
\pmatrix{a^{i}(t; \{ u^{k}\} )\cr a_{D,i}(t; \{ u^{k} \}) \cr} =
e^{{{\pi it}\over{h^{\vee}}}}
\pmatrix{a^{i}(\{ e^{-{{2\pi itd_{k}}\over{h^{\vee}}}} u_{k} \} )\cr
a_{D,i}(\{ e^{-{{2\pi itd_{k}}\over{h^{\vee}}}} u_{k} \} ) \cr}}
where $h^{\vee}$ is the dual Coxeter number. For $G = SU(r+1)$,
$h^{\vee} = r+1$.
Here the functions $a^{i}, a_{D,i}$, $i=1, \ldots, r$ are the central
charges of $\CN=2$ susy algebra. They are given by the periods of a certain
meromorphic differential $\eta$ on the Riemann surface
$C_{u_{1}, \ldots,
u_{r}}$ of genus $r$. For the group $SU(r+1)$ the curve is:
\eqn\crvsun{z +
{1\over{4z}} = x^{r+1} + \sum_{k=1}^{r} u_{k} x^{r+1-k} } and the
differential \eqn\dfrnt{\eta = x {{dz}\over{z}}} One can choose a set of $A$
and $B$ cycles in $H_{1}(C_{\vec u}; \IZ)$ which form a base with canonical
intersections. Then $$ a^{i} = \oint_{A_{i}} \eta, \quad a_{D,i} =
\oint_{B^{i}} \eta, \quad A_{i} \cap B^{j} = \delta^{j}_{i} $$ We can also
write down a formula for the pair contact term between the $2$-observables
$\CO^{(2)}_{u_{k}}$ and $\CO^{(2)}_{u_{l}}$ in the first non-vanishing
order. It is:  \eqn\cntctrnii{\CF_{u_{k},u_{l}} = {{\p u_{k}}\over{\p a^{i}}
} {{\p u_{l}}\over{\p a^{j}}} {{\p {\rm log}{\Theta}}\over{\p \tau_{ij}}}}
where
\eqn\tht{\Theta = \sum_{\lambda \in \Lambda} \exp \left( 2\pi i \langle
\lambda, \tau \lambda\rangle  + \pi i \langle \lambda, \rho \rangle \right) }
with $\Lambda$ being the set of weights and $\langle, \rangle$ the
restriction of the Killing form on $\liet$ - the Cartan subalgebra of $\lieg$.
The matrix $\tau_{ij}$ is the period matrix of $C_{\vec u}$ in the
given basis of $A$ and $B$ cycles. The vector $\rho$ equals the half
the sum of positive
roots.

\sssec{Observables \quad as \quad nilpotent \quad deformations.}
The advantage of \expn\ is the possibility to have nilpotent
parameters $t^{k}$. In the context of gauge theory these parameters
are the elements of the cohomology of the space-time manifold $\Sigma$.
The function $\CF$ takes now
values in the graded (super)commutative algebra
$\CV = H^{*}({\Sigma}, {\IC})$.
Let $e_{\alpha}$ be the base of $\CV$. The product of the classes
$a$ and $b$
is denoted as $a \cdot b$.
The times are denoted as $T^{k,\alpha}$, where $k = 1, 2, \ldots$ and
$\alpha$ runs over the base of $\CV$.
The corresponding $\CV$-valued
Hamiltonians $H_{k,\alpha}$ are of the form
$H_{k}(a, a_{D}) e_{\alpha}$ with $e_{\alpha} \in \CV$.
We may immediately apply \expn\ and \cnctrmi\ to get:
\eqn\cnctrmii{\eqalign{{{\p \CF(a, a_{D})}\over{\p T^{k, \alpha}}} &= -
H_{k}(a, a_{D}) e_{\alpha} \cr
\CF_{(k,\alpha), (l, \beta)} &=
\Biggl[ kl u^{k+l-2} \left(
{{{\rm d} u - a {{du}\over{da}}}\over{a_{D}{{da}\over{du}} -
a {{da_{D}}\over{du}}}} \right) \Biggr]
e_{\alpha} \cdot e_{\beta}
\cr}}

The meaning of the nilpotent deformations along $H^{*}({\Sigma})$ is clear.
They simply correspond to the observables added to the action.
The $p$-observable corresponds to $H^{4-p}({\Sigma})$.
The meaning of \cnctrmii\ is simply the {\it pair contact term}
between $p$ and $q$-observables, which  is $4- p - q$-observable.

The fact that the form of the contact terms is essentially the same
for all $p$ and $q$ is formulated as the
following {\it universality principle}:

{\it The contact term between
the
observables
$\CO_{\phi_{1}}^{(p)}$ and
$\CO_{\phi_{2}}^{(q)}$
is equal to $\CO_{C(\phi_{1}, \phi_{2})}^{(p+q-4)}$,
with some universal (i.e. independent of $p$ and $q$)
$C({\phi}_{1}, {\phi}_{2})$.}

\medskip
\sssec{The \quad principle \quad of \quad holomorphic  \quad integration} is
rather methodological.  It states that in order to figure out the structure
of the contact terms and to compute the correlation functions it is
sufficient to look at the expression for the effective action (with all
observables included) and evaluate it on the harmonic forms only.
Also, one
needs only the minimal number of fields entering the twisted $\CN=2$
multuplet (see below).
It is analogous to the methods used in \ttstar.
This principle {\it does not} explain why the
contribution of the moduli space of vacua vanishes for the manifolds
$\Sigma$
with $b_{2}^{+}>1$.
Nevertheless it proves to be useful in the course of
getting the right expressions for the observables.
To make the statetement
properly we need to reinspect the twisted supersymmetry transformations.
The following chapter is devoted to this question.

\newsec{Geometry of twisted supersymmetry.}

We start with reminding the structure of
the twisted vector multiplet,
$Q$-symmetry and observables. We discuss the
subtleties concerning preserving duality invariance and present the
formulae for the modular covariant $Q$-operator in the effective
low-energy abelian gauge theory.
We discuss $Q$-invariant observables
in the ultraviolet non-abelian theory and their
low-energy abelian counterparts.
We show that the proposed picture of $\Gamma$-invariant
Lagrangian submanifolds of $\IC^{2r}$ naturally arises in the
attempt to define the low-energy effective measure.

\subsec{Microscopic pure gauge theory. $Q$ operator and observables.}

The field content of the twisted $\CN=2$ vector multiplet is the gauge field
$A = A_{\mu}dx^{\mu}$,
the complex scalar $\phi$ and its conjugate ${\bar \phi}$ and the fermions:
the one-form $\psi$, the scalar $\eta$ and the self-dual two-form $\chi$.
All fields take values in the adjoint representation of the gauge group.
The $Q$-transformation has the form;
\eqn\susy{\eqalign{ Q \phi = 0, \quad  Q {\bar \phi} & = \eta,
\quad Q \eta = [\phi, \bar \phi] \cr
Q \chi = H, \quad  & Q H= [\phi, \chi] \cr
Q A = \psi, \quad & Q \psi = d_{A}\phi \cr}}
Here $H$ is an auxiliary bosonic self-dual two-form field with values
in the adjoint representation.
The operator $Q$ squares to the gauge transformation generated by $\phi$.

The theory has a set of natural observables.
Start with invariant polynomial $\CP$ on the algebra $\lieg$.
Let $C^{k}, k=0, \ldots 4$ be the closed $k$-cycles in the space-time
manifold $\Sigma$. Their homology cycles are denoted as
$[ C^{k} ] \in H_{k} (\Sigma; {\IC})$.
The observables form the descend sequence:
\eqn\dscndsqnc{\eqalign{
\CO^{(0)} =  {\CP} ({\phi}) & \quad \{ Q, \CO^{(0)} \} = 0 \cr
d \CO^{(0)} = \{ Q , \CO^{(1)}  \} & \quad ( \CO^{(1)}, [ C^{1} ] )
= \int_{C^{(1)}} \CO^{(1)} =
\int_{C^{1}} {{\p {\CP}}\over{\p \phi^{a}}} \psi^{a} \cr d \CO^{(1)}
= \{ Q, \CO^{(2)} \} & \quad
( \CO^{(2)}, [ C^{2} ] )
= \int_{C^{2}} \CO^{(2)} =
\int_{C^{2}} {{\p {\CP}}\over{\p \phi^{a}}} F^{a}  + {\half} {{\p^{2}
{\CP}}\over{\p \phi^{a} \p \phi^{b}}} \psi^{a} \wedge \psi^{b} \cr}}
and so on.

In particular, the top degree observable has
the form:
$$
\CO^{(4)}_{\CP} =
{\half} {{\p^{2} \CP}\over{\p \phi^{a} \p \phi^{b}}} F^{a}  F^{b} +
{1\over{3!}} {{\p^{3}\CP}\over{\p \phi^{a} \p \phi^{b} \p \phi^{c} }}
F^{a} \psi^{b}  \psi^{c} + {1\over{4!}} {{\p^{4} \CP}\over{\p \phi^{a}
\p\phi^{b} \p \phi^{c} \p \phi^{d}}} \psi^{a} \psi^{b} \psi^{c} \psi^{d}
$$
It enters the Seiberg-Witten low-energy effective
action, where all the fields are specialized to be abelian. In general, the
whole action $S$ equals the sum of the
$4$-observable, constructed
out of the prepotential $\CF$ and the $Q$-exact term:
\eqn\actn{S = \CO^{(4)}_{\CF} + \{ Q, R \} . }
The standard choice is
\eqn\actnst{\CF = \left( {{i\theta}\over{8\pi^{2}}} + {1\over{e^{2}}} \right)
{\Tr}{\phi^{2}}, \quad R = {1\over{e^{2}}}
{\Tr} \left(\alpha \chi  F^{+} - \beta \chi H +
\gamma D_{A}\bar\phi \star \psi  +
\delta \eta \star  [\phi, \bar\phi] \right) }
for $\alpha=\beta=\gamma=\delta = 1$.

The gauge invariant observables annihilated by $Q$ form a special
class of operators. Their correlation functions do not
change under a small variation of metric on the four-manifold $\Sigma$.
Let us denote
\eqn\cmlg{\CV = \oplus_{p=0}^{4} H^{p}({\Sigma}, {\IC})}

In the Donaldson theory ($G = SU(2)$ or $G = SO(3)$) one's aim is to compute:
\eqn\corr{\langle \exp ( (\CO^{(2)}_{u}, w)  + \lambda \CO^{(0)}_{u} )
\rangle}
where $w \in H^{2}(\Sigma, {\IR})$,
$\CO^{(0)}_{u} =  u \equiv {\Tr} {\phi}^{2}$,
\eqn\obsrv{(\CO^{(2)}_{u}, w )  = -{1\over{4\pi^{2}}}
\int_{\Sigma} {\Tr} ({\phi}F +{\half} {\psi}{\psi}) \wedge w}

From the point of view of topological theory the full
set of correlators of interest is given by the generating
function of correlators of all $p$-observables:
\eqn\corri{\CZ( \{ T^{k,\alpha} \}) =
\langle  e^{ T^{k,\alpha}(\CO^{(4-d_{\alpha})}_{u^{k}}, e_{\alpha})} \rangle}

\sssec{Choices.} In principle
the dependence on the coefficients $(\alpha,\beta,\gamma,\delta)$
which were introduced in \actnst\ might be non-trivial.
Nevertheless the following general remarks are in order. There exist
the following non-anomalous symmetries of the measure:
\eqn\smtrsi{\eqalign{
(\chi, H) & \mapsto t_{1} (\chi, H)\cr
(\eta, \bar\phi) & \mapsto t_{2}(\eta, \bar\phi)\cr}}
Hence the
correlation functions depend only on the orbit
of $(\alpha, \beta, \gamma, \delta)$  under the action of the group
of rescalings:
\eqn\smtrsii{(\alpha, \beta, \gamma, \delta) \mapsto (t_{1}^{-1}\alpha,
t_{1}^{-2}\beta, t_{2}^{-1}
\gamma, t_{2}^{-2}
\delta)}
One momentarily notices that the points $(1,1,1,1)$ and $(1,0,1,0)$
belong to the different orbits. The orbit of $(1,1,1,1)$ corresponds
to the $\CN=2$ twisted theory, while $(1,0,1,0)$ has a transparent
meaning from the point of view of Donaldson theory. Indeed,
if $\beta=\delta=0$ then the action \actn\
yields the Gaussian integral  which localizes onto the space of
solutions of the following system:
\eqn\sstm{\eqalign{F^{+}_{A} \quad & =\quad 0\cr
d^{+}_{A}\psi = 0, \quad &\quad d_{A}^{*}\psi = 0\cr
\phi \quad  = & {1\over{\Delta_{A}}} [ \psi, \star \psi] \cr}}
where $\Delta_{A} = d_{A}^{*}d_{A}$ is the gauge-covariant
Laplacian on the scalars. The rest of the fields is integrated out.
We assume that the gauge -fields are irreducible, i.e. there
is no normalizable solutions to the equation $d_{A}\phi=0$.

The first equation implies that $A$ is an anti-self-dual gauge field.
The second equation implies that $\psi$ is the tangent vector
to the moduli space $\CM$ of the ASD gauge fields. The third
equation means that $\phi$ is certain $\lieg$-valued
two-form on $\CM$. More thorough study (which involves
gauge fixing and Faddeev-Popov ghosts $c$)
identifies $\phi + \psi + F$ with the curvature of the universal
connection $A + c$ in the universal
bundle $\CE$ over  $\CM \times \Sigma$.

The solutions to \sstm\ when substituted into
\dscndsqnc\ produce  representatives
of  cohomology classes of the moduli space $\CM$. The correlation functions
are the integrals of the cohomology classes over the fundamental cycle
of the compactified moduli space $\bar \CM$. It is non-trivial
to check that the representative constructed with the help of
\sstm\ actually determines a cohomology class of $\bar \CM$.
For the observables constructed out of ${\Tr}{\phi}^{2}$ it has been
done by indirect methods (cf. \DoKro).
For the higher observables ${\Tr}{\phi}^{k}$ it is not known
to authors.

``\sssec{Quantum \quad multiplication.}'' Classicaly, in $SU(2)$ theory
one certainly has the equality:
\eqn\cle{{\Tr}{\phi}^{2r} = 2^{1-r} \left( {\Tr}{\phi}^{2} \right)^{r}}
When the expressions \sstm\ are substituted in \cle\
one gets the equality of the representatives of the cohomolgy classes
of $\biggl[ {\Tr} {\phi}^{2r} \biggr]$ and
$2^{1-r} \biggl[ {\Tr} {\phi}^{2} \biggr]^{r}$ {\it on} $\CM$. However
the
divergencies which  appear
near the boundary $\bar \CM \backslash \CM$
of $\CM$ may spoil the validity of \cle\ in $H^{*}(\bar\CM)$.
Thus we expect that {\it a priori} the observables
constructed out of ${\Tr}{\phi}^{2r}$ may differ from the cohomology
class of $2^{1-r}\biggl[{\Tr}{\phi}^{2}\biggr]^{r}$ where the
power is understood as a multiplication in the  cohomology
of $\bar \CM$. We return to  this subtlety in the chapter $5$.

\subsec{Macroscopic theory. Electric-magnetic duality.}

We shall make use of the low-energy effective theory, whose  action
on $\IR^{4}$
has been computed in \SeWi, and certain
aspects of it for the general four-manifold $\Sigma$
have been worked out in \wittabl\ and also recently in \grwt.

The low-energy theory contains $r$
$\CN=2$ vector multiplets, which
are defined up to $\Gamma$- transformation, where
$\Gamma$ is a subgroup of $Sp_{2r}({\IZ})$, e.g ${\Gamma}(2)$ or
${\Gamma}_{0}(4)$ for $r=1$.
Let us denote the scalar components of the multiplet, which are monodromy
invariant (up to a sign) at $u^{k} = \infty$ by $a^{i}$.
Then the $S$-dual ones will be denoted as $a_{i,D}$.
The low-energy effective couplings are denoted as:
\eqn\lecpl{
\tau_{ij} (a) =
\left( {{4\pi i }\over{g_{eff}^{2}}} +
{{\theta_{eff}}\over{2\pi}}\right)_{ij} =
 {{\p a_{i,D}}\over{\p a^{j}}}
= \tau_{ij, 1} +i {\tau}_{ij, 2} \equiv \Re\tau_{ij} + i \Im\tau_{ij}
}

Our problem here is to write down the $Q$ transformations for the fields
entering the twisted vector multiplet. Naively we might simply
adopt the transformations \susy\ from the previous section. The trouble
is that they are valid only for the theory with field independent
coupling constant. It is known from the studies of
two dimensional topological strings that the deformation of the theory
by the observable of top degree (two in two dimensional case)
changes the transformations of the fields.

Let us present the correct transformations and then discuss them:
\eqn\susyi{\eqalign{Q \psi^{i} = da^{i} \quad & \quad Q a^{i} = 0 \cr
Q ( {\Im\tau} \chi )_{i} = ( {\Im\tau} H)_{i}
\quad & \quad Q ( {\Im\tau} H )_{i} = 0 \cr
Q {\bar a}^{\bar i} = \eta^{\bar i} \quad & \quad Q \eta^{\bar i} = 0\cr}}
We introduce a notation, inspired by the analogy with
type $\bf B$ sigma model:
\eqn\tta{\theta_{i} = \Im\tau_{i\bar j}\chi^{\bar j}, \quad
{\bar F}_{i} = \Im\tau_{i\bar j}H^{\bar j}}

We also need the transformations for the gauge field. We shall write them a bit
later.

Notice that the transformations \susyi\ are consistent with the following
action of the modular group on fields:
\eqn\mdlri{\eqalign{\psi^{i} \mapsto \psi_{i,D}   =& \tau_{ij} \psi^{j} \cr
a^{i} \mapsto a_{i, D} \quad  \quad \tau_{ij} & \mapsto \tau_{D}^{ij}
= -({\tau}^{-1})^{ij}\cr
\chi^{\bar i} \mapsto \chi_{\bar i, D} =
\bar\tau_{\bar i\bar j} \chi^{\bar j} \quad
\eta^{\bar i} \mapsto \eta_{\bar i, D} &= \bar\tau_{\bar i\bar j} \eta^{\bar j}
 \quad H^{\bar i} \mapsto H_{\bar i, D}
= \bar\tau_{\bar i\bar j} H^{\bar j}\cr}}

The $Q$- transformations of the gauge fields also must be
consistent with the electric-magnetic duality. It is not clear a priori
that such a $Q$-action  exists, since the duality is a non-local
operation on the fields, while the $Q$ is a local one.
One may try to imitate
the twisted version of the supersymmetric duality transformation
presented in \SeWi. We do it in the next subsection,
while here simply present the result.
Introduce two more fields: the two-form $F^{i}$ and one-form $A_{i,D}$ with
the following $Q$-transformations:
\eqn\susyii{Q A_{i, D} = \tau_{ij} \psi^{j} \qquad  Q F^{i} = d\psi^{i}}
Then one has two options. Either
${\widetilde \CA_{i, D}}
= \left(A_{i, D}, \psi_{i,D}, a_{i, D}, \bar a_{\bar i, D}, \chi_{\bar i, D},
H_{\bar i, D}, \eta_{\bar i, D}\right)$ is
treated as a twisted multiplet with the standard $Q$-action,
or ``on-shell'' $F^{i} = dA^{i}$ and then
${\widetilde \CA}^{i}
= \left( A^{i}, \psi^{i}, a^{i} , \bar a^{\bar i},
\chi^{\bar i}, \eta^{\bar i}, H^{\bar i}\right)$ form the twisted
multiplet. The passage from ${\widetilde \CA}^{i}$ to
${\widetilde \CA_{\bar i,D}}$
can be done by a Gaussian integration and is discussed in some details
in the next two sections.

\subsec{Holomorphic  approach.}
Here we state the last
principle of the computations.
This principle allows one to check the modular invariance of the
measure in a relatively simple
setting, where all the $Q$-exact (see below) terms
are ``thrown out''and one works with (formal)
contour integrals.  Most of the constructions can be done working
with the ``holomorphic'' fields $a, \psi, A$ only. The only
trouble with such a prescription is the absence of Laplacians
and non-definiteness of the topological term $F \wedge F$.
The first problem is avoided in  certain cases by working
with cohomology (with harmonic forms) while the
second may be treated via analytic continuation. In any case
such an approach is useful in getting the right structures.
Once it is done one may introduce the quartet
of the fields ${\bar a}, \eta, \chi, H$ and justify the
constructions by working with the standard positive-definite actions.

\subsec{``Off-shell''holomorphic formulation}

Consider the {\it short superfield}:
\eqn\spfld{\CA_{i,D} = a_{i,D} + \psi_{i,D} + F_{i,D}}
where $dF_{i,D} = 0$.
The operator $Q$ acts as follows:
\eqn\susyiv{Q a_{i,D} = 0, \quad Q \psi_{i,D} = da_{i,D},\quad
QF_{D} = d\psi_{i,D}}
We impose (by hands) the condition that $F_{i,D}$ represents the
integral cohomology class of the space-time manifold $\Sigma$.
Thus,
$$
\CA_{i,D} \in \Omega^{0}(\Sigma)_{B} \oplus \Omega^{1}(\Sigma)_{F} \oplus
\Omega^{2}_{\IZ} ({\Sigma})_{B}
$$
Here $\Omega^{2}({\Sigma})_{\IZ}$ is the space of closed
two-forms with periods in $2\pi i\IZ$.
The indices $B, F$ denote the bosonic and fermionic fields respectively.
The superfield $\CA_{D,i}$ obeys
the condition $(Q-d)\CA_{D,i} = 0$.
One may
also fulfill the condition of $Q-d$-closedeness by
introducing a complete
set of $p$-forms which we call {\it the long superfield}:
\eqn\lngsp{\eqalign{ &
\matrix{\CA^{i} & = & a^{i}  + &\psi^{i} + &F^{i} + &\rho^{i} + &D^{i}\cr
  &  & 0 \quad & 1 \quad & 2 \quad & 3 \quad & 4 \quad\cr}\cr
& \CA^{i} = \sum_{p=0}^{4} \CA^{i,p} \in V = \oplus_{p=0}^{4} \Omega^{p}({\Sigma})\cr}}
and $Q$ acts as follows:
\eqn\susyiii{Q\CA^{i,p} = d\CA^{i,p-1}}
Let $\CF_{D}$ be a holomorphic function on $\IC^{r}$.
The ``action''
\eqn\aci{S = \int_{\Sigma}\CF_{D} (\CA_{D})} is clearly
$Q$-invariant.
The long superfield $\CA$ allows reparameterizations:
\eqn\dffm{\CA^{i} \mapsto {\widetilde{\CA^{i}}} (\CA)}
induced by the holomorphic maps $a^{i} \mapsto {\tilde a}^{i}(a^{k})$.
Let $\CL \subset \IC^{2r}$ be a $\Gamma$-invariant
Lagrangian subvariety.
Let $u^{k}, k=1, \ldots r$ be the generators of
the ring $\CW_{\CL}$ of  globally defined $\Gamma$-invariant
holomorphic functions on $\CL$. Extend them to the long superfields
$\CU^{k}, k=1, \ldots, r$. Define the measure
\eqn\msr{[\CD\CU] = \prod_{k=1}^{r} du^{k} d\psi_{u}^{k} dF^{k}_{u}
d\rho^{k}_{u}dD^{k}_{u}}
where $(Q - d) \left( {u}^{k} + \psi^{k}_{u} + F^{k}_{u} + \rho^{k}_{u} +
D^{k}_{u} \right) = 0$.

The duality transformation proceeds as follows:
introduce both $\CA_{i,D}$ and $\CA^{i}$ and consider the action
\eqn\acii{\eqalign{\quad  {\pi i} S^{\prime} = \int_{\Sigma}  &
{\CA}^{i}{\CA}_{i,D} - \CF({\CA}^{i})\cr
=
\int_{\Sigma} F^{i}F_{i,D} + \rho^{i} \psi_{i,D}
 +  D^{i} (a_{i,D} - &  \p_{i} \CF)
- {\half} {\p_{ij}^{2} \CF} (F^{i}F^{j} +
2\rho^{i}\psi^{j})\cr
- {1\over{2}}{\p^{3}_{ijk} \CF} F^{i}\psi^{j}\psi^{k} &
-{1\over{24}} {\p_{ijkl}^{4} \CF}   \psi^{i}\psi^{j}\psi^{k}\psi^{l}\cr}}
Let us consider the following (formal) path integral:
\eqn\frmlpi{
\int \CD \CU^{i} \CD \CA_{i,D} e^{-S}.}
The measure $\CD\CA_{i, D}$
is defined canonically. The dependence of the
measure on $\CA^{i}$ on the choice of the measure
$du^{1} \wedge \ldots du^{r}$ is
completely parallel to  anomaly in Type $\bf B$ sigma models
in two dimensions \ttstar\Witr.
The integral over $\CA_{i,D}$ (together with summation over
the fluxes of $F_{i,D}$) forces $D^{i}, \rho^{i}$ to vanish,
while $F^{i}$ becomes a curvature of a connection $A^{i}$.
We hope that the reader will not confuse
the fields $D^{i}, \rho^{i}$ etc. entering $\CA^{i}$ with
the fields $D^{i}_{u}, \rho^{i}_{u}\ldots$ which enter
$\CU^{i}$. Of course, there is a simple formula which expresses
$D^{i}_{u}, \rho^{i}_{u}, \ldots$ in terms of $D^{i}, \rho^{i}, \ldots$.
As a result one gets a measure
$$
{\Det}_{ij} \left( {{\p u^{i}}\over{\p a^{j}}} \right)^{\chi \over{2}}
\prod_{k=1}^{r} d a^{k} d\psi^{k}  dF^{k}
$$

On the other hand, performing
the integral over $\CU$ gives us:
\eqn\dle{\eqalign{a_{i,D}  = {{\p \CF}\over{\p a^{i}}}, \quad & \quad
\psi^{i}  = \left({\tau}^{-1}\right)^{ij}\psi_{j,D} \cr
F^{i}  =  \left({\tau}^{-1}\right)^{ij} & \left(
F_{j, D}   - \half ({\tau}^{-1})^{lm} ({\tau}^{-1})^{kp}
(\p^{3}_{lkj}\CF) \psi_{m,D}\psi_{p,D}
\right)\cr}}
with
$$
\tau_{ij} = {{\p^{2} \CF}\over{\p a^{i}\p a^{j}}}
$$
The determinants in this case are slightly more involved:
\eqn\mdlr{\eqalign{
{\Det}_{ij} \left( {{\p u^{i}}\over{\p a^{j}}} \right)^{\chi \over{2}} &
\left( {\rm Det}\tau\right)^{-{\rm dim}\Omega^{0} + {\rm dim}\Omega^{1}  - {\half}
{\rm dim}\Omega^{2}}  = \cr
({\Det}\tau)^{-{{\chi}\over{2}}} &
{\Det}_{ij} \left( {{\p u^{i}}\over{\p a^{j}}} \right)^{\chi \over{2}} = \cr
{\Det}_{ij}& \left( {{\p u^{i}}\over{\p a_{D,j}}} \right)^{\chi \over{2}}\cr}}
Without the
factor ${\Det}_{ij} \left( {{\p u^{i}}\over{\p a^{j}}} \right)^{\chi \over{2}}$
the duality transformation would be anomalous.
This anomaly was already observed in
\wittabl\ (for $r=1$).
The ``action'' \acii\ turns into
\aci\ with the substitution of $\CF$ by $\CF_{D}$,
which is the Legendre transform of $\CF$.

\subsec{Justification of the action}
In this section we pass from the holomorphic approach to
the ``harmonic'' one. We introduce the regulator fields
$H, \chi, \eta, \bar a$ and make the manipulations of
the previous section well-defined.
We first write down the expression for four-observable:
\eqn\actni{\CO^{(4)} = {\half}{\tau}F \wedge F + {\half}
{{\p\tau}\over{\p a}} F \psi^{2} +
{1\over{24}}
{{\p^{2}\tau}\over{\p a^{2}}} \psi^{4}
+
F F_{D} }
where we write $F_{D}= dA_{D}$ in order to stress the fact that
$F_{D}$ may be closed, but not exact form with integral periods.
Now let us add to \actni\ a $Q$-exact term, which would enforce
electric-magnetic duality\foot{We should warn the reader
that we merely point out the appearing structures without
exact coefficients}:
\eqn\acntii{L = {i\over{4}}\CO^{(4)} + \{ Q, R_{0} \} \quad R_{0} =
{\tau}_{2} \left( \chi ( F^{+} -  H) + d\bar a \star \psi \right)
+ {1\over 2}{{d\tau_{2}}\over{da}} \psi^{2} \chi +
{1\over 6} {{d\tau_{2}}\over{d\bar a}} \chi^{3}
}
Expanding $\{ Q, \ldots \}$ out we get:
\eqn\actniii{\eqalign{L & = {i\over{8}}  \tau F^{2} + FF_{D} +
{\tau}_{2} \left( H (F^{+} - H) + da \star d\bar a \right) + \cr
&
+ {\tau}_{2} \left( \chi (d\psi)^{+} + \eta d^{*}\psi \right) + \cr
& + {i\over{8}} {{d\tau}\over{da}} F\psi^{2} +
{{d\tau}\over{da}} \chi (da \wedge \psi) +
H
\left( {{d\tau_{2}}\over{d\bar a}} ({\half}\chi^{2} + \chi \eta)  +
{\half}{{d\tau_{2}}\over{da}} \psi^{2}  \right) \cr
& + {i\over{96}} {{d^{2}\tau}\over{da^{2}}} \psi^{4} -
{\half}{{d{\rm log}\tau_{2}}\over{d\bar a}} {{d\tau}\over{da}} \chi \eta \psi^{2}
-{1\over{12}}{{d^{2} (\tau_{2}^{-2})}\over{d\bar a^{2}}} \eta
(\tau_{2}\chi)^{3}
\cr}}
Performing gaussian integration over $H$  we get:
$$
H = {\half} F^{+} + {1\over{\tau_{2}}}
\left( {{d\tau_{2}}\over{d\bar a}} ({\half}(\chi^{2})^{+} + \chi \eta)  +
{{d\tau_{2}}\over{da}} (\psi^{2})^{+}  \right)
$$
and
\eqn\actniv{\eqalign{-i \CL &= {\half} ( \tau (F^{-})^{2} - \bar\tau (F^{+})^{2} )
+ {\tau}_{2} (\chi (d\psi)^{+} + \eta d^{*}\psi + da\star d\bar a)\cr
&+ {\half} {{d\tau}\over{da}} F(\psi^{2})^{-} +
+ {{d\tau}\over{da}} \chi (da\wedge \psi) + FF_{D} +\cr
&+ F^{+}  {{d\tau_{2}}\over{d\bar a}} ({\half}(\chi^{2})^{+} + \chi \eta)  +
\ldots\cr}}
where $\ldots$ denote the quartic fermionic terms.
\medskip
{\quad \it Duality}
\medskip
\noindent
The duality is manifested in two ways one may treat the fields
$F$ and $A_{D}$. The first option is to integrate $A_{D}$ out and to
sum over all line bundles thus ensuring that $F$ is a closed
two-form with integral periods. Then $F = dA$, where $A$ is a
connection in some line bundle over $X$. The $Q$ transformation for
$A$ can be read off from \susyii : $Q A = \psi$.

The second option is to integrate out $F$. We get:
\eqn\fout{\eqalign{F^{-} &= -{1\over{\tau}} \left( F_{D}^{-} +
{{d\tau}\over{da}} (\psi^{2})^{-} \right)\cr
F^{+} & = -{1\over{\bar \tau}} \left( F^{+}_{D} +
{{d\tau_{2}}\over{da}} \eta\chi \right)\cr}}

Both the action \actniv\  and the $Q$ transformations
become identical to those we get in the previous case with the
replacements of all the fields by their duals. In particular the
transformation law for $A_{D}$ is $Q A_{D} = \psi_{D}$ as expected.

As an illustration of how this works let us look at the quartic
fermionic terms, involving $\psi^{4}$:

\eqn\ffrmntrm{\xi_{\psi^{4}} = {1\over{24}} {{d^{2}\tau}\over{da^{2}}} \psi^{4}
- {1\over{\tau_{2}}} \left({{d\tau_{2}}\over{da}}
(\psi^{2})^{+}
\right)^{2}
}
As a result of integration over $F$ this term changes by a piece
$\delta \xi_{\psi^{4}} =
- {1\over{\tau}} \left({{d\tau}\over{da}}
(\psi^{2})^{-}
\right)^{2}$. The sum $\xi_{\psi^{4}} + \delta\xi_{\psi^{4}}$
is equal to \ffrmntrm\ after the substitution $a \to a_{D}$ etc.

\newsec{\bf Correlation functions}

In the next sections we explain the origin of the
principles and derive the formulas for the correlators.
We also test them in certain simple cases, where alternative
calculations are available.

\subsec{Observables and Contact terms: First appearance}

We discuss the issue of observables first in holomorphic setting.
Namely, we study formally the deformations of four observable
by adding $p$-observables. We immediately face the necessity of
introduction of the
contact terms in order to preserve modular invariance.
We then prove that the deformations along the nilpotent
directions, studied in the previous section contain
those terms.
In the next section we shall supply the expressions
for the deformed action, containing the anti-holomorphic
fields.

The first question is to determine the appropriate deformation
space. Since one may add all $p$-observables to the action it
is natural to assume that the tangent space to the space
of deformations is the space $\CV_{\CL,\Sigma}$
of $\Gamma$-invariant
$\CV_{\Sigma} \equiv H^{*}({\Sigma}, \IC)$-valued holomorphic
 functions $f$ on $\CL$.

Given such a function
$f(a)$ one may analytically continue it to a holomorphic
function on $\CV$ ($d$-exact terms do not
matter) and form the action:
\eqn\dfac{S = \int_{\Sigma} -\CA\CA_{D} + \CF(A)}

The total deformation can be
expanded in a series:
\eqn\ttldfr{\CF = \CF_{0} + \sum t^{(p)}\CF_{(p)} +
\sum t^{(p_{1})}t^{(p_{2})} \CF_{(p_{1}p_{2})} + \ldots, \quad f =
t^{(p)}\CF_{(p)}}
One may get a set of recursion relations between the terms
$\CF_{(p_{1}\ldots p_{k})}$ by examining the duality properties of \ttldfr.

Write the deformed $\CF$ as a sum: $\CF = \sum_{p=0}^{4} \CF_{p}$,
$\CF_{p} \in H^{p}({\Sigma})$. The action
density  $\CF(\CA)  - \CA\CA_{D}$
equals ($\p = \p / \p a$):
\eqn\exi{\eqalign{S  = a_{D} D + & \psi_{D} \rho + F_{D} F  +\cr
 + \p\CF_{0} D + {\half} \p^{2}\CF_{0}  ( F^{2} + 2\psi \rho) +
& {\half} \p^{3}\CF_{0} F\psi^{2} + {1\over{24}} \p^{4}\CF_{0} \psi^{4}\cr
 + \p\CF_{1} \rho + \p^{2}\CF_{1} & F \psi + {1\over{6}} \p^{3}\CF_{1}
\psi^{3} + \cr
 + \p\CF_{2} F + & {\half} \p^{2}\CF_{2} \psi^{2} + \cr
 + \p \CF_{3}& \psi + \cr
 + & \CF_{4} \cr}}

If we integrate out $a_{D}, \psi_{D}, F_{D}$ first, then
we get the standard action in the presence of observables.
If $D, \rho, F$ are integrated out then the rest of the fields
assumes the values:
\eqn\exii{\eqalign{
a_{D} &= \p\CF_{0}\cr
\psi_{D}  = \p^{2}\CF_{0} & \psi + \p\CF_{1}\cr
F_{D}  = \p^{2}\CF_{0} F + {\half}\p^{3}\CF_{0} & \psi^{2} + \p^{2}\CF_{1}
\psi + \p \CF_{2}\cr}}
and the new action is equal to ($\tau = \p^{2}\CF_{0}$):
\eqn\exiii{\eqalign{
S_{D}  = - \int_{\Sigma}
{1\over{2\tau}} ( F_{D} - {\half}\p^{3}\CF_{0} & \psi^{2} -
\p^{2}\CF_{1}\psi - \p\CF_{2})^{2}\cr
+ {1\over{24}}
 \p^{4}\CF_{0}\psi^{4} + {1\over{6}} \p^{3}\CF_{1}\psi^{3} & + {\half}
\p^{2}\CF_{2}\psi^{2} + \p\CF_{3}\psi + \CF_{4}\cr}}
We have already taken
care of the determinants by the virtue of the factor
$\left({{du}\over{da}}\right)^{{\chi}\over{2}}$ (cf. \mdlr).
The new action can be
interpreted as the old one written in new coordinates $a_{D} = \p\CF_{0},
\psi_{D} = \tau \psi$ and with the new function $\CF_{0}^{D}$ such that
${{\p^{2}\CF_{0}^{D}}\over{\p a_{D}^{2}}} = - {1\over{\tau}}$
iff the following equations are obeyed:
\eqn\master{\CF_{p}(a_{D})  - \CF_{p}(a) =
\sum_{l=2}^{4} {{(-1)^{l-1}}\over{l!}}
\sum_{i_{1} + \ldots i_{l} = p} {{\p^{l-2}}\over{\p a_{D}^{l-2}}}
{1\over{\tau}}
{{\p \CF}_{i_{1}} \ldots {\p\CF}_{i_{l}}}}
modulo terms which do not affect the observables, e.g.
$\CF_{3} \sim \CF_{3} + {\rm const}$.
Some of the equations \master\ can be solved explicitly.
In particular,
if all $\CF_{l}$ for $l < p$ vanish then
$\CF_{p}$ must be modular invariant: $\CF_{p} = g_{p}(u)$.
Also, if $\CF_{1} = 0$, then $\CF_{2}$ and $\CF_{3}$ are modular
invariant $\CF_{2,3} = g_{2,3}(u)$, while
\eqn\cffr{\CF_{4} = g_{4}(u) - (\p g_{2}(u))^{2} \p_{\tau} {\rm log}
\theta(\tau)}
Here $g_{p}(u)$ are modular invariant functions,
\eqn\thta{\eqalign{\theta ({\tau}) & = \theta_{00}({\tau},0), \cr
\theta_{\alpha\beta}({\tau}, z)  = \sum_{n\in\IZ} & (-)^{n\alpha}
e^{{\pi i} \left({\tau}
(n + {\beta}/2)^{2} + 2(n + {\beta}/2)z\right)}, \quad \alpha,\beta = 0,1 \cr}}
Although the solution of the equations \master\ seems to be a complicated
problem in general the formalism of Hamilton-Jacobi equations
\hmjcb\ automatically takes care of it  since the
modular invariance is build in it.
Indeed, if we forget about the nilpotent terms in $f$ then the condition
on $\CF_{0}$ is simply the requirement that it defines a new
$\Gamma$-invariant
Lagrangian submanifold $\CL$. The observables modify the equation
of the Lagrangian submanifold to \exii. One may think of \exii\
as of some kind of supermanifold in the space with coordinates
$a, a_{D}, \psi, \psi_{D}, F, F_{D}$. Our super-Hamilton-Jacobi
equations \shmjb\ are simply the generalizations of the standard
canonical formalism for such supermanifolds.

The general problem is to understand how to extend a function $f$
on $\CL$, perhaps with values in $\CV_{\Sigma}$ to a globally
wel-defined Hamiltonian on $\IC^{2r}$. Our proposal is to use the
homogeneity properties of the asymptotically free theories. In particular,
we extend the function $u$ to the homogeneous function
of degree $2$. Of course, in the case of rank higher then one this
condition does not fix the continuation uniquely. But it is
quite satisfying to see that in the case $r=1$ this principle leads
to the sensible results consistent with ghost number anomalies.

\subsec{The tests of the universality and homogeneity}

In this section we
test the proposed principles.

\sssec{Ghost \quad number \quad anomaly}

The first test deals with the deformation of the microscopic
theory by the four-observable constructed out of
$-{1\over{8\pi^{2}}}{\Tr}\phi^{2}$.
Consider the $SU(2)$ theory with $N_{f}$
massless hypermulitplets
in the fundamental representation.
The deformation amounts to the multiplication of the $k$-instanton
contribution to any correlation function by the factor $e^{2\pi i tk}$.
This can be accomplished by multiplying every operator $\CO$
of the ghost charge $\Delta_{\CO}$ by the factor\foot{The factor
is motivated by the ghost number anomaly $2(4-N_{f})k - {3\over{2}}{\chi} -
{{N_{f}+3}\over{2}} \sigma$ in the case of trivial $Spin^{c}$ structure,
see below}
$\exp ( {{\pi i t}\over{4-N_{f}}} {\Delta}_{\CO} )$ and the whole
correlation function by the factor
\eqn\fctr{
\exp ( {{3(\chi + \sigma) + N_{f} \sigma}\over{2(4-N_{f})}} {\pi i}t )}
(neglecting for the moment the subtleties related to the choice of
$Spin^{c}$ structure, see below). Since the effective
measure contains a factor $\left( {{du}\over{da}} \right)^{\chi \over{2}}$,
the $\chi$-dependent part of \fctr\ can be absorbed
into the redefinition of $a$:
\eqn\absb{a \mapsto  e^{{2\pi i t}\over{4-N_{f}}} a }
(one should remember that the measure has an anomaly
under the transformation \absb\ which is proportional to
${\chi + \sigma}\over{2}$). The remaining  part
${{2 + N_{f}}\over{4(4-N_{f})}}$
is absorbed in the redefinition of the
$\sigma$-dependent factor in the measure: $\Delta^{\sigma \over{8}}$,
since $\Delta$ is the polynomial in $u$ of degree $N_{f} + 2$.
Nevertheless, as $\Delta$ does not coincide with $u^{N_{f}+2}$ the last
rescaling maps $\Delta(u)$ to another polynomial\foot{which
can be obtained by rescaling $\Lambda_{N_{f}}$} in $u$. To get around
this point we ``undo'' the transformation on $u$ inside the
integral using the invariance of integral under the changes
of variables. As a result the function $a(u)$ has changed:
\eqn\afl{a (u) \mapsto e^{{2\pi i t}\over{4-N_{f}}} a
(e^{-{4\pi it}\over{4-N_{f}}} u ) }
Also, all the $p$-observables which involve $a$ or its derivatives
have changed.
We claim that \afl\ is the result of the
evolution with respect to the degree $d=2$ homogeneous
Hamiltonian
$H(a, a_{D})$, whose restriction on $\CL$ is $u$.
Indeed, since the Hamiltonian is the conserved quantity in the
course of evolution the following identity holds:
\eqn\idnt{u = H( a(u), a_{D}(u) ) = H
\left( e^{{2\pi it}\over{4-N_{f}}} a (e^{-{4\pi it}\over{4-N_{f}}} u ),
e^{{2\pi it}\over{4-N_{f}}} a_{D} (e^{-{4\pi it}\over{4-N_{f}}} u ) \right)}
To check that the parameter $t$ actually coincides with the
evolution time (it could have been some function of the latter) it
is sufficient to compare the first derivatives. Let the Hamiltonian
$H$ be the degree $d=2$
homogeneous extension of the function $\gamma \cdot u$ where $\gamma$ is
some constant.
As we explain momentarily the equations of motion \hmem\ yield:
\eqn\fstdr{{{da (u,t)}\over{dt}}  = {{\p H}\over{\p a_{D}}}
 = \gamma {{a - 2u{{da}\over{du}}}\over{W(u)}} }
where
\eqn\wrnsk{W(u) = a {{da_{D}}\over{du}} - a_{D} {{da}\over{du}}}
The equations \fstdr\wrnsk\ follow from the two conditions \ext:
\eqn\exti{\eqalign{
{{da}\over{du}} {{\p H}\over{\p a}} & + {{da_{D}}\over{du}}
{{\p H}\over{\p a_{D}}} = \gamma\cr
a {{\p H}\over{\p a}} &+ a_{D} {{\p H}\over{\p a_{D}}} = 2 \gamma u\cr}}
where all derivatives are restricted onto $\CL$.
So it is enough to find $\gamma$ such  that
$$
W(u) = \gamma {{4 - N_{f}}\over{2\pi i}}
$$
Since in all cases with $N_{f} < 4$ both $a$ and $a_{D}$ are the solutions
of the corresponding Picard-Fuchs equations of the form:
$$
{{d^{2}}\over{du^{2}}} \pmatrix{a \cr a_{D}\cr} = {{f(u)}\over{\Delta(u)}}
\pmatrix{a \cr a_{D}\cr}
$$
for appropriate $f(u), \Delta(u)$
the quantity $W(u)$
being
the Wronskian
is actually $u$-independent.
To calculate its value it is sufficient
to check the large $u$ asymptotics in the normalization
$a (u) \sim {\half} \sqrt{2u} + \ldots,
a_{D}(u) \sim {{N_{f}-4}\over{2\pi i}} a(u) {\rm log} u + \ldots$.
Using the formula (4.4) of the second
paper in \SeWi\   we
get that $\gamma = -\half$.
Thus the evolution with respect to the Hamiltonian $-{\half} u$
leads to the expected dependence on the instanton charge.

\sssec{Contact \quad terms \quad of \quad two-observables}.

The paper
\gz\ computes the Donaldson invariants of $\IP^{1} \times \IP^{1}$
and other rational surfaces. It is possible
to extract the expression for the contact term between the two
$2$-observables constructed out of $u$:
\eqn\cnttrm{C_{2,2}(u,u) = \left( {{du}\over{da}} \right)^{2} G({\tau}),}
where
\eqn\efrlti{G({\tau}) = - {1\over{24}} \left(
E_{2}({\tau}) - 8u \left( {{da}\over{du}}\right)^{2} \right)}

We use the formulae:
\eqn\efrlt{\eqalign{a = &
{{2E_{2} + \theta_{01}^{4} + \theta_{00}^{4}}\over{6\theta_{00}\theta_{01}}} \cr
{{da}\over{du}} & = {{\theta_{00}\theta_{01}}\over{2}} \cr
u = & \left( {{\theta_{00}^{4} + \theta_{01}^{4}}\over{
2 \theta_{00}^{2} \theta_{01}^{2}}}\right) \cr}}
It is easy to check that our expression
\eqn\cttrm{{1\over{4}} \left( a{{du}\over{da}} - 2u \right)}
coincides with \cnttrm\ for the pure $SU(2)$ theory.
It is also easy to check that for the theory with $N_{f}$ flavors
our proposal \cnctrmii\ yields for $p=2$:
\eqn\cttrmi{C_{2,2} (u,u) = {1\over{4-N_{f}}}
\left( a{{du}\over{da}} - 2u \right)}
which again coincides with
$$
- {1\over{24}} E_{2}({\tau})\left( {{du}\over{da}} \right)^{2} +
{{u}\over{3}}
$$
of \grwt (provided that one performs the important shift $u \mapsto
u + {2\over{3}}$ which is not detected just by the modular invariance
and asymptotics at infinity requirements of \grwt).

\sssec{Derivation \quad from \quad blow \quad up.} In order
to derive the contact term between the two two-observables $\CO^{(2)}_{u}$
in the theory with massless quarks we perform the
blow up of the manifold $\Sigma$ at the point of
the intersection of the two-surfaces $C_{1}$ and $C_{2}$.
The next arguments are presented for the case of the gauge group
$G = SU(N)$ but are easily generalized for any simple-laced
gauge group.
We assume that the intersection is transverse and that it contributes
$+1$ to the intersection index $C_{1} \cap C_{2}$. The blow up
can be achieved by gluing to $\Sigma$ a copy of the
projective plane $\overline{\IC\IP}^{2}$ with the orientation
opposite to the one induced from the complex structure:
$\tilde \Sigma = \Sigma \# \overline{\IC\IP}^{2}$.
The homology lattice  $H_{*}(\tilde \Sigma)$
of $\tilde\Sigma$ is that of $\Sigma$ plus a factor of
$\IZ$. The intersection form is simply
$$
(,)_{\tilde \Sigma} = (,)_{\Sigma} \oplus (-1)
$$
as the exceptional divisor $e$ (the two-sphere inside
$\overline{\IC\IP}^{2}$)
has self-intersection $-1$.
Under the isomorphism $H_{*}(\tilde\Sigma) = H_{*}(\Sigma)
\oplus {\IZ}$
the inverse images of the cycles in $\Sigma$ belong to the component
$H_{*}(\Sigma)$ of $H_{*}(\tilde\Sigma)$. We shall denote them
by the same letters as the cycles in $\Sigma$.
To derive the contact term we compare the correlation functions
\eqn\crfi{
\langle \int_{C_{1}} \CO^{(2)}_{u} \int_{C_{2}} \CO^{(2)}_{u}
\ldots\rangle_{\Sigma}}
and
\eqn\crfii{\langle \int_{\tilde C_{1}} \CO^{(2)}_{u}
\int_{\tilde C_{2}} \CO^{(2)}_{u}
\ldots\rangle_{\tilde \Sigma}}
where the cycles $\tilde C_{1}, \tilde C_{2} \in H_{*}({\tilde\Sigma})$
do not intersect each other in the vicinity
of $p$ and are given by the formulae:
\eqn\lft{{\tilde C}_{k} = C_{k} - e,\quad
 \# {\tilde C}_{1}\cap {\tilde C}_{2} =
\# C_{1}\cap C_{2} - 1}
First of all:
\eqn\crfiv{
\langle \int_{C_{1}} \CO^{(2)}_{u} \int_{C_{2}} \CO^{(2)}_{u}
\ldots\rangle_{\Sigma} =
\langle \int_{C_{1}} \CO^{(2)}_{u} \int_{C_{2}} \CO^{(2)}_{u}
\ldots\rangle_{\tilde \Sigma}}
This follows from the topological invariance of the correlators:
the point of blow up can be taken far away from the intersection
point of $C_{1}$ and $C_{2}$. The very existence of the blowup
of the zero size should not affect the correlation functions
of the operators which do not contain integration over the
glued ${\overline{\IC\IP}}^{2}$.
The next step is the exploring of the structure of the moduli space
of instantons on $\tilde \Sigma$. Let us take the metric
on $\tilde \Sigma$ to be such that the glued   ${\overline{\IC\IP}}^{2}$
is very far away from the rest. By conformal transformation
we may view the manifold $\tilde \Sigma$ as the bouquet
of $\Sigma$ and ${\overline{\IC\IP}}^{2}$ glued at the point
$p$. This picture is similar to the familiar representation
of the moduli space of holomorphic bundles on two dimensional
surface (see, e.g. \nhol). The instanton charge may be divided
between $\Sigma$ and ${\overline{\IC\IP}}^{2}$.
We represent the moduli space of instantons on $\tilde\Sigma$
as the quotient of the disjoint union of the products of the
framed moduli spaces\foot{The framed moduli space is
the quotient of the space of
solutions of instanton equations by the group of punctured
gauge transformations} of instantons on $\Sigma$ and
${\overline{\IC \IP}}^{2}$:
\eqn\rpr{\CM_{k, \tilde \Sigma} = \left(
\amalg_{l=0}^{k} \CM_{l, \Sigma; P} \times
\CM_{k-l, {\overline{\IC \IP}}^{2} ; P}\right) /G}
where $P$ denotes the point where the allowed gauge transformations
must be equal to one. The dimensions of the framed moduli
spaces for the group $G = SU(N)$ are equal to:
\eqn\dmns{\eqalign{{\rm dim} \CM_{l, \Sigma; P} & = 4N l  -
{{N^{2}-1}\over{2}} (\chi +\sigma -2)\cr
{\rm dim} \CM_{k-l, {\overline{\IC \IP}}^{2} ; P} & = 4N (k-l) \cr}}
The correlation functions in the theory are the integrals of certain
cohomology classes over the moduli spaces of instantons.
To get the different correlation function the observable associated
with the cycles in $\overline{\IC \IP}^{2}$ must have the ghost
charge at least $4N$. One and two $2$-observables
constructed out of $u^{k}$ for any $k=1, \ldots , N-1$
have the total ghost charge less or equal to $4N-4$.
Therefore, the correlation function does not change if we
replace $C_{k}$ by ${\tilde C}_{k}$
(see \frmor\fintstern for the mathematical proof of this result
for $G=SU(2)$)\foot{For general simple groups the number
$4N$ is replaced by $4h^{\vee}$ where $h^{\vee}$ is the dual Coxeter number,
while the maximal ghost charge of two $2$-observables is $2(2h -2)$
since the highest degree of the invariant polynomial on $\lieg$ is $h$.
For simply-laced groups $h^{\vee} = h$.}
The net result of our
manipulations is the replacement of the intersecting cycles on
the manifold $\Sigma$ by the non-intersecting cycles on the manifold
$\tilde \Sigma$.
Physically the crucial fact is that under the blowup of
a point $P$ a new two-cycle
$e$
appears and it leads to the possibility for the gauge field
to have a flux through it. In the low-energy effective theory
the insertion of the new two-cycle must be reflected in the
new factor in the Maxwell partition function, which is
the sum over all line bundles on $\overline{\IC\IP}^{2}$
in the presence of two $2$-observables $\CO^{(2)}_{u_{k}}$
and $\CO^{(2)}_{u_{l}}$ integrated
over $e$. This partition function is simply:
\eqn\cnttrmii{{{\p u_{k}}\over{\p a^{i}}}
{{\p u_{l}}\over{\p a^{j}}} {{\p}\over{\p \tau_{ij}}}
{\rm log} \Theta(\tau)}
In the rank one case it reduces to
\eqn\cnttrmiii{\left({{du}\over{da}}\right)^{2} {{\p}\over{\p \tau}}
{\rm log}{\theta}(\tau)}
after the shift $\tau \to \tau + 1$. The origin of the sign $(-1)^{\langle
\lambda, \rho \rangle}$ in $\Theta({\tau})$ is the same curious
minus sign of \wittabl. It can be derived by
examining the sign of fermion determinants by means of index
theorem if the fact that $c_{1}({\IC\IP}^{2}) = 3$
is taken into accout. By the way, the appearence of this
minus sign is  quite
analogous to the shift $\lambda \to \lambda + \rho$ in two dimensional gauge
theories \Witdgt\btverlinde\gerasimov.

One may check that this expression is again equivalent to
\cnttrm. The reasoning above does not immediately
carry over to the case of arbitrary
$2$-observables constructed out of $P_{1}(u)$ and $P_{2}(u)$.
However, assume the validity of the ``K\"unneth'' formula
inside the correlator:
$$
\langle \int_{C} {\CO}^{(2)}_{P(u)} \ldots \rangle =
\langle {\CO}^{(0)}_{dP(u)/du} \int_{C} {\CO}^{(2)}_{u}\ldots \rangle +
\langle \sum_{i,j} \eta^{ij}_{C}
\int_{L^{i}} \CO^{(1)}_{dP(u)/du} \int_{L^{j}} \CO^{(1)}_{u} \ldots
\rangle
$$
with $\Delta (C) = [pt] \otimes C \oplus C \otimes [pt] \oplus
\eta^{ij}_{C} L^{i} \otimes L^{j}$, $L^{i} \in H_{1}({\Sigma})$.
Hence we reduce this case to the one already studied.

Finally, notice that the argument with the counting
of the dimensions of framed moduli space
suggests that the contact term between two $2$-observables
$\CO^{(2)}_{u}$ in the $SU(2)$ theory with $N_{f}=1$ coincides
with \cnttrmiii. However, two $2$-observables may saturate
the dimension bound $4$ in the theory with $N_{f}=2$. Indeed,
the formula \cttrmi\ differs from \cnttrmiii\ in this case by
a one-instanton contribution. We hope to elaborate further
on this subject elsewhere.

\sssec{Blowup \quad formula.} If the $2$-observable
$\CO^{(2)}_{g}$ correspodning to the exceptional divisor is inserted in
the correlation functions then the Maxwell
partition function with the contact term included changes by a factor:
\eqn\blpfct{{{\theta_{00} (\tau, {{dg}\over{da}})}\over{
\theta_{00} (\tau, 0)}}
e^{-\left({{dg}\over{da}}\right)^2 {{\p}\over{\p \tau}}
{\rm log} \theta_{00}(\tau,0)}}
This explains the origin of the blowup factor in the formula \frmlii.
It is also equivalent to the formula of R.~Fintushel and R.~Stern \fintstern.
In the case of the theory with matter one must take into account
the curious minus sign of \wittabl\ and the change in the graviational
renormalization factor.

\sssec{Comments \quad on \quad the \quad last \quad
derivation \quad of \quad the \quad contact \quad term.}
As an attempt to formulate the
appropriate analogue of Gromov-Witten
paradigm we propose to think of the computation of
the  correlation function as of the two step procedure.
Let $g_{1}, \ldots, g_{k}$ be the zero-observables.
First of all, the topological field theory prepares
the cohomology class
\eqn\gw{\CI_{k}(g_{1} \otimes \ldots \otimes g_{k}) \in
H^{*} ( \bar M_{k, \Sigma} )}
where $\bar M_{k,\Sigma}$ is the suitable compactification
of the
moduli space of $k$ distinct points  on $\Sigma$  (an analogue
of Deligne-Mumford compactification).\foot{As an example
one may keep in mind Fulton-MacPherson's space, which is the resolution
of diagonals in $\Sigma^{k}$ for complex
$\Sigma$}.    We expect that the map
$\CI_{k}$ from the $k$'th tensor power of the space $\CH$ of
$Q$-cohomologies in the zero-observable sector
 to the cohomology of the moduli space $\bar M_{k,\Sigma}$
obeys a certain list of axioms, analogous to those listed in \konman.

Next, the actual correlation function is defined as long as
the cycle $\Theta_{\vec C} \in H_{*}(\bar M_{k,\Sigma})$ is chosen:
\eqn\corrfn{\langle
\int_{C_{1}^{l_{1}}} \left( g_{1} \right)^{(l_{1})} \ldots
\int_{C_{1}^{l_{k}}} \left( g_{k} \right)^{(l_{k})} \rangle
= \int_{\Theta_{\vec C}} \CI_{k} ( g_{1} \otimes \ldots \otimes g_{k} )}
Here ${\Theta}_{\vec C}$ is the cycle in $\bar M_{k, \Sigma}$
constructed as follows. Let $\pi: \bar M_{k,\Sigma} \to \Sigma^{k}$ be
the tautological map. Let
$\Delta$ be the big diagonal
where at least for one pair $i \neq j$, $x_{i} = x_{j}$.
The preimage of the
cycle $C_{1} \times \ldots \times C_{k} - \Delta$ is a locally finite
chain in $\bar M_{k,\Sigma}$. Its closure in $\bar M_{k,\Sigma}$
is the cycle
$\Theta_{\vec C}$.

Now suppose that all $C_{\alpha}$'s have dimensions zero and two.
It means that the representative in $H_{*}(\bar M_{k,\Sigma})$ of
$\Theta_{\vec C}$ can be chosen in such a way that its projection
$\pi ({\Theta_{\vec C}})$ intersects the big diagonal $\Delta$
only at the stratum of the lowest codimension (this is fancy
way of saying that the cycles $C_{\alpha}$ can be chosen either
intersecting each other
transversely or not intersecting at all).

Hence we must understand the topology
of $\bar M_{k,\Sigma}$ in the vicinity
of these strata. In the example of \fulmac\
it is the blowup of the diagonal $\Delta \subset \Sigma \times \Sigma$.
Since the blowup can be perfomed in the real category it is plausible
that the proper moduli space $\bar M_{k,\Sigma}$ does look
like a blowup of the diagonal near this region.
Fix one point $P \in \Sigma$. Its preimage under the forgetful map
$\bar M_{2,\Sigma} \to \bar M_{1, \Sigma} = \Sigma$ is the manifold
$\Sigma$ blown up at the point $P$. If we interpret the fiber of the
forgetful map over the point $P$ as the ``effective space-time manifold
in the presence of the observable inserted at the point $P$ (it may
be a density of some operator)'' then the appearence
of the gauge theory on the blowup becomes natural.
However,  we don't know whether these ideas can be easily generalized
to the theories with matter.

Loosely speaking, one may say that while the origin of the non-perturbative
corrections to the effective coupling comes from the
point-like instantons, the contact terms between non-local operators
come from the point-like monopoles created at the intersection points.

We shall discuss  the compatibility  of the contact terms
in the four dimensional theory and the contact terms
in the two dimensional type $\bf B$ theory in the corresponding
chapter $7$.

\subsec{Observables and contact terms: second act}

The next issue is to find the expressions for the observables consistent
with the duality in
the presence of $Q$-exact regulators.
In this section
we consider the non-trivial case of the two-observable.
Here we also meet the difficulty which we mentioned at the end
of the discussion of microscopic theory.
Two-cycles on a four-manifold typically intersect each other.
Therefore we expect to have contact terms associated with any
such intersection.

The obvious problem is that the $2$-observable,
being holomorphic in $a$ (and $\tau$) under the duality
transformation can only
by multiplied by $\tau$. Let $w$  represents the cohomology
class Poincare dual to $C^{2}$. It can be  decomposed as
$w_{+} + w_{-}$ - the sum of the self-dual and anti-selfdual parts,
thereby
$$
(\CO^{2}_{u}, w)
\sim \int F^{+}\wedge v_{+} + F^{-} \wedge v_{-} = 2{\pi} (m_{+}, v_{+}) +
2{\pi} (m_{-}, v_{-}) $$
where $v = {{du}\over{da}} w$.
The equations \fout\ suggest
that under the duality the curvature $F$ ``transforms'' as:
$$
F_{D}^{+} \sim  \bar \tau F^{+}, \quad F_{D}^{-} \sim  \tau F^{-}
$$
thereby establishing  a
contradiction between the modular invariance of $(\CO^{2}_{u} , w)$ and
its holomorphy.

The idea  is to examine the supersymmetry transformations \susy. On the
equations
of motion $H \sim F^{+}$ and therefore the problematic part of
the $2$-observable is $Q$-exact on shell.
It appears reasonable to eliminate this
part of the observable by adding something $Q$-exact to the action.
To do it carefully we write everything with the help of the auxiliary field
$H$ and eliminate it only when the final expression for the action
including the observables is constructed.
\medskip
{\quad \it Modified  action.}
\medskip
\noindent
We deform $L_{0}$ by adding to it the naive expression for two-observable:
\eqn\actdf{S_{0} = L_{0} +
\int_{\Sigma} w_{\alpha} \wedge \left( {{dg^{\alpha}}\over{da}} F +
{\half}{{d^{2}g^{\alpha}}\over{da^{2}}} \psi\psi \right)}
We also add to \actniii\ a $Q$-exact term:
\eqn\mdfac{S =   S_{0} +  \{ Q, \CR_{w,g} \} }
with
$\CR_{w,g} = \CR_{0}  - 2{{dg^{\alpha}}\over{da}} (\chi_{+}, w_{\alpha, +})$.
Computing $\{  Q, \ldots \}$ and integrating out $H$ we get:
\item{(i)} a factor $(2\tau_{2})^{\half b_{2}^{+}}$
(this is regularization independent part of the determinant -
see \wittabl\ for thorough discussion);
\item{(ii)} effective $w$-dependent interaction:
\eqn\mdfacii{ {{dg^{\alpha}}\over{da}} (w_{\alpha,-}, F)  +
{{dg^{\alpha}}\over{da}}
(w_{\alpha,+}, {{d{\rm log}{\tau}_{2}}\over{d\bar a}} \eta\chi ) +
{1\over{\tau_{2}}} (w_{\alpha,+} {{dg^{\alpha}}\over{da}}, w_{\beta,+}
{{dg^{\beta}}\over{da}} ) }
\medskip
\sssec{Contact \quad term \quad again.}
The next question is whether the modification of the naive
observable by the $Q$-exact term provides the desired modular
invariance.
To check this we must compare two results:
an integration over $F$ with fixed $A_{D}$ or integration
over $A_{D}$ (with summation over all line bundles understood)
with fixed $F$ which turns out to be $F = dA$.

The first integration yields:
\eqn\fouti{\eqalign{F^{-} &= -{1\over{\tau}} \left( F_{D}^{-} +
{{d\tau}\over{da}} (\psi^{2})^{-} + w_{\alpha, -} {{dg^{\alpha}}\over{da}} \right)\cr
F^{+} & = -{1\over{\bar \tau}} \left( F^{+}_{D} +
{{d\tau_{2}}\over{da}} \eta\chi \right)\cr}}
with the resulting action for $A_{D}$ and the rest of the fields
equal to the sum of  \actniv\ and
\eqn\mdfaciii{\eqalign{ {{dg^{\alpha}}\over{da}} (w_{\alpha,-}, F)  +
{{dg^{\alpha}}\over{da}}
(w_{\alpha,+}, {{d{\rm log}{\tau}_{2}}\over{d\bar a}} \eta\chi ) & +
{1\over{\tau_{2}}} (w_{\alpha,+} {{dg^{\alpha}}\over{da}}, w_{\beta,+}
{{dg^{\beta}}\over{da}} ) + \cr
&
+ {1\over{\tau}} (w_{\alpha} {{dg^{\alpha}}\over{da}}, w_{\beta}
{{dg^{\beta}}\over{da}} ) \cr}}
On the other hand, the integration over $A_{D}$ would simply produce
\mdfacii\ as $w$-dependent piece of the resulting action.

The difference is therefore (we temporarily do not discuss
the prefactors which come from the evaluations
of determinants):
\eqn\anmly{{1\over{\tau}} (w_{\alpha}, w_{\beta})
{{dg^{\alpha}}\over{da}}
{{dg^{\beta}}\over{da}} }
This ``anomaly'' clearly has to do with the intersections of
the two-cycles, since it is proportional
to
$(w_{\alpha}, w_{\beta})$. This is precisely the extra terms
which we saw in \exiii\ and whose cancelation
was encoded in the equations \master.
As we mentioned in the previous
section, when the cycles $C^{k}$ intersect the truly
$Q$-closed observables in the effective theory
are modified by the additions of
contact terms, attributed to the intersections of the
cycles. In our case the contact term can be guessed from the requirement
that it must be holomorphic, almost modular
invariant and vanish in the perturbative regime
$u \to \infty$ where everything can be calculated using
Feynmann diagrams\foot{We did perform this computation}.

These conditions yield the operator
(up to modular invariant operator $\CO_{\alpha\beta}^{\prime}$):
\eqn\cntctrm{\CO_{\alpha\beta} =
{{dg^{\alpha}}\over{du}}
{{dg^{\beta}}\over{du}} G(u, \tau) }
which is to be inserted at the point $p_{\alpha\beta} =
C_{\alpha} \cap C_{\beta}$. Here:
\eqn\gdva{
G(u, \tau) = -{1\over{24}} \left(
E_{2}(\tau) \left({{du}\over{da}}\right)^{2} - 8u\right)
= {1\over{\pi i}}
\left({{du}\over{da}}\right)^{2}{\p_{\tau}} {\rm log} {\theta}({\tau})}
The first form of the contact term was presented in \grwt.
The second form is valid in the pure Donaldson theory and
its r\^ole is already explained in the section devoted to the
blowups. The third form \cttrm\ was discussed in the previous section.
In order to normalize $G(u,\tau)$ properly we
use the fact that
as $u \to \infty$, $\tau \to i\infty$ and
$a \sim {1\over{\sqrt{2}}}\sqrt{u}$.

\subsec{Macroscopic measure}

As the last
check of our assertions let us try to calculate the $u$-plane integral,
contributing to
\eqn\prblm{
\langle  e^{p{\CO}^{(0)}_{u} + (\CO^{2}_{u},w)} \rangle_{\Sigma}}
We denote the contribution of the
$u$-plane to this correlator by $Z_{f}$.
We set $b_{2}^{+}=1$.

The measure which we get  has the form:
\eqn\ansmsr{
d{\mu} = ({\tau}_{2})^{-b_{2}^{+}/2}
\CD a \CD {\bar a}
\CD {\chi} \CD  {\eta} e^{S + b(u){\chi}(X) + c(u) {\sigma}(X) +
p u} }
(to compare with \wittabl\ notice that
the powers of ${\tau}_{2}$ coming from the kinetic term of
the scalars $a, \bar a$ are cancelled
by the similar terms coming from the kinetic term of
$\eta, \psi$ by supersymmetry).

Here $b(u), c(u)$ are the gravitational renormalization coefficients
computed for the low-energy $SU(2)$ theory in \wittabl:
\eqn\grvrn{e^{b(u)\chi + c(u)\sigma} =
\left( (u^{2}-1) {{d\tau}\over{du}} \right)^{{\chi}\over{4}}
(u^{2}-1)^{{\sigma}\over{8}}}
We can rewrite it using the formulae of \gz:
\eqn\grvrni{
{\theta}({\tau})^{-\chi}
\left( {{(du)^{3}}\over{(da)^{2}d\tau}}\right)^{{\chi + \sigma}\over{4}}}

In the form \grvrni\
the gravitational renormalization factor is not valid in arbitrary
abelian twisted $\CN=2$ theory, but has more
transparent physical meaning in the Seiberg-Witten
theory. The modular non-invariant part
$\theta^{-\chi}$ has the following origin.
Suppose  one performs a blowup of the manifold $\Sigma$ at some
point $p$, i.e. glues $\bar\IP^{2}$ to it. If the size of
the glued piece is small and no operators
are inserted at the point of the blowup then their correlations
functions must be unchanged. On the other hand the partition
function of Maxwell theory gets extra factor $\theta({\tau})$
from summing over the magnetic fluxes through the
two-sphere created in the blowup.
The Euler characteristics $\chi$ increases by one and therefore
the factor $\theta^{-\chi}$  cancels the extra piece
of the Maxwell partition function.

The factor
$\left( {{(du)^{3}}\over{(da)^{2}d\tau}}\right)^{{\chi + \sigma}\over{4}}$
is modular invariant. It accounts for the  $U(1)_{R}$  anomaly
in the perturbative $u \to \infty$ regime. It follows
easily from the Picard-Fuchs equations \pfmd
 (cf. \sty\matone\ for the case of pure $\CN =2$ theory)
that the  factor
$\left( {{(du)^{3}}\over{(da)^{2}d\tau}}\right)^{{\chi + \sigma}\over{4}}$
equals
$\left( {{\Delta}\over{f}} \right)^{{\chi + \sigma}\over{4}}$.

There is another gravitational correction to the effective action,
described in \wittabl, namely, if the manifold $X$ is not
spin, then there is a term:
$$
(-1)^{(w_{2}(X), m)}
$$
We can get rid of it in the course of the study of $SU(2)$ theory by
the shift $\tau \to \tau +1$ thanks to Wu formula
$$
(m,m) \equiv (w_{2}(X),m) {\rm mod}2
$$
for any $m \in H^{2}(X; {\IZ})$ (see the discussion under the formula
\cnttrmiii).

Putting all things together we arrive at
the following measure on the $u$-plane:
\eqn\msri{\eqalign{ d \mu = &d{\mu}_{0}  \quad {\Phi}(u , w^{2}) \cr
 {{d{\mu}_{0}}\over{2{\pi}i }} = {{{\CD a \CD {\bar a}
\CD {\hat \chi}}}\over{{\tau}_{2}^{b_{2}^{+}/2}
{\theta}({\tau})^{\chi}}}        &
\left( {{(du)^{3}}\over{(da)^{2}d\tau}}\right)^{{\chi + \sigma}\over{4}}
{{d{\tau}_{2}}\over{d\bar a}}
e^{{{{\pi} (v_{+}, v_{+})}\over{2 \tau_{2}}}+ p u} {\Theta} ({\tau},w)\cr
{\Theta} ({\tau}, w) = \sum_{m \in \Lambda}
\left(  m_{+} + {{v_{+}}\over{2i{\tau}_{2}}} \right) &   \eexp \bigl[
(v_{-},m_{-}) +
i{{\tau}_{2}\over{2}} \langle m, m\rangle
- {{\tau_{1}}\over{2}} (m,m) \bigr] \cr} }
Here
\eqn\frfctri{{\Phi}(u , w^{2}) =
e^{{2\pi}^{2}G ({\tau}, u) (w,w)}}

\sssec{Sample \quad computation \quad on \quad}$\Sigma = \IP^{1} \times \IP^{1}$.
The cohomology lattice of $\Sigma$ is two dimensional:
$H^{2} = {\IZ}e_{b} \oplus {\IZ}e_{f}$ and the self(anti-self)dual
components of the element $m = m_{1} e_{b} \oplus m_{2} e_{f}$ are:
\eqn\prjct{m_{\pm} = {{m_{1}}\over{2R}} \pm m_{2}R }
Here
$2R$ is the ratio of the areas of the base $\IP^{1}$ and fiber  $\IP^{1}$.
The K\"ahler form decomposes as:
\eqn\klfrm{\omega = t\left(\sqrt{2R} e_{b} \oplus {1\over{\sqrt{2R}}}
e_{f}\right)}
where $t$ measures the overall size of the manifold, while
$R$ corresponds to its ``shape''\foot{The strange notation $2R$ is
motivated by the analogy of the
integral we are computing to the one-loop computation
of the  partition
function of the string on a circle of radius $R$ \polch\hm}.
The computations of the
correlation functions of zero- and two-observables can be performed in the
large $t$ limit, where all fields can be replaced by the zero modes, i.e. by
the harmonic forms.  In particular we may drop all the fields $\psi$. Also,
the curvature of the gauge field $F$ equals $2\pi$ times the harmonic
representative of the integral cohomology element $m$.

Collecting all the terms in the measure we arrive at the formula:

\eqn\msrp{d \mu = {{d a d{\bar \tau}}\over{{\tau}_{2}^{1/2}
{\theta}({\tau})^{4}}}
\left( {{(du)^{3}}\over{(da)^{2}d\tau}}\right)
e^{{{{\pi} v_{+}^{2}}\over{2 \tau_{2}}} + 4 \pi^{2} w_{1}w_{2} G({\tau}, u) + pu}
{\Theta} ({\tau},w)}
where the function ${\Theta}  ({\tau}, w)$ is given by:
\eqn\thmsp{ \sum_{m_{1}, m_{2} \in \IZ}
\left(  {{m_{1}}\over{2R}} + m_{2}R +
{{v_{+}}\over{2i{\tau}_{2}}} \right)  \eexp \bigl[
-v_{-}({{m_{1}}\over{2R}}-m_{2}R) - {\tau}_{1} m_{1}m_{2} +
i{\tau}_{2}({{m_{1}^{2}}\over{4R^{2}}} + m_{2}^{2}R^{2}) \bigr]  }
Here
$$
v_{\pm} = {{v_{1}}\over{2R}} \pm v_{2}R
$$
and $v_{1} = {{dH}\over{da}} w_{1}$, $v_{2} = {{dV}\over{da}} w_{2}$
for the holomorphic functions $H$ and $V$ corresponding to the
``horizontal'' (base $\IP^{1}$) and ``vertical'' (fiber $\IP^{1}$)
two-observables.
It is convenient to perform Poisson resummation in $m_{2}$ (or $m_{1}$).
Denoting the dual variable by the same letter and omitting
the unimportant overall factors we get:
\eqn\msrpi{\eqalign{d\mu  = {{du \wedge d\bar\tau}\over{R^{2}\tau_{2}^{2}}}
e^{v_{1}v_{2}{\hat G} + pu}   {{du}\over{da}}
\sum_{m_{1}, m_{2}} & \left( m_{2} - m_{1}\tau - 2v_{2}R^{2} \right)
\times\cr
\times \eexp \bigl[ {i\over{4R^{2}\tau_{2}}}
\vert m_{2} - m_{1}\tau \vert^{2} &
+ {{iv_{-}}\over{2\tau_{2}R}} (m_{2} - m_{1}\bar\tau) \bigr] \cr}}
where
$$
\hat G =
-{{\pi^{2}}\over{6}} \left( {\hat E_{2}} -
8 u \left( {{da}\over{du}}\right)^{2}\right),
\quad {\hat E_{2}}  = E_{2} - {{3}\over{\pi \tau_{2}}}
$$
The sum over $(m_{1}, m_{2})$ is performed as follows. First, the
contribution of $m_{1} = m_{2} =0$ equals
\eqn\msprii{\pbar
\left( {{du}\over{w_{1}}} e^{v_{1}v_{2}{\hat G}} \right)}
Now assume  $m_{1}^{2} + m_{2}^{2} \neq 0$. Let $N$ denotes
the maximal common divisor of $m_{1}$ and $m_{2}$ is both  of them
are not zero. Otherwise $N$ denotes the one of numbers
$m_{1}$ and $m_{2}$ which does not vanish. We fix its sign
by the requirement $N<0$ if $m_{1}>0$.
Define the element of the quotient
$$
g = \pmatrix{ a & b\cr c & d\cr} + \IZ \pmatrix{ c & d\cr c & d\cr} \in SL_{2}({\IZ}) / {\IZ}
$$
by the conditions:
$m_{1} = - N c$, $m_{2} = Nd$, $c>0$.
The sum over the rest of $m_{1}$ and $m_{2}$ is performed in two steps.
First we sum over $g$ by unfolding the integration region\foot{One must
average over the quotient $SL_{2}({\IZ})/{\Gamma}^{0}(4)$
but we skip this step as it is straightforward in this case.} and
replacing it by the strip $\CH /{\IZ}$. The integrand for given
$N$ equals:
\eqn\mspriii{\pbar\left(
{{(du)^{2}}\over{N da + w_{1} dH}}  e^{v_{1}v_{2}G + p u}
\eexp \bigl[  {{i}\over{4R^{2}\tau_{2}}} (N + v_{1})(N - 2v_{2}R^{2}) \bigr]
\right)}
The final step is the integration by parts. The boundary of the integration
region has various parts. In the limit $R \to 0$ the only
surviving contribution is that of $\tau_{2} \to \infty$.
One may drop the non-holomorphic part of the $\hat G$ function
there. To prove this one integrates along the contour $\tau_{2} = const$
first (integration across the strip of \borch) and then
takes the limit $\tau_{2} \to \infty$. The integration
across the strip simply picks up a residue.

One should understand that the integral we were evaluating
is not absolutely converging. The last steps of its computation
involved a particular {\it prescription}. In fact, what is
crucial is the choice of the particular coordinate near the
cusp which determines the sequence of contours. Our
choice $\tau_{2} = const$ is dictated by the metric on the
Coulomb branch $\tau_{2} da d\bar a$ but there may be
other choices as well. Their study is important in the problem
of deformation of theory \krichever.

Thus, we arrive at the expression:
\eqn\fnmsr{\sum_{N \in \IZ}
{{(du)^{2}}\over{N da + w_{1} du}}  e^{v_{1}v_{2}G + pu}
}
The series in $N$ may be summed up to
the $cotan$ function \gz\grwt\ but we
shall keep it in this form as it is this form which reveals
the similarity to the Landau-Ginzburg model of chapter $7$.

\subsec{Contour  integral representation  of  the
correlators. }

To finish with the derivations of the formulae
for the correlation functions of all $p$-observables
announced in the beginning of the paper we need
a contour integral representation for the contribution of
singular points on the $u$-plane to the correlator.

Introduce the following Seiberg-Witten function -
\eqn\swf{\CS\CW (T) = \sum_{\ell} SW(\ell) e^{\CF(a + \psi  + \ell ; T)}}
Here $SW(\ell)$ is the
Seiberg-Witten invariant defined in the section $2$.
Then the contribution of the
singularity where $a=0$ on the $u$-plane to
the correlation function can be expressed
as the residue:
\eqn\rsdcf{\langle \ldots \rangle = \int [d\psi] \oint {{da}\over{a}} a^{{\chi + \sigma}\over{4}}
\Delta^{\sigma \over{8}} \left( {{du}\over{da}} \right)^{{\chi}\over{2}}
\CS\CW (T)}
To prove this one notices that the effective couplings
and the gravitational renormalization function
change upon including the massless field of the hypermultiplet.
This change is taken into account by the multiplication
of the effective measure by the factor:
\eqn\olp{a^{{(\ell,\ell)- {\sigma}}\over{8}} =
a^{{{{\chi}+{\sigma}}\over{4}} + {{d_{\ell}}\over{2}}}}
In the chapter $6$ we discuss this issue in more details.
The rest is done by the remark that $a + \psi + \ell$ is the
decomposition of the universal curvature form over the
product $\CM ({\ell}) \times \Sigma$, therefore the integral
over $\CM({\ell})$ is computed by picking out of the effective
measure including the factor \olp\ the term proportional
to $a^{{d_{\ell}}\over{2}}$ and replacing it by $SW(\ell)$. This
is what \rsdcf\ does\foot{One does not have to worry
about the higher observables since in this case
the moduli space $\CM({\ell})$ is compact and the K\"unneth
formula works, see below}. The formula similar to
\rsdcf\ (for all higher times set to zero
and only with $2$-observables) is written in \grwt\
where it was derived by  the analysis of the jumps
of the Coulomb branch contribution to the correlation
function on the manifolds with $b_{2}^{+}=1$ and $b_{2}^{-}  \geq 9$.
These jumps must be cancelled by the similar jumps
of the contributions of the monopole and dyon points,
anticipated in \wittmon. The comparison gives the formula
of the form similar to \rsdcf\ which by universality extends to other
cases ($b_{2}^{+} > 1$) as well.

\newsec{Generalizations I. Theories with  matter and higher dimensional
theories.}

The pure $\CN=2$ super-Yang-Mills theory has been shown \twisted\Witr\ to
be equivalent (after appropriate twisting) to the theory, computing the
Donaldson invariants of a four-fold. Donaldson theory studies
the intersections
of  certain homology classes of the moduli space of instantons.

The idea of physical approach is the following. The twisted
theory, when formulated on a general four-fold has an unbroken scalar
supercharge $Q$ which squares to zero on the gauge-invariant observables.
The action of the theory is $Q$-closed, and its
metric-dependent part is $Q$-exact. Therefore,
the correlation functions of $Q$-closed operators are metric-independent and
define the invariants of a smooth structure of manifold (for a review, see \CMR\ ).
Now, the important piece of the action to
look at is $$ {1\over{e^{2}}} \Vert F^{+} \Vert^{2} + \ldots$$
where $\ldots$ denote fermionic part, making the whole thing
$Q$-exact. By taking the limit $e^{2} \to 0$ one realizes that the
$Q$-closed observables correlators are nothing but the
integrals over instanton moduli spaces $\CM_{k}$. This argument shows
that in any $\CN=2$ twisted gauge theory the correlators
of $Q$-closed operators are certain integrals over $\CM_{k}$.

\subsec{Observables and K\"unneth formula}

The $Q$-cohomology in the sector of zero-observables
is provided by the space of all gauge invariant functions of $\phi$,
i.e.
the functions of $u \sim {\Tr}{\phi}^{2}$. Since the field
$\phi$ is the two-by-two traceless
matrix all the higher casimirs
${\Tr}{\phi}^{2r}$ are up to a multiple the powers of $u$.
Let us introduce the notation:
\eqn\chr{
Ch_{r} = {\Tr} \left( {{\phi}\over{2\pi i }} \right)^{r}}

Then the property of $\phi$ to be the traceless two-by-two matrix
translates to the equation:
\eqn\rl{Ch_{2r+1} = 0, \quad Ch_{2r} = 2^{1-r} Ch_{2}^{r},
\quad r = 0, 1,\ldots}
These equations are the classical relations in the cohomology
of $BG$ for $G = SU(2)$. In the quantum theory the
relations \rl\  might be modified. To explain what we mean let us
discuss the meaning of the observables \chr\ in the gauge theory.

One usually assumes the existence of the universal instanton $\CE$
over the product $\bar\CM_{k} \times \Sigma$ of the
compactified moduli space of instantons and the space-time manifold $\Sigma$.
The object $\CE$ is the bundle over the interior $\CM_{k}$ of
$\bar\CM_{k}$ corresponding
to the ordinary gauge connections. We assume that there are no reducible
connections. The trouble hides at the ``boundary'' of
$\bar\CM_{k}$, where the point-like instantons are situated. K.~Uhlenbeck's
compactification \uhlnb\ which simply
adds the point-like instantons
does not allow for the universal instanton to
exist as a bundle over the totality of $\CM_{k} \times \Sigma$.
\foot{One may hope to extend $\CE$ as a (perhaps, perverse)
sheaf with the condition of being torsion free and to fit
into exact sequence of sheaves:
$0 \to \CE_{0} \to \CE \to \CS \to 0$,
where $\CS$ is supported at the point-like instantons,
$\CE_{0}$ is the continuation of the universal bundle to $\CM$
by zero to the boundary and $\CE$ is the desired universal sheaf.
In some case this construction is known to work (Gieseker compactification).
The important fact which follows from the existence of the sheaf
$\CE$ is the fact that the Chern classes $Ch_{r}({\CE})$ are defined.
We {\it conjecture} that the observable
\eqn\obsrv{
\sum_{p=0}^{4} Ch_{r}^{(p)}\otimes e_{p} \in H^{*}({\CM} \times \Sigma)}
is nothing but the class $Ch_{r}({\CE})$ (the usage of complex
language is not important here).
As it was already noticed in the physical context in \avatar\
and subsequently widely used in the studies of $D$-branes
the relations \rl\ which are certainly obeyed for a rank two
bundle with trivial determinant may be violated if
the  object $\CE$ is a sheaf. The simplest type of such violation
is the relation between $Ch_{4}$ and ${\half} Ch_{2}^{2}$:
from the sequence  one gets the relation between the Chern characters.
\eqn\vltn{Ch_{4} = {\half} Ch_{2}^{2} + \delta_{1}}
where $\delta_{1}$ is the Poincare dual to the submanifold
$\CM_{k-1} \times (\Delta \subset \Sigma\times \Sigma) \subset \CM_{k}$
of the
point-like instantons of charge $1$.}
Nevertheless, the bundle $\CE$ does exists over the complement
$\CM_{k}^{\circ}$ to
the submanifold $S$ of codimension $8$ in $\bar\CM_{k} \times \Sigma$. The
submanifold $S$ is the union of
the product $\CM_{k-1}\times (\Delta \subset \Sigma \times \Sigma)$
and higher codimension strata of Uhlenbeck's compactification.
The complement to this submanifold  is the space of pairs
$(A_{k}, x)$ where $A_{k}$ is the instanton connection of charge $k$
and $x \in \Sigma$ or $(A_{k-1}, y,x)$ where $A_{k-1}$
is the instanton connection of charge $k-1$ and $x \neq y \in \Sigma$.
Here $y$ is the center of the point-like instanton. The fact that the
running point $x$ does not hit the center of the point-like
instanton is the origin of the existence of the continuation of the universal
instanton. The existence of $\CE$ over $\CM_{k}^{\circ}$ allows
to define the following element $\Theta$
of $H^{4}(\bar\CM_{k} \times \Sigma)$. Its integral over a four-cycle
represented by the simplices, which do not intersect $S$
(such a representative exists for dimensional reasons) is given by the
integral of the second Chern class of $\CE$. In other words,
the imbedding $\CM_{k}^{\circ} \to \bar \CM_{k} \times \Sigma$
induces the isomorphism of the fourth cohomology
and the class $\Theta$ is the class whose pull-back
coincides with $c_{2}({\CE}) \in H^{4}({\CM}_{k}^{\circ})$.

To summarize, we expect that the correlation functions
of $Ch_{2r}^{(p)}$ differ from those
of $2^{1-r}\left( Ch_{2}^{r} \right)^{(p)}$. To test that the
correlation functions of the latter are computed correctly we check
the validity of ``K\"unneth'' formula:
\eqn\knfrl{\langle \int_{C} \CO_{P_{1}P_{2}}^{(d_{C})} \ldots \rangle =
\langle \sum_{A, B \in H_{*}({\Sigma})} \eta_{C}^{AB}
\int_{A} \CO_{P_{1}}^{d_{A}} \int_{B} \CO_{P_{2}}^{d_{B}} \ldots\rangle}
where the structure constants $\eta^{AB}_{C}$ are the coefficients
of the decomposition of the diagonally embedded class
$C \in H_{d_{C}}({\Sigma})$
into
$H_{*}({\Sigma} \times {\Sigma})$ (hence the name of the formula):
$$
\Delta(C) = \eta^{AB}_{C} A \otimes B, \quad A \in H_{d_{A}}({\Sigma}),
\quad etc.
$$
Here $P_{1,2}$ denote the polynomials in $u$. The formula \knfrl\
has to hold if the construction of the descendants goes
via  decomposing the class
$$
P_{1}(Ch_{2}) P_{2}(Ch_{2}) \in H^{*}({\CM}\times \Sigma) \approx
H^{*}({\CM}) \otimes H^{*} ({\Sigma})
$$

The formula \knfrl\ is not guaranteed to hold true in any
topological field theory. For example, it is not valid
in two dimensional type $\bf A$ sigma model. The interested reader
may try to compute a few correlators of $(\omega^{2})^{(2)}$ in
$\IC\IP^{2}$ sigma model using \knfrl\ and compare them with
the highly non-trivial results of \konman.

\sssec{Check \quad on} $\IP^{1} \times \IP^{1}$.
We check \knfrl\ in the case of $\Sigma = \IP^{1} \times \IP^{1}$.
We have $H^{*}({\Sigma}) = {\IC}\langle 1, e_{b}, e_{f}, e_{\Sigma} \rangle$
with the multiplication law:
$$
e_{\alpha} \cdot 1 = e_{\alpha}, \quad e_{b} \cdot e_{f} = e_{f} \cdot
e_{b} = e_{\Sigma}, \quad e_{\alpha}^{2}=0, e_{\alpha} \neq 1
$$
By generalizing \fnmsr\ to the case of
two-observables constructed out the polynomials
$H(u) = \sum_{k} H_{k}u^{k}$
(corresponding to the large $\IP^{1}$) and $V(u) = \sum_{k} V_{k}u^{k}$
(corresponding
to the small $\IP^{1}$) we get the formula (the times
$T^{k, 1}$ are implicitly included):
\eqn\fnmsri{Z = \langle e^{\int_{\Sigma} e_{f} \wedge \CO^{(2)}_{H} +
e_{b} \wedge \CO^{(2)}_{V} + e_{\Sigma} \wedge \CO^{(0)}_{\Pi}} \rangle  =
\sum_{N \in \IZ}
\oint {{(du)^{2}}\over{Nda + dH}} e^{\Pi (u) + {{V^{\prime}H^{\prime}}\over{4}}
(a{{du}\over{da}} - 2u)}}
For the polynomial $P = \sum_{k} P_{k} u^{k}$ we denote by
$\nabla_{P,\alpha}$ the differential
operator $\sum_{k} P_{k} {{\p}\over{\p T^{k,\alpha}}}$.
The formula \knfrl\ is equivalent to the identity:
\eqn\vrs{{\nabla}_{P_{1}P_{2}, 1} Z =
\Bigl[ {\nabla}_{P_{1}, 1} \nabla_{P_{2}, e_{\Sigma}} +
{\nabla}_{P_{1}, e_{b}} \nabla_{P_{2}, e_{f}} + (1 \leftrightarrow 2) \Bigr]
Z }
The left hand side of \vrs\ equals:
\eqn\vrsl{\eqalign{&
\sum_{N \in \IZ} \oint {{(du)^{2}}\over{Nda + dH}} e^{\Pi (u) +
{{V^{\prime}H^{\prime}}\over{4}}
(a{{du}\over{da}} - 2u)} \times \cr
&\Bigl[ {{V^{\prime}H^{\prime}(P_{1}P_{2})^{\prime}}\over{4}}
{{\p a}\over{\p t_{1}}} {{du}\over{da}} -
{{d \left({{\p a}\over{\p t_{1}}} (P_{1}P_{2})^{\prime} \right)}\over{da}}
\left( a {{du}\over{da}} {{V^{\prime}H^{\prime}}\over{4}} +
{{Nda}\over{Nda + dH}} \right) \Bigr]\cr}}
where
$$
{{\p  a}\over{\p t_{1}}} = {1\over{4}} \left( a - 2u {{da}\over{du}} \right)
$$
The difference
$$
\Bigl[ {\nabla}_{P_{1}P_{2}, 1} -
{\nabla}_{P_{1}, 1} \nabla_{P_{2}, e_{\Sigma}} -
{\nabla}_{P_{2}, 1} \nabla_{P_{1}, e_{\Sigma}} \Bigr] Z
$$
as it follows immediately from \vrsl\ is equal to:
\eqn\vrsli{
- 2 \sum_{N \in \IZ}
\oint {{(du)^{2}}\over{Nda + dH}} e^{\Pi (u) + {{V^{\prime}H^{\prime}}\over{4}}
(a{{du}\over{da}} - 2u)} {{du}\over{da}} {{\p a}\over{\p t_{1}}}
  P_{1}^{\prime} P_{2}^{\prime} \Bigl[ a {{du}\over{da}} +
{{Nda}\over{Nda +dH}} \Bigr]}
which coincides with $\Bigl[ \nabla_{P_{1}, e_{b}} \nabla_{P_{2}, e_{f}} +
 \nabla_{P_{1}, e_{f}} \nabla_{P_{2}, e_{b}} \Bigr] Z$
(the terms in the square brackets in \vrsli\ come from the
term $V^{\prime}H^{\prime}$ in the exponential and $Nda + dH$
in the denominator of \fnmsri\ respectively).

Hence we conclude that in the infrared theory in the presented formalism
the observables $(u^{r})^{(p)}$ correspond to
$(Ch_{2}^{r})^{(p)}$ and a priori not to $(Ch_{2r})^{(p)}$.

\subsec{Theory with matter as integration of the equivariant Euler class}

To get a hint on what the correlation functions of the
descendents of $Ch_{r}$ might be let us study
the theory with fundamental matter, e.g. $N_{f}$
hypermultiplets in fundamental representation.
Its interpretation is the following.

Let $\CM_{k}$ be the uncompactified moduli space of
instantons on $\Sigma$. We again assume that
there are no reducible connections.
Let us pick for any point $m \in \CM_{k}$ its
representative gauge field $A_{m}$. Let $\CE$ be the rank two complex
bundle over $\CM_{k} \times \Sigma$ with connection $\CA$, whose
$\Sigma$ components coincide with $A_{m}$ at $\{ m \} \times \Sigma$.
This connection exists if
the only non-trivial cohomology group in Atiyah-Hitchin-Singer
complex is $H^{1} \approx T_{m}M_{k}$. In fact, if the component
of the connection $\CA$ along $\CM_{k}$ is denoted as $c$, then
the following equations hold:
\eqn\sstmi{\eqalign{Q A_{m}  & =
dm^{i} a_{i} + d_{A_{m}} c \cr
\psi =  \psi^{i} a_{i}\quad & (d_{A_{m}}^{+} \oplus d^{*}_{A_{m}} )
a_{i} = 0\cr
\phi = & Q c + {\half} [ c, c] \cr
Q \equiv & dm^{i}{{\p}\over{\p m^{i}}}\cr}}
Let $\CE_{m}$ be the restriction
of $\CE$ on $\{ m \} \times \Sigma$. We have: $c_{1}(\CE_{m}) = 0$,
$ch_{2} = c_{2}(\CE_{m}) = -k$.

Let $S^{+} \otimes \CF$
be a $Spin^{c}$ bundle over $\Sigma$,
i.e. rank two complex
bundle $S^{+} \otimes \CF$ whose projectivization
coincides with the projectivization of the spinor bundle $\IP(S^{+})$
(the latter exists even is $S^{+}$ doesn't). Strictly speaking
the  line bundle $\CF$ may not
exist. However its square $\CF^{2} = L_{c}$ is the honest
line bundle $L_{c}$
on $\Sigma$, whose first Chern class is a lift of $w_{2}({\Sigma})$:
$$
c_{1}(L_{c}) \equiv w_{2}({\Sigma}) {\rm mod}\quad 2
$$
Going from one $Spin^{c}$ structure to another one is equivalent
to multiplication of $L_{c}$ by a square of  a line bundle.
Similarly one defines $S^{-} \otimes \CF$.

There exists a Dirac-like operator $D_{A_{m}}$  which maps
the sections of $S^{+} \otimes \CF \otimes \CE_{m}$
to the sections of $S^{-} \otimes \CF \otimes \CE_{m}$.
Its index bundle $E_{k}$, $(E_{k})_{m}
= {\rm Ind} D_{A_{m}} = {\rm Ker}D_{A_{m}} -
{\rm Coker}D_{A_{m}}$
defines an element of the $K$-group of $\CM_{k}$.
The rank of $E_{k}$ is given by the index theorem:
\eqn\rin{{\rm rk}E_{k} =  \int_{\Sigma} Ch({\CE}_{m}) Ch(\CF)
{\widehat A}_{\Sigma} =
- 2k + 4(ch_{2}({\CF}) - {{\sigma}\over{8}})}
(since for $k >>0$ this quantity is negative it is $-E_{k}$
who has positive dimension). The reason for
the coefficients $2$ and $4$ in \rin\  is the fact that the
bundle $\CF$ viewed as a real bundle has the rank two
while the bundle $\CE_{m}$ has rank four.
The trouble is that when we
approach the point-like instanton the rank of $-E_{k}$ drops
by two.
The idea is to extend
$-E_{k}$ to the compactification stratum $\CM_{k-1} \times \Sigma$
corresponding to the point-like instantons of  charge one as
follows.
Physically the extra two zero modes
of the Dirac operator are localized at the point $P$
where the instanton
is going to shrink to zero and become singular in the limit.
The rest of the zero-modes can be mapped (by a gauge transformation)
to the honest  zero modes of the instanton of charge $k-1$
(which exists by Uhlenbeck's theorem  \uhlnb).

Thus we expect that the zero modes
bundle over the stratum $\CM_{k-1} \times \Sigma$ of codimension $4$
can be decomposed as $-E_{k-1} \oplus \CT_{\Sigma}$ where $\CT_{\Sigma}$
is some rank two vector bundle over $\Sigma$, whose
existence follows from the excision principle \DoKro.
In fact, let $\bar\CM_{k}$ be
the compactification of the moduli space of instantons
by adding the ideal instantons:
$$
\bar\CM_{k} = \CM_{k} \cup \CM_{k-1} \times \Sigma \cup \ldots \cup
\CM_{k-l} \times S^{l}\Sigma \ldots
$$
Let $\Delta_{l} \subset \Sigma \times S^{l}\Sigma$ be the subspace,
consisting of the pairs $(x, I = \sum_{i} \nu_{i} [x_{i}])$, s.t.
$x = x_{i}$ for some $i$. Then it is possible
 to extend the bundle $\CE_{k}$ to the complement in $\bar\CM_{k} \times
\Sigma$ to the submanifold $\bar \Delta_{l} = \CM_{k-1} \times \Delta_{1}
\cup \ldots \cup \CM_{k-l} \times \Delta_{l} \cup \ldots$:
\eqn\extn{\matrix{\bar \CE_{k}\cr \downarrow \cr \bar\CM_{k} \times
\Sigma \backslash \bar\Delta_{l}\cr}}
It seems that the index bundle of the Dirac operator coupled to
$\bar\CE_{k}$ actually extends to the bundle $\bar E_{k}$
over the whole $\bar\CM_{k}$.
Then the Chern classes of the bundle
$\bar E_{k}$ are the cohomology classes of $\bar\CM_{k}$.
Let us denote them as: $\sigma_{l} = c_{l}(\bar E_{k})$.
The theory above easily generalizes to the case of $N_{f} > 1$ copies
of $\CF$. More precisely, let
$S^{+} \otimes \CF^{\prime}$ be a $Spin^{c}$ bundle and $F$ a
$Spin(2N_{f})$ bundle over $\Sigma$. Their tensor product we
denote as $S^{+} \otimes \CF$ and symbolically $\CF = \CF^{\prime} \otimes
F$. These ``equations'' are useful when computing Chern classes.

The bundle $E_{k} = {\rm Ind}D_{\CE \otimes \CF}$ has the rank
$$
-2N_{f}k - {{\sigma}\over{2}} + 4\int_{\Sigma} ch_{2}({\CF})
$$
The bundle $\CF$ is acted on by the global group $Spin(2N_{f})$.
Let us denote the scalar generator of the equivariant cohomology
of $Spin(2N_{f})$ by $\vec m = (m_{1}, \ldots, m_{N_{f}})$.
It belongs to the Cartan subalgebra of $Spin(2N_{f})$
which is $N_{f}$ dimensional. The bundle $E_{k}$ is also
$Spin(2N_{f})$ equivariant and one may compute its
equivariant Chern character using index theorem (we do it
in the next section).

 We {\it claim} that the correlation functions in the theory with
massive matter
are given by
\eqn\crmt{
\langle \CO_{1} \ldots \CO_{p} \rangle_{N_{f}, \vec m} =
\sum_{k} \Lambda_{N_{f}}^{(4-N_{f})k} \left( \prod_{i=1}^{N_{f}} m_{i}
\right)^{{\rm rk} \bar E_{k}}
\int_{\bar\CM_{k}} \left( \prod_{i=1}^{N_{f}}
\left(\sum_{l} {{{\sigma}_{l}}\over{m_{i}^{l}}} \right)\right)
\omega_{1} \wedge \ldots \wedge \omega_{p}}
with $\omega_{k}$ being the ordinary Donaldson  classes.

In other words, {\it the inclusion of the matter
amounts to inserting into integral over $\bar \CM_{k}$ of the equivariant
Euler class of the Dirac index bundle. }

Therefore the correlation functions in the theory with massive
matter are the generating functions for the intersection numbers
of the standard Donaldson observables and the Poincare duals to
the Chern classes of the various vector bundles on $\CM_{k}$.

The twisted theory with matter has been considered by several authors
\lbstmrn\parkii\grwt. The relation of the studies of non-abelian
monopoles \lbstmrn\parkii\ to our approach is simply
the localization with respect to the global group $Spin(2N_{f})$.
Consideration of the general $Spin(2N_{f})$ bundles is the generalization
of the approach of \parkii\ where several $Spin^{c}$ structures were
studied (in our langauge it
corresponds to the  global group broken to its maximal torus).
\subsec{Subtleties of the theories with matter}

The construction with general bundle $\CF$ may seem to be too abstract.
There are at least five reasons why these
theories may (and should!) be studied.

\item{1.} If $\Sigma$ is not spin for $N_{f} >1$ one has
to study varios options: the $Spin^{c}$ structures for
different quarks may be chosen differently. Clearly this is just the
study of the bundles with the structure group being the
maximal torus of $Spin(2N_{f})$.
The twisted theory depends
on the choice of $Spin^{c}$ structure.
If the gauge group
contained a $U(1)$ factor, which acts non-trivially
on ${\rm det}({\CE})$ then we would sum over all such line bundles
and henceforth over all $Spin^{c}$ structures.
This is the case of $U(1)$ theory \wittmon. However in the $SU(2)$
case this is not true. Therefore, the theories
with matter are labelled by some extra discrete data, such as a choice of
$\CF$.

\item{2.} The second  occasion where the different $F's$ may appear is
the study of $D3$ probes in $F$-theory compactifications.
The group $Spin(2N_{f})$ is the gauge group of the $D7$ background
branes. The additional branes may create a non-trivial
instanton background for the global group on the $D3$ probe.

\item{3.} The third reason is the possibility
to learn about the topology of the universal ``bundle''
and the ``quantum cohomology'' of the classifying space $BG$.
Indeed, one may compute the Chern character of $\bar E_{k}$ and express it
through the characteristic classes of the would-be-universal instanton.
Given the knowledge of the correlation functions in the theory
with matter (see \grwt\ for the first steps in this direction)
one may deduce the intersection numbers of the Chern classes
of $\bar E_{k}$ using \crmt. The standard formulae,
relating the Chern character of $E_{k}$
and topology of $\CE$ to the Euler class of $E_{k}$ become the
perturbative expressions for the correlation function in the
physical language.  The subtle issues of the compactification of
the moduli space, extension of $E_{k}$ to the compactification
strata and defining the Euler class are resolved by summing
up the ``instanton corrections'' to the perturbative answer.
This point is completely analogous to the examples, studied in
\diss\fivedim\cssev\ so we just sketch the line of arguments.
The family index theorem gives\foot{Riemann-Roch-Grothendieck
would give the same answer with the replacement
$\CF \to \CF \otimes K_{\Sigma}^{\half}$}:
\eqn\rrg{\eqalign{ Ch\left( E_{k}  \right) = &
\int_{\Sigma} Ch({\CE}) Ch ({\CF}) {\hat A}_{\Sigma} = \cr
\int_{\Sigma} {\Tr} \left( e^{{1\over{2\pi i}} (\phi + \psi + F)} \right)
&\left( 2N_{f} + c_{1}({\CF}) +
ch_{2}({\CF})\right)
 {\widehat A}_{\Sigma}\cr}}
where we have used the standard decomposition of the curvature
of the universal instanton \Witr. To convert the Chern character
into the equivariant Euler class we use the following trick:
\eqn\rglrz{{\rm log} {\rm Eu}_{m} =
\sum_{l,i} {\rm log}(x_{l} + m_{i}) =
\lim_{s\to 0} {1\over{{\Gamma}(-s)}}
\left( {1\over{s}} \int_{0}^{\infty} {{dt}\over{t^{1+s}}}
\sum_{i} \sum_{l} e^{t(x_{l}+m_{i})} - {{A}\over{s}}\right) }
where ${A}\over{s}$ is the divergent part, contributing to the
perturbative beta function.
The sum $\sum e^{t(x + m)}$
is obtained from \rrg\ by multiplying
$\phi$ by $t$, $\psi $ by $t^{1/2}$
and by shifting $\phi$ by $m$. One should also
remember that \rrg\ gives
the Chern character of the alternated sum of the bundles:
$$
H^{0} - H^{1} + H^{2}
$$
while we need the expression for the Euler class
of $H^{1}$ (if all other cohomology groups vanish). This is another
way of saying that we need the bundle $-E_{k}$ rather then $E_{k}$.
So we
change the sign in \rrg\ and get:
\eqn\rglrzi{{\rm log} {\rm Eu}_{m}  = \int_{\Sigma} {\CO}^{(4)}_{f}
-  {{\sigma}\over{8}} \CO^{(0)}_{h}}
with
\eqn\elrcl{\eqalign{f = {\half}  \sum_{k=1}^{N_{f}}
{\Tr} \left({{{\phi} + m_{k}}\over{2\pi i}}\right)^{2} {\rm log}
({\phi} + m_{k} )  \sim &
\sum_{n=1}^{\infty} \sum_{i=1}^{N_{f}} \left( {{2\pi i}\over{m_{i}}}
\right)^{2n} {{Ch_{2n+2}}\over{n(n+1)(2n+1)}} \cr
h =  -  {\Tr} \biggl[  {\rm log} & ( m_{k} + {\phi}) \biggr] \cr
A =  \int_{\Sigma} {\CO}^{(4)}_{f_{a}} - {{\sigma}\over{4}},
& \quad f_{a} = 2{\Tr}\phi^{2} \cr}}

\item{4.} The forth reason to study more
general $\CF$'s is a possibility
to have vectormultiplets charged under $G$. We write a more general formula,
which would contain
the contributions both from the hypermultiplets and vectormultiplets:
\eqn\gnrfl{\langle \omega_{1} \ldots \omega_{l} \rangle_{V,H} =
\sum_{k} \int_{\bar\CM_{k}} \omega_{1} \wedge \ldots \wedge \omega_{l}
{{\prod_{i=1}^{N_{h}} \prod_{l_{i}} (x_{l_{i}} + m_{i})}\over{
\prod_{j=1}^{N_{v}} \prod_{l_{j}} (x_{l_{j}} + m_{j})}}}
where the notations are self-explanatory.

The  generalization of the theory containing
both hyper- and vectormultiplets
charged under the gauge group is the following.
Let $\CF$ be any bundle over $\Sigma$ and $R$ some representation of
$G$. It may be
the case that the complex of bundles (sheaf)
$$
\Omega^{0} \otimes R(\CE) \otimes \CF   \to
\Omega^{1} \otimes R(\CE) \otimes \CF  \to
\Omega^{2,+} \otimes R(\CE) \otimes \CF
$$
has non-trivial cohomology in various dimensions.
Denote the cohomology groups as
$\CE^{i}_{R,\CF} = H^{i}(\Sigma, {\CE} \otimes \CF)$.
The ranks of the bundles
$\CE^{i}_{R, \CF}$ over $\CM_{k}$ are locally constant and one may define
the ratio:
\eqn\gnrfli{
{{{\rm Eu}_{m}({\CE}^{1}_{R,\CF})}\over{
{\rm Eu}_{m}({\CE}^{0}_{R, \CF}) {\rm Eu}_{m}({\CE}^{2}_{R,\CF})}} }
The hope is that the ratio \gnrfli\ defines a cohomology
class of the compactification $\bar \CM_{k}$.

\item{5.} The proper understanding of
these issues would allow to check independently
the solutions of Seiberg and Witten \SeWi\ of
the $\CN=2$ $SU(2)$
theories with matter in the fundamental and adjoint
representations.
The theory with adjoint matter can be treated similarly. One
just plugs into \gnrfli\ the representation
$R = ad(E)$.

\subsec{Higher dimensional theories}

The torodial compactifications
of the higher dimensional theories are the particular
examples of this construction \diss.  In these cases an infinite
set of vector multiplets forms the tower of Kaluza-Klein states.
For example, a five dimensional theory
on a circle can be considered as a deformation of the four dimensional theory, the
radius $R$ of the fifth dimension being the deformation parameter.
The analogue of \crmt\ looks in this case as:
\eqn\intpri{\langle \CO_{1} \ldots \CO_{p} \rangle =
\sum_{k} \int_{\bar\CM_{k}} \omega_{1} \wedge \ldots \wedge \omega_{p} \wedge
{\hat A} ({\CM}_{k}) }
To make contact with \gnrfl\ we expand $\hat A$-genus as:
\eqn\exnah{{\hat A} ({\CM}_{k}) = \prod_{j=1}^{\infty} \prod_{l}
{1\over{x_{l} + {{2\pi j}\over{R}} } } }
Finally, one may replace the circle by a two-torus ${\bf T}^{2}$
and study the compactification on $\bf T^{2}$ of the six dimensional
$\CN=1$ theory.  Then \intpri\ generalizes to
\eqn\intprii{\langle \CO_{1} \ldots \CO_{l} \rangle =
\int_{\CM_{k}} \omega_{1} \wedge \ldots \ldots \omega_{l} \wedge E_{q,y} ({\CM}_{k}) }
where $E_{q,y}({\CM}_{k})$ is the combination of the characteristic classes of
$\CM_{k}$ entering its elliptic genus, $q = e^{2\pi i \tau}$ is the modular parameter
of the two torus, $y$ is a $U(1)$ Wilson loop, measuring the fermion number.

The formula \gnrfl\ is useful in the study
of the breaking of the gauge group $G$ to its subgroup $H$ by
the vacuum expectation value of the adjoint scalar $\phi$
which commutes with $H$. In this case the supersymmetric
configurations are the solutions to the equations:
$$
F_{A}^{+}=0, \quad A \in \lieh, \quad d_{A}\phi =0
$$
The moduli space $\CM_{H}$ of irreducible solutions of these equations
are the instantons of the group $H$. The standard localization
arguments show that the instanton measure is modified
by the equivariant Euler class of the normal bundle to $\CM_{H}$
viewed as a submanifold in $\CM_{G}$. Since the masses of the vector
multiplets  are determined by the eigenvalues of $\phi$ the
localization formulae of \atbott\ gives precisely \gnrfl.

\sssec{Four \quad dimensional \quad Verlinde \quad formula.}
Consider the five dimensional example of the previous section.
We may think of it as of the deformation of the four dimensional
theory by adding an infinite number of four-observables
to the action \diss\fivedim.
In fact,
\eqn\rglrzii{{\rm log} {\hat A}({\CM}_{k})  = \int_{\Sigma} {\CO}^{(4)}_{f}
-  {{\chi + \sigma}\over{4}} \CO^{(0)}_{h}}
with
\eqn\fddfr{\eqalign{f =  {1\over{R^{2}}} {\Tr} & \left( {\rm Li}_{3}(e^{R\phi})
+  {{\phi^{2}}\over{2}} {\rm log}
(-R^{2}e^{3} \phi^{2}) \right) = \cr
= \sum_{k=1}^{\infty} & {{B_{2k}R^{2k} Ch_{2k+2}}\over{k (2k+2)!}}\cr
h = \sum_{k=1}^{\infty} & {{B_{2k}R^{2k} Ch_{2k}}\over{k (2k)!}}\cr}}
with $B_{2k}$ being the Bernoulli numbers:
$B_{2} = {1\over{6}}, B_{4} = - {1\over{30}}, B_{6} = {1\over{42}}$ etc.
As explained in  \avatar, the correlation function
of the exponential of two-observable of ${\Tr} \phi^{2}$
in this theory gives rise to the four-dimensional
analogue of Verlinde formula:
\eqn\vrlnd{\int_{\bar \CM_{k}} e^{c_{1}(L)} {\hat A}({\CM}_{k})}
where $L$ is a line bundle over $\bar \CM_{k}$\foot{If the manifold
$\Sigma$ is not spin one replaces $L$ by $L \otimes K^{\half}$
where $K$ is the canonical bundle of some almost
complex structure on $\Sigma$. This replacement changes
$\hat A \to Td$}.
From the field theory point of view \vrlnd\ is the
expectation value of Chern-Simons observable \avatar\diss\fivedim\cssev.
One may, of course, study the simpler observables - the
descendants of the Wilson loops. These correspond to the
computations of the integrals like:
\eqn\stndfr{\langle \exp \left(\int_{C} \CO^{(2)}_{{\Tr} g} +
\CO^{(0)}_{{\Tr} g}   ( 1 - {\rm genus}(C)) \right) \rangle =
\int_{\bar \CM_{k}} e^{Ch(\CE_{C})} {\hat A}({\CM_{k}) }}
where
$$
g = P\exp\oint (A_{5} + i\varphi) dx^{5}
$$
$A_{5}$ is the fifth-component of the gauge field,
$\varphi$ is the scalar in the $\CN=1$ $d=5$ multiplet,
the bundle $\CE_{C}$ is the Dirac index bundle of the restriction
of the Dirac operator coupled to the universal instanton to
the two dimensional surface $C \subset \Sigma$.

\newsec{Generalizations II. Macroscopic theories}

In this chapter we briefly discuss the subtleties of the
low-energy effective description of the theories considered in the previous
chapter. For simplicity we study the rank one case only. Thus
the Lagrangian manifold $\CL$ is one-dimensional and is described
by the one-dimensional family of elliptic curves:
\eqn\odfml{y^{2} = 4x^{3} - g_{2}(u) x - g_{3}(u)}
with the holomorphic symplectic form on the total space of fibration:
\eqn\smplfrm{\omega = {{dx \wedge du}\over{y}}}
There is a classification of possible singularities of
elliptic fibrations due to Kodaira.
Its physical interpretation has been  discussed
 in \GaMorSe.
We are mainly interested in the singularities of $A_{k}$ and $D_{k}$ type,
which correspond to the appearence of massless particles.

In either $A$ or $D$ case the coupling constant
(elliptic modulus) exhibits a logarithimic singularity
(we denote by $z$ a local coordinate near the singularity,
the point of singularity being $z=0$) :
\eqn\singlrt{\eqalign{ A_{k} \quad {\rm case}: \quad & \tau (z) =
(k+1) {{{\rm log}(z)}\over{2\pi i}}  + \ldots, \quad M = \pmatrix{1 & k+1\cr 0 &1\cr} \cr
 D_{k} \quad {\rm case}: \quad & \tau (z) = (k-4){{{\rm log}(z) }\over{2\pi i}} + \ldots, \quad M = \pmatrix{ -1 & 4-k \cr 0 & -1\cr} \cr}}
The $A_{k}$ singularity corresponds to the $U(1)$
theory with $k+1$ massless hypermultiplets
of charge $1$, while
$D_{k}$ corresponds to the $SU(2)$ gauge theory with $k$
massless hypermultiplets in the fundamental representation.

The issue is to find the proper analogue of the numbers $SW(\ell)$
which correspond to the $A_{0}$ case.
Of course, for the $A_{k}$ and $D_{k}$ case we know the answer.
One simply should study the moduli spaces of the generalized
monopole equations and their intersection theory produces
the analogues of the Seiberg-Witten invariants. This approach looks
like  a difficult problem, though. However, there is a trick
which reduces everything to the already studied $A_{0}$ case.

Consider the $A_{k}$ case first.
The monopole equations have the schematic form:
\eqn\mneqi{\eqalign{&
F^{+}_{ij} = -{i\over{2}} \sum_{\alpha = 1}^{k+1}
{\bar M}_{\alpha}\Gamma_{ij} M^{\alpha}\cr
									 & \sum_{i} \Gamma^{i}D_{i} M^{\alpha} = 0\cr}}
where the index $\alpha$ runs from $1$ through $k+1$. Let
$\CM_{A_{k}}(\ell)$ be
the moduli space of solutions of \mneqi\ for $[F] = 2\pi i \ell \in H^{2}(
\Sigma, 2\pi i\IZ)$.
Clearly it is acted on by the global symmetry group $SU(k+1)$. The cohomology
of $\CM_{A_{k}}$ can be studied with the help of
equivariant cohomology of $\CM_{A_{k}}$ with respect to the
action of $SU(k+1)$. In particular, the integration
over $\CM_{A_{k}}$ of $SU(k+1)$-invariant classes
is reduced to the fixed points of the action of the
generic element of the maximal torus of $SU(k+1)$.
These are, in turn,
the solutions to the $A_{0}$ equations, which are obtained from
\mneqi\ by setting $M^{\alpha} = 0$ for
$\alpha \neq \beta$.
The contribution of the fixed point set $\CM^{\beta}_{A_{0}}$
is the integral over $\CM ({\ell})$ of
a ratio of the evaluation of the corresponding
equivariant class at $\CM_{A_{0}}^{\beta}({\ell})$
and the equivariant Euler class
of the normal bundle to $\CM_{A_{0}}^{\beta}({\ell})$ in
$\CM_{A_{k}}({\ell})$.
To be specific consider the first Chern class $c_{1}({\CL})$ of the
line bundle $\CL$  associated to the $U(1)$ bundle
$\CM_{A_{k}}({\ell}, P) \to \CM_{A_{k}}({\ell})$
where $\CM_{A_{k}}({\ell}, P)$ is the framed moduli space.
The class $c_{1}({\CL})$ is $SU(k+1)$-invariant. In fact, its
representative
can be rather explicitly written:
\eqn\con{c_{1}({\CL}) = {1\over{ - \Delta + \bar M_{\gamma}
M^{\gamma}}} (\delta \bar M_{\alpha} \wedge \delta M^{\alpha} )}
where $\Delta$ is the scalar Laplacian, all functions
are evaluated at the point $P$ and again we
are not careful about the exact coefficients.
Let ${\rm diag}(m_{1}, \ldots, m_{k+1})$ be the scalar generator
of the
$SU(k+1)$
equivariant cohomology, i.e. $m_{1} + \ldots + m_{k+1} = 0$.
The equivariant extension of \con\
is given by the form:
\eqn\eex{c_{1}({\CL}) + \sum_{\alpha = 1}^{k+1} m_{\alpha} H_{\alpha}^{\alpha}}
where\foot{The operator
$(-\Delta + \bar M M)$ has positive spectrum and is invertible
iff there are no abelian instantons}
\eqn\hml{H_{\alpha}^{\beta}  =
{1\over{ - \Delta + \sum_{\gamma} \bar M_{\gamma} M^{\gamma}}}
{\bar M}_{\alpha} M^{\beta} (P)}
The collection of the Hamiltonians $H_{\alpha}^{\beta}$ forms
the moment map for the $SU(k+1)$ action on $\CM_{A_{k}}({\ell})$
if \con\ is treated as a symplectic form. In order to check
that the form \eex\ is equivariantly closed one must keep
in mind that the variations $\delta A$ and $\delta M^{\beta}$ of
the solutions to \mneqi\ obey in addition the gauge fixing condition:
\eqn\gfx{d^{*} \delta A + \delta \bar M_{\gamma} M^{\gamma} - \bar M_{\gamma}
\delta M^{\gamma} = 0}
(which was used in deriving \con\ ).
In particular, one gets that the infinitesimal global rotation
generated by $H_{\alpha}^{\beta}$ goes  together with the compensating
gauge transformation:
$$
\Delta M^{\gamma} = \delta^{\gamma}_{\alpha} M^{\beta} - M^{\gamma}
{1\over{ - \Delta + \bar M_{\gamma} M^{\gamma}}}
{\bar M}_{\alpha} M^{\beta}
$$
The fixed points of the flow generated by $H_{\alpha}^{\alpha}$
are easily seen from this to have $M^{\gamma} = 0$ for $\alpha \neq \gamma$.
The value of the Hamiltonians $H_{\gamma}^{\beta}$ is computed from
\hml:
\eqn\vlhml{H_{\gamma}^{\beta} \vert_{\CM^{\alpha}_{A_{0}}} = \delta_{\gamma}^{\alpha}
\delta^{\beta}_{\alpha}}
These remarks are useful for computing the equivariant Euler class
of the normal bundle $\CN$ to $\CM_{A_{0}}^{\beta}$ in  $\CM_{A_{k}}$.
It is easy to see that the normal bundle is spanned by the solutions
to the equation
$$
D_{A} M^{\gamma} = 0
$$
with $\gamma \neq \beta$. There are ${{k}\over{4}} \left( (\ell, \ell) - \sigma
\right)$ solutions to this equation. We must understand the topology
of $\CN$. Clearly,
\eqn\dcmps{\CN = \CL \vert_{\CM_{A_{0}}} \otimes \IC^{{{k}\over{8}} \left( (\ell, \ell) - \sigma
\right)}}
as the vector bundle over $\CM_{A_{0}}$. As the equivariant bundle it
decomposes further with the result:
\eqn\dcmp{{\rm Eu}_{m} (\CN) =
\prod_{\beta \neq \alpha} \left(
c_{1}({\CL}) + m_{\alpha} - m_{\beta} \right)^{{1\over{8}} \left( (\ell, \ell) - \sigma
\right)} }
Hence
\eqn\fnl{\int_{\CM_{A_{k}}({\ell})}
e^{c_{1}({\CL}) + \sum_{\alpha} m_{\alpha} H_{\alpha}^{\alpha}} =
\sum_{\alpha} \int_{\CM_{A_{0}}} {{e^{c_{1}({\CL}) + m_{\alpha}}}\over{\prod_{\beta \neq \alpha} \left(
c_{1}({\CL}) + m_{\alpha} - m_{\beta} \right)^{{1\over{8}} \left( (\ell, \ell) - \sigma
\right)}}} }
Finally, the right hand side of \fnl\ is simplified using
the notation $SW({\ell})$ introduced in the chapter $2$ as follows:
\eqn\fnli{SW({\ell}) \oint e^{a} da P(a)^{{1\over{8}} \left( (\ell, \ell) -
\sigma
\right)} \sum_{\alpha = 1}^{k+1} \left( a - m_{\alpha}
\right)^{{{\chi + \sigma}\over{4}} -1}}
where $P(a)  = \prod_{\alpha} (a - m_{\alpha})$.
The limit $m \to 0$ exists and it implies that the integral
of $e^{c_{1}({\CL})}$ over  $\CM_{A_{k}}(\ell)$ is equal to
$(k+1)$ times $SW({\ell})$.

\sssec{Apologies.} Of course, the trick with the equivariant cohomology
$H^{*}_{SU(k+1)}({\CM}_{A_{k}})$ is nothing but the deformation of
the theory by giving bare masses to the hypermultiplets.
It serves as a further justification of the analysis which led
to \rsdcf.

\sssec{Higher \quad critical \quad points.} It is straightforward
to generalize this idea to the case of $A,D,E$ singularities.
In either case one has a global symmetry group $\IG$ of the type
$A, D,E$
and the corresponding moduli space $\CM_{\IG}$
(which remains a mystery
for the $E_{k}$ case) is acted on by $H$.
The study of $H_{\IG}^{*}({\CM}_{\IG})$
is done by the localization techniques. It is equivalent to the
unfolding
of the singularity by turning on the masses.
In the $D_{k}$ cases the global group is $SO(2k)$. By turning on $k$
masses one gets various critical points, some of which are
``mutually non-local''. The analysis of our paper carries over to
these
cases.
It would be interesting to see whether one can learn anything about
the spaces $\CM_{E_{k}}$ using our techniques \ganor. We hope to return to
this subject elsewhere.

\sssec{Nontrivial \quad bundles} $\CF$. It is interesting to
generalize
the formulae for the correlation functions in the theories
with matter to the case of non-trivial bundles $\CF$,
say with $c_{2}({\CF}) \neq 0$ (for $N_{f} > 1$).
Our proposal is the following. The formula \swf\
involves the function $\CF$. It is the function of the masses
$m_{i}$ and can be expanded in $m_{i}$ as a series. Moreover,
the coefficients of the expansion are the functions which are
invariant under the action of the Weyl group of $Spin(2N_{f})$.
Write
$$
\CF = \sum_{l} \CF_{l} p_{l}(m)
$$
where $p_{l}$ run through a basis in the space of Weyl invariant
functions on the Cartan subalgebra of $Spin(2N_{f})$.
These functions are in one-to-one correspondence with the
universal characteristic classes of the $Spin(2N_{f})$ bundles.
Replace the function $p_{l}(m)$
by the corresponding Pontryagin class. We get another
 function $\tilde\CF$
with values in $H^{*}(\Sigma)$. Substitute it into \swf\
and compute the contour integral \rsdcf. We conjecture
that this procedure yields the correct answer. The motivation
is clear - the masses and the curvature of the bundle $\CF$
are related by the $Q$-symmetry. Our formalism allows the
analytic continuation in the fermionic directions.

\sssec{Chern-Simons  \quad observable.} This is the issue
of the low-energy expression for the observable in the five dimensional
theory compactified on a circle containing Chern-Simons functional.
To explain what the problem is let us recall the expression
for the two-observable in the holomorphic approach:
\eqn\tobs{\CO^{(2)}_{u} = {{du}\over{da}} F + {\half}
{{d^{2}u}\over{da^{2}}}
\psi^{2}}
The difficulty with Chern-Simons terms is that they are not
the descendants of the gauge-invaraint functionals and moreover
give rise to the gauge-invariant observables only in
the exponential with quantized coefficients.
We try:
\eqn\try{e^{\int_{C \times S^{1}} CS(A) + \ldots }
\sim e^{\int_{C} z F + {\half} {{dz}\over{da}} \psi^{2}}}
Here $z$ is a section of a certain local system over the
Coulomb branch of the theory.
The expression \try\ allows for $z$ to be defined up to
the shifts  (which are remnants of the five-dimensional
gauge invariance):
\eqn\gge{z \to z + n, \quad n \in \IZ}
The question is
whether the modular invariance can be preserved. In fact, under the
holomorphic modular transformation the
factor
$$
e^{{{z^{2}}\over{\tau}} \# (C \cap C) }
$$
appears. The contact term $z^{2} E_{2} + \ldots$
will not work because it violates the gauge invariance \gge.
But the following contact term works:
$$
e^{{{z^{2}}\over{\tau}} \# (C \cap C) } \mapsto
\left( {{{\theta}_{00} ({\tau}, z)}\over{\theta_{00} (\tau)}}
\right)^{\# (C \cap C)}
$$
At the moment we do not know how to complete this program
by including the full set of fields $\chi, \eta$ etc.

\newsec{Two dimensional analogies}

\subsec{Reflections on type $\bf A$ sigma models}

The analogue of the theory with matter in two dimensions exists.
Consider the type $\bf A$ sigma model  with target space $B$.
Let $E$ be rank $r$ complex
hermitian vector bundle over $B$. The group $G = U(r)$ acts on $E$,
preserving
the hermitian metric on the fibers. The sigma model with the
target $E$ is probably ill-defined since the space is non-compact.
Nevertheless, consider the following {\it equivariant} model.
First we present its mathematical definition and then
discuss its physical realization.
Let $\CM_{\beta; n}$ denote the moduli space of degree
$\beta \in H_{2}(B; {\IZ})$ stable
(pseudo)-holomorphic maps
of the $n$-punctured worldsheet $\Sigma$ to $B$. There are the
bundles
(sheaves, complexes of sheaves, ...) $\CE^{i}$ ($i=0,1$) over
$\CM_{\beta;n}$
which are defined as follows. The fiber $\CE^{i}_{f}$ over the
map $f$ is the $i$'th cohomology group of the pullback
of $E$:
\eqn\shvs{\CE^{i}_{f} = H^{i}({\Sigma}, f^{*}E)}
More invariantly the sheaves $\CE^{i}$ are defined as
$\CE^{i} = R^{i}({\rm fgt}_{n+1})_{*} {\rm ev}^{*}_{n+1}E$, where
${\rm ev}_{n+1}: \CM_{\beta; n+1} \to B$ is the evaluation at
the $n+1$'st point and ${\rm fgt}_{n+1}: \CM_{\beta; n+1} \to
\CM_{\beta; n}$
is the map which forgets the position of the $n+1$'st point on
$\Sigma$.
The group $G$ acts on $\CE^{i}$ naturally. Let
$Eu_{m}(\CE^{i})$ denote the $G$-equivariant Euler classes of the
bundles
$\CE^{i}$.
Let $\omega_{1}, \ldots, \omega_{n}$ be the natural cohomology
classes of $\CM_{\beta; n}$, which were defined in the sigma model
with the target $B$. For $\phi_{1}, \ldots, \phi_{n} \in H^{*}(B)$
we write $\omega_{i} = {\rm ev}_{i}^{*}\phi_{i}$.
We {\it define} the correlation function
in the sigma model with the target $E$ the following integral:
\eqn\dfnttn{\langle \int_{\Sigma} \phi_{1}^{(2)} \ldots
\int_{\Sigma} \phi_{n}^{(2)} \rangle_{E, m, \beta} =
\int_{\CM_{\beta; n}} {{Eu_{m}(\CE^{1})}\over{Eu_{m}({\CE}^{0})}}
\omega_{1}
\wedge \ldots \omega_{n}}
{\bf Remarks}. \item{1.} The definition \dfnttn\ is motivated by the
constructions of  \givental\konman\konenu\ and \phases\plesmor.

\item{2.} If $\CE^{1} = 0$ then the formula \dfnttn\ can be
interpreted
as a fixed locus contribution to the correlation function
in the theory on the target space $X$, which is
$G$-space and the fixed locus contains $B$. In this case the bundle
$E$ may be interpreted as a normal bundle to $B \subset X$ ,
while the bundle $\CE^{0}$ is the normal bundle to the
submanifold $\CM_{d}$ in the moduli space of the stable
maps $\Sigma \to X$.

\item{3.} If $\CE^{0} = 0$ then the formula \dfnttn\ can be
interpreted
in the limit $m=0$ as a computation of the number of the curves
in $E$ in the generic almost complex structure, whose homology
class is determined by $\beta$.

In any case one may think of the set of correlation functions
\dfnttn\ as of the ``theory'' interpolating between the sigma model
on $B$ and the sigma model on $E$. In particular, by choosing
the bundle $E$ in such a way, that $c_{1}(E) = - c_{1}(B)$
one gets a non-compact Calabi-Yau manifold. In this case the
correlation functions \dfnttn\ are the ``regularized''
correlators in the theory on $B$, where the regularization
is achieved by embedding the theory into the
ultra-violet finite (= conformal) theory on Calabi-Yau.
The phase diagram
and the beta function of the two dimensional theory can be
recovered from the computation of the powers of $m$ entering the
correlation function.

The leading power of $m$ is given by
the index formula:
\eqn\indx{{\rm rk}{\CE}^{0} - {\rm rk}{\CE}^{1} = \int_{\beta}
c_{1}(E) +
{{\chi}\over{2}} r }
where $\chi$ is the Euler characteristics of $\Sigma$.
Thus,
\eqn\asym{\langle \CO_{1} \ldots \CO_{l} \rangle_{E, m, \beta}
\sim_{m \to \infty} m^{d + \chi r/2} \langle \CO_{1} \ldots \CO_{l}
\rangle_{B, \beta} }

If $\CE^{0} = 0$, then the
correlation function \dfnttn\ is the polynomial in $m$
of the degree ${\rm rk}\CE^{1}$. If $\CE^{1} = 0$ then the correlator
\dfnttn\ is the polynomial in $m^{-1}$ of the degree
${\rm rk}\CE^{0}$.

\sssec{Physical \quad realization.}  To realize the correlation
functions \dfnttn\ physically we start with the set of
fields $x^{\mu}, \rho^{\mu}, \psi^{i}_{\bar z}, \psi^{\bar i}_{z}$
of the type $\bf A$ sigma model with the target space $B$
and add the fields $X^{\mu}, \epsilon^{\mu}, \Psi^{i}_{\bar z},
\Psi^{\bar i}_{z}$
which represent the maps to $E$. Also we introduce the auxiliary
fields $p^{i}_{\bar z}$, $p^{\bar i}_{z}$ and $P^{i}_{\bar z}$,
$P^{\bar i}_{z}$ (which are usually omitted). The $Q$-operator
representing the
equivariant derivative with respect to $G$ reads as follows:
\eqn\susyv{\eqalign{Q x^{\mu} = \rho^{\mu} \quad & \quad Q \rho^{\mu}
= 0\cr
Q \psi^{i}_{\bar z} = p^{i}_{\bar z} \quad & \quad Q p^{i}_{\bar z} =
0\cr
Q X^{M} = \epsilon^{M} \quad & \quad Q \epsilon^{M} = \phi^{A}
T_{A,N}^{M} X^{N}\cr
Q \Psi^{I}_{\bar z} = P^{I}_{\bar z} \quad & \quad Q P^{I}_{\bar z} =
\phi^{A} T_{A,J}^{I} \Psi^{J}_{\bar z} \cr}}
modulo the terms involving various connections which can be
trivially
reconstructed on the general covariance principles.
The action of the theory is given by the sum of the pull-back of
the equivariant
extension of the symplectic form $\varpi_{E}$ on $E$ and the
standard $Q$-exact
terms:
\eqn\actnv{\eqalign{S = \int_{\Sigma} \omega_{\mu\nu} dx^{\mu}
\wedge dx^{\nu} + &
\int_{\Sigma} \{ Q, \epsilon^{M} X^{N} \omega_{MN} d^{2}z \} + \cr
+
\{ Q, \int_{\Sigma} \Psi^{I}_{\bar z} \left( \p_{z} X^{\bar I} -
P^{\bar I}_{z} \right) & +
\psi^{i}_{\bar z} \left( \p_{z} x^{\bar i} - p^{\bar i}_{z} \right) +
c.c.  \} \cr}}

The formula \dfnttn\ arises as a standard result of the expansion
around the zero locus of the $Q$-transformation.

Concluding this section let us make the following remarks.
Historically
the first supersymmetric models which were used to compute certain
topological invariants of manifolds or operators were precisely
of this type. Namely, the relevant $Q$-transformation
squares to the global symmetry transformation \pauldan\aniemi.
Of course, the mathematical apparatus of equivariant cohomology
corresponds precisely to that kind of constructions \atbott.
Recently the aspects of such models and their relation to
the twisted $\CN=2$ algebras with central extension was pointed out
in \lbstmrn. The compactifications of higher dimensional theories
on tori were also interpreted along these lines \diss\fivedim\cssev.

The models  \susyv\actnv\ are useful in understanding
the local mirror symmetry \locmir\instsm.

\subsec{Comparison with topological type $\bf  B$ sigma model}

\sssec{Supersymmetry \quad algebra.}
Let us look once again at the transformations \susyi\susyii.
Introduce the auxiliary fields:
\eqn\wittgauge{\theta_{i} = \Im \tau_{i\bar j} \chi^{\bar j},
\quad {\bar F}_{i} = \Im \tau_{i\bar j} H^{\bar j}}
Also, change the notations $\psi^{i} \to \rho^{i}$, $a^{i} \to X^{i}$,
$\bar a^{\bar i} \to \bar X^{\bar i}$.

Then the algebra \susyi\susyii\ assumes the form:
\eqn\susynew{\eqalign{
Q F^{i} = d\rho^{i}, \quad & \quad Q\rho^{i} = dX^{i}\cr
Q \theta_{i} = \bar F_{i} \quad & \quad Q \bar X^{\bar i}=
\eta^{\bar i}\cr}}
and $Q^{2} =0$. Except for the piece with $A_{D}$
(where $Q^{2}$ does not vanish) this algebra is identical
to the algebra of sigma model of type $\bf B$ \Witr
(before eliminating the auxiliary fields $F, \bar F$).

This analogy will turn out to be fruitful in the next sections.

\sssec{Target \quad space.}

The target space of the sigma model which is relevant for
four dimensional gauge theory can be worked out as follows.
Suppose the space-time manifold $\Sigma$ contains a two-cycle
with zero self-intersection. Then $\Sigma$ may be represented
as a fibration over a two-dimensional base $B$. Let us study
the limit of the metric on $\Sigma$ in which the size of the
generic fiber $F$ of the fibration is much smaller then
that of $B$. In this limit the gauge degrees of freedom
along the fiber $F$ are almost frozen.

Let us study this effective two dimensional theory.
For simplicity we consider the pure Yang-Mills case only.
The generalization involving matter fields or more complicated
backgrounds are straightforward. The theory turns
out to be equivalent to Landau-Ginzburg model formulated on
a target space $\CM$ defined below.

The target space $\CM$ is the moduli space of the
following triples $(E, C, P)$, where $E$ is elliptic curve,
$C$ is a one-cycle, $C \in H_{1}(E,{\IZ})$ and $P$ is a
cyclic subgroup of $E$ of the order $4$.

There exists bundle $\CS \to \CM$, whose fiber at the triple
$(E, C, P)$ is the curve $E$ itself. The total space
$S$ is endowed with holomorphic symplectic form $\omega$.

The model has superpotential $W$, whose
differential is defined canonically by the formula:
\eqn\sprptnl{dW_{(E,C,P)} = \int_{C} \omega}

The space $\CM$ is not connected. Its connected components
are labelled by an integer $N \in \IZ$: $\CM = \amalg_{N} \CM_{N}$.
The component $\CM_{N}$ consists of the triples $(E, C, P)$
where $C$ is $N$ times the primitive element of $H_{1}(E, {\IZ})$.
All components $\CM_{N}$ with $N \neq 0$ are isomorphic
to each other and to the strip $\CH / 4{\IZ} = \{ \tau \vert
\tau_{2} > 0 \}
/ ( \tau \sim \tau + 4)$. The component $\CM_{0}$ is isomorphic
to $\CH / {\Gamma}^{0}(4)$.

The holomorphic functions on $\CM$ all come from the
functions $f(u)$, where $u$ is the canonically
defined function on $\CM$, whose description follows.
The curve $E$ with the subgroup $P$ is identified with
the double cover of $\IP^{1}$ which can be written as follows:
\eqn\crv{y^{2} = x ( x^{2} - ux + {1\over{4}} )}
The subgroup $P$ is generated by the points
$x = {\half}, y = \pm \half\sqrt{1-u}$.

The space $\CS$ also splits as $\CS = \amalg_{N} \CS_{N}$,
with $\CS_{0}$ being the surface in $\IC^{3}$:
\eqn\srfce{y^{2} = x(x^{2}-ux + {1\over 4})}
The symplectic form $\omega_{0}$ on $\CS_{0}$ can be written
explicitly
as follows:
\eqn\smplfrm{\omega_{0} = {{dx \wedge dy}\over{x^{2}}} }
There exists a covering  $p_{f}: \CM \to \CM_{0}$, defined by
forgetting
$C$. It extends to the
map of the spaces $\CS \to \CS_{0}$.
The  form $\omega$ on $\CS$ is the pullback of $\omega_{0}$
with respect to $p_{f}$.

The superpotential $W$ is defined in the coordinates $u$ by
the integral:
\eqn\spr{W = N a(u) \equiv N \int_{-1}^{+1} {{y dx}\over{x^{2}}} }
It vanishes on $\CM_{0}$.

The data defining the Landau-Ginzburg model contains
the K\"ahler metric on $\CM$,
the choice of distinguished coordinates near infinities of $\CM$
along with the superpotential $W$. In our case the metric is  fixed by
special geometry:
\eqn\mtrc{g_{u\bar u} du d\bar u = \tau_{2} da d\bar a}
The distinguished coordinates are exactly the special coordinates
(each cusp goes together with a set of canonical special coordinates
corresponding to the vanishing cycles of  the fiber).
The distinguished coordinates allow to define the principal
value integrals  over $\CM$. We have met this necessity while computing
Couomb branch integrals above.

Now we claim that the gauge theory on $\Sigma$, whose intersection
form contains a factor
\eqn\fcth{H = \pmatrix{0 &1\cr 1& 0\cr}}
is equivalent in certain chamber to the type $\bf B$
topological sigma model
with the worldsheet $B$ and the target space $\CM$.
The surface $B$ in question may be singular. It is a representative
of a class $x \in H_{2}({\Sigma})$, whose self-intersection zero.
This class together with another null
class $x^{\prime}, (x^{\prime})^{2} =0$ span the two
dimensional lattice $H$ \fcth. Hence $x \cdot x^{\prime} = 1$.
We call $x^{\prime}$ the class of a fiber.

The chamber we mentioned corresponds to
the limit of the metric on $\Sigma$ in
which the size of $B$ is much larger then that of
the rest of $H_{2}({\Sigma})$. In computing a particular
combination of observables of finite ghost number one needs
some particular inequality to be fulfilled.
The ratio of the classes must be larger then a certain number
depending on the net ghost number of the
observables.

The fluxes of the gauge field strength span the lattice
$H \oplus \Gamma$,
where $\Gamma$ denotes the rest of $H^{2}({\Sigma}, {\IZ})$.
Let us replace the summation over the fluxes through $B$ by the
integral over the auxiliary field $F$ with the constraint
$\int_{B} F \in 2\pi i \IZ$ imposed by means of the delta function:
\eqn\dlt{\sum_{m \in \IZ} e^{ m \int_{B} F}}
The part of the effective action, containing  $F$
is simply:
\eqn\actnf{\int_{B} F (m + n\tau + {{dh}\over{da}} ) +
\{ Q, \ldots \} }
where $h$ is the function of $u$, whose $2$-observable
is to be integrated over $B$. The number $n$ is the flux
through the cycle, which intersects $B$ (since the
intersection form is $H \oplus \Gamma$ there is only one
such cycle). From the point of view of the theory
on $B$ the field $F$ is nothing but the auxiliary
field of the type $\bf B$ sigma model, while $m, n$ are
some extra labels. If $m=n=0$ we get a sigma model
with the target being the $u$-plane with the
superpotential $W_{0} = h$. If $m^{2} + n^{2} \neq 0$ then summation
over
all pairs $(m,n)$ with a given maximal common divisor $N$ is
equivalent to unfolding the quotient by the group $\Gamma^{0}(4)$.
Thus we get a target space $\CM_{N}$ and the superpotential
$W_{N} = N a + h$.  In the absence of the observable associated
with $h$ we get the announced superpotential \spr.
The rest of $\Gamma$ is encoded in the peculiar zero-observables
which are to be inserted at some points of $B$.
The generalization to the case of higher rank groups $G$ is straightforward
and yields the formula \sprptnl\ for the superpotential.

We conclude that the origin of the disconnectness of the target
space is the presence of the discrete degrees of
freedom in the gauge theory.

Given the identification of target spaces one may wonder
about the geometrical meaning of the various observables
of the gauge theory in terms of sigma model.
Recall that in the type $\bf B$ sigma model with the
compact target space $\bf X$ the $Q$-invariant observables
are in one-to-one correspondence with the elements of the
cohomology groups:
\eqn\tbobs{\CH = \bigoplus_{p,q} H^{p}({\bf X}, \Lambda^{q}T_{\bf X})}
In our case the target space is non-compact. This is not a completely
unknown case since the Landau-Ginzburg models are usually formulated
on the non-compact spaces and the appropriate cohomology theory is
provided by the cohomology of the operator
$$
Q = \pbar + \p W \wedge
$$
while the rest of the \tbobs\ are  the
deformations of the operator $Q$ (in particular
$H^{0}(\Lambda^{0}T_{\bf X})$
deforms $W$ itself, $H^{1}(\Lambda^{1}T_{\bf X})$ deforms
the complex structure on $\bf X$, $H^{0}({\Lambda}^{2}T_{\bf X})$
yields the deformation quantization of $\bf X$ \konrec\ and so on).

In the gauge theory on $\Sigma$ the observables correspond to the
(co-)homology of $\Sigma$ while in the sigma model they come from
the (co-)homology of $B$. For simplicity assume that the manifold
$\Sigma$ projects to $B$. The fibers
over the different points on $B$ need not to be the same.
The projection $p: \Sigma \to B$ induces a
pushforward map in cohomology $p_{*}: H^{*}({\Sigma}) \to H^{*}(B)$.
The imbedding  $i: B \to \Sigma$ yields another map
$i^{*}: H^{*}({\Sigma}) \to H^{*}(B)$. The net effect of these
fancy operations is simply the correspondence $p_{*}$ between
the $4$, $3$, $2$ observables in gauge theory and
one set of $2$, $1$, $0$ observables in sigma model (we call them
{\it vertical} observables) and the correspondence $i^{*}$
between
$2$, $1$, $0$ observables in gauge theory and another
set of $2$, $1$, $0$ observables in sigma model (which we
call {\it horizontal}).

The horizontal $0$ -observable is simply the
holomorphic function $h$  of $u$. The standard descend applied
to the function $h$ produces $1$ and $2$ -observables which turn out
to be horizontal $1$ and $2$ observables.

The vertical $0$-observable inserted at the point
$P \in B$ is the result of integration
of $2$-observable $\CO^{(2)}_{v}$
of gauge theory along the fiber $p^{-1}(P)$. Here $v$ denotes a
holomorphic function of $u$ whose second descend gives rise to
$\CO^{(2)}_{v}$.
It can be expressed in terms of the standard
fields of the sigma model (modulo irrelevant terms):
\eqn\vrtzr{{1\over{\tau_{2}}}
{{dv}\over{da}} {{d\bar\tau}\over{d\bar a}} \theta \eta }
where $\theta$ is the integral of the two-form field
$\tau_{2}\chi$ of the gauge theory along the fiber.
The structure $\eta\theta$ corresponds to the
elements of $H^{1}(\Lambda^{1}T_{\bf X})$ (cf. the indices in
\susynew).
Thus the vertical zero-observable is the Beltrami differential
responsible for the deformations of the complex structure of the
target space. Its descendants produce the deformation of the theory.

This identification allows to perform one extra check
of the formula for the contact terms by comparing it
with the known expressions for the contact terms
$$
C(\phi_{1} , \phi_{2}) =  G {1\over{Q}}
\biggr[ 1 - \Pi \biggr] \left(\phi_{1}\phi_{2} \right)
$$
where $\Pi$ is the projector onto the ``harmonic'' representative.
in the Landau-Ginzburg theory \contchl.
Unfortunately in this case it is not really an independent test,
since we
must rely on the requirement of modular invariance (it enters the
definition of the target space).

\sssec{Contact \quad terms.}
It turns out that the analogy is not as simple as it may seem
from the previous arguments. The subtle point is the behavior
of the observables under the deformations of the theory.

The two-dimensional theory is deformed by adding two-observables
to the action. These observables come from $p$-observables
in four dimensional theory, integrated over $p-2$-cycles
in the fiber. The zero-observables come from $p$-observables
in four dimensions, integrated  along $p$-dimensional
cycles in the fiber. Now, if $p+q < 4$ then in four dimensions
the corresponding cycles do not intersect each other and
no contact terms appear. The distinction between
different types of observables is not clear to
the two-dimensional observer. Therefore the sigma model
must be treated not as the conventional type $\bf B$ model
(although we don't know how to deal with the conventional sigma model
on the non-compact target space in the absence of
sufficient superpotential) but rather as a secretly
four dimensional theory. The structure
of contact terms mimics the geometry and the topology of
the compactified space of distinct points on $\Sigma$.

Nevertheless the analogy with two dimensional model suggests
the principle of universality of contact terms.

Before stating it in the full generality let
us consider the abstract two dimensional topological
theory, whose correlation functions are computed
as the integrals of Gromov-Witten invariants.
Consider the correlator of $0$ and $2$-observables
on the two-torus $E$:
\eqn\zt{
\langle \CO^{(0)}_{\phi_{1}} \int_{E} \CO^{(2)}_{\phi_{2}} \rangle
}
Consider also the correlator of two one-observables:
\eqn\oo{
\langle \int_{A} \CO^{(1)}_{\phi_{1}} \int_{B} \CO^{(1)}_{\phi_{2}}
\rangle}
where $A$ and $B$ are the basic elements of $H_{1}(E)$.
These correlators are equal as both can be represented as the
integrals
of
$$
GW_{1,2}(\phi_{1} \otimes \phi_{2} )
$$
over the generic
fiber of the projection $\bar\CM_{1,2} \to \bar\CM_{1,1}$.

Now to be specific let consider the
example of Landau-Ginzburg theory, where
it is believed that
\eqn\tz{
\langle \CO^{(0)}_{\phi_{1}} \int_{E} \CO^{(2)}_{\phi_{2}}
\rangle_{W} =
{{d}\over{dt}}\vert_{t=0}\langle \CO^{(0)}_{\phi_{1}}
\rangle_{W + t\phi_{2}} +
\langle \CO^{(0)}_{C_{W;2+0}(\phi_{1},\phi_{2})} \rangle_{W}}
with
$$
C_{W;2+0}(\phi_{1}, \phi_{2})
$$ being the contact term between the zero-observable $\phi_{1}$ and
the two-observable $\phi_{2}$.
One may rewrite \tz\ explicitly as:
\eqn\tzi{\oint {- d{\phi_{1}d\phi_{2} +
(dX)^{2} W^{\prime\prime} C_{W;2+0}(\phi_{1}, \phi_{2})}\over{dW}} }
On the other hand,       the correlator \oo\ also has the contour
integral representation:
\eqn\ooi{\oint {{- d{\phi_{1}d\phi_{2} +
(dX)^{2} W^{\prime\prime} C_{W;1+1}(\phi_{1}, \phi_{2})}\over{dW}}} }
Here the first term appears as a result of saturating the fermionic
zero modes in the one-observables $\p \phi_{i} \psi$, while
the second is the contact term.
Comparing the expressions \ooi\ and \tzi\ we arrive at the
conclusion: ``1+1 = 2+ 0'', or more drastically
that the contact term is universal.

\sssec{Flow \quad to \quad two-dimensional \quad Yang-Mills \quad
theory.}
The trick with replacing the summation over the magnetic fluxes
by an integral over the constrained field $F$ has been widely
used in the context of two-dimensional gauge theories
(see, for example \BlThlgt). The fact that this established
an equivalence with the type $\bf B$ sigma model has not
been really appreciated, though (the papers
\ken\wittgr\ point out the relation to Landau-Ginzburg theory).
Now we can easily formulate the sigma model description of
the two dimensional topological Yang-Mills theory.
It is again a sigma model with non-compact disconnected target space
$\CM$,
with the components labelled by the dominant weights $\lambda$
of the group $G$ (cf. \Witdgt).  The component
$\CM_{\lambda}$ with $\lambda \neq 0$
is isomorphic to the Cartan subalgebra ${\liet}_{\IC}$ of
the complexified Lie algebra $\lieg_{\IC}$: $\CM_{\lambda}
\approx {\liet}_{\IC}$. The component $\CM_{0}$ is the set of
semi-simple
conjugacy classes in ${\lieg}_{\IC}$: $\CM_{0} \approx
{\liet}_{\IC}/W$.
The model has the superpotential, which is equal to
\eqn\sprtn{W_{\lambda} = \langle \lambda, X \rangle, \quad X \in
\CM_{\lambda}}
Upon the deformation by the two-observable constructed out of
the gauge-invariant function $F$ the superpotential
changes to
\eqn\sprtni{W_{\lambda} = \langle \lambda, X \rangle + F(X)}
To complete the description of the model one needs
the expression for the holomorphic volume
form $dQ$ on $\CM$. It is induced from the
Haar measure on $G$. The latter yields $W$-invariant holomorphic
volume form on $\liet_{\IC}$.

As  a check of these assertions one may re-derive
E.~Witten's results \Witdgt\ by examining the handle glueing
operator:
\eqn\hndlgl{H =
{\Det} {{\p^{2} W_{\lambda}}\over{\p X^{i} \p X^{j}}}
\left({{dX^{1} \wedge \ldots \wedge dX^{r}}\over{dQ}}\right)^{2}
}

It is interesting to note that the results for the computations
of the Donaldson invariants on the manifold
$\IP^{1} \times B$\foot{This
case
wasn't studied in \bjsv\ precisely because of the
difficulties of the Donaldson theory in the $b_{2}^{+}=1$ case}
where $B$ is the Riemann surface are equal to those of the
two dimensional
gauge theory on $B$ in the limit $\Lambda \to 0$. Of course
this is an expected result since in this limit the
four dimensional instantons are forced to have vanishing
instanton charge and become simply flat connections on $B$.
In this sense the four dimensional perspective
provides one more ``deformation''
of the intersection theory on the moduli space
of flat connections on $B$. It is interesting to see
whether it coincides with the quantum cohomology\foot{
To compare it with the results of
\bjsv\ one needs to pass through an infinite
number of walls}

\subsec{Sigma model on $BG$?}
The diagram \grwtn\
seems to exist with the target space $T$ being
$BG$ - the classifying space of the (finite-dimensional)
gauge group. This idea is based on two arguments. The first argument
compares the space $\CB^{*}$ of gauge equivalence classes of
irreducible
connections  with the classifying space of the group of the gauge
transformations. The latter is homotopy equivalent to the
space of maps of $\Sigma$ into $BG$ \atbotti.
 The idea that the gauge theory shares many similar properties
with sigma model is by no means new \polyakov, but
here we suggest more concrete realization of
this idea at least in the topological setting.

The standard obstructions in defining the sensible
(topological) sigma model
with the worldsheet of dimension higher then two are
that there are no natural first order
conditions to impose on the map
which would yield  a finite-dimensional moduli space \gromov.
However, in the context
of the gauge theory there exists such a condition - the
instanton equation. The  instantons form a submanifold
$\CM^{+}$ in the infinite-dimensional manifold $\CB^{*}$.
Now, given a homotopy
$$
h: \CB^{*} \to  {\rm Map}({\Sigma}, BG)
$$
one obtains the ``submanifold of special maps ''
$h({\CM}^{+}) \subset {\rm Map}({\Sigma}, BG)$.
All the computations of the correlation functions in the
topological gauge theory can be translated into the
sigma model language. It would be interesting to pursue this
idea futher.

To conclude we restate one of the puzzles which we tried to solve in
this
paper. There are at least three definitions of the higher
characteristic
classes of the universal instanton over $\bar \CM \times \Sigma$.
One is the continuation of the harmonic representative
${\Tr} \left( {\Delta_{A}}^{-1} [ \psi , \star \psi ] \right)^{2r}$
given by the ultraviolet gauge theory to the compactification
of the moduli space of instantons. The second uses
the construction of the universal sheaf (in the complex
situation). The last simply takes the four dimensional
class $\Theta$ and raises it to the $r$'th power.
We have seen that the infrared theory is best suited for the
last definition. Nevertheless, the study of the theory
with matter suggests that the first two definitions
are also  relevant. It would be highly desirable
to understand the relation between these three and apply it
to the theories, which are not well-studied yet, such as
the compactifications of five- and six-dimensional theories
and $F$-theory vacua.
It would also be interesting to compare the formalism of the
Hamilton -Jacobi equations which we used in deriving
the contact terms to the Whitham hierarchies of \gmmm.

\centerline{\bf Acknowledgments}
We thank G.~Moore for many discussions on Donaldson theory
and related matters over the years. S.Sh. is grateful to E.~Witten
for important discussion reviving the interest in the subject.
We are grateful to A.~Gerasimov for useful comments.
We also thank A.Gorsky, A.~ Morozov and A.~ Mironov
for discussions on Whitham hierarchies.
N.~N. also thanks A.~Lawrence and C.~Vafa
for numerous
discussions on local mirror symmetry
and N.~Seiberg for the discussion on contact terms.

The research of A.~L.~ is supported partially by DOE under grant
DE-FG02-92ER40704, by
PYI grant PHY-9058501 and RFFI under grant 96-01-01101.
The research of N.~N.~ is supported by Harvard Society of Fellows,
partially by NSF under  grant
PHY-92-18167,  and partially by RFFI grant 96-02-18046.
The research of S.Sh. was supported by DOE grant DE-FG02-92ER40704,
by NSF CAREER award, by OJI award from DOE and by A.P.~Sloan
Foundation.
A.~L.~ and N.~N.~ are also partially supported by
grant 96-15-96455 for scientific schools.

\listrefs
\bye